\documentclass{aa}
\usepackage{txfonts}
\usepackage[dvips]{color}
\usepackage{graphicx}
\usepackage{natbib}
\bibpunct{(}{)}{;}{a}{}{,}
\usepackage{captcont}
\usepackage{subfigure}
\usepackage{hyperref}
\usepackage{sidecap}
\usepackage{rotating}
\usepackage{longtable}
\usepackage{caption}

\begin{document}

\title{The (w)hole survey: an unbiased sample study of transition disk candidates based on Spitzer catalogs}
\titlerunning{The (w)hole survey}
\author{N. van der Marel\inst{1,2} 
\and B.W. Verhaar\inst{1}
\and S. van Terwisga\inst{1}
\and B. Mer{\' i}n\inst{3}
\and G. Herczeg\inst{4}
\and N.F.W. Ligterink\inst{1}
\and E.F. van Dishoeck\inst{1,5}
}
\institute{Leiden Observatory, Leiden University, P.O. Box 9513, 2300 RA Leiden, the Netherlands
\and Institute for Astronomy, University of Hawaii, 2680 Woodlawn Drive, Honolulu HI 96822, USA
\and European Space Astronomy Centre (ESA), P.O. Box 78, 28691 Villanueva de la Ca\~{n}ada, Spain
\and Max-Planck-Institut f\"{u}r Extraterrestrische Physik, Giessenbachstrasse 1, 85748 Garching, Germany
\and Kavli institute, Peking University, Beijing, China
}
\date{2016}

\abstract{Understanding disk evolution and dissipation is essential for studies of planet formation. Transition disks, i.e., disks with large dust cavities and gaps, are promising candidates of active evolution. About two dozen SED-selected candidates have been confirmed to have dust cavities through millimeter interferometric imaging, but this sample is biased towards the brightest disks.
The \emph{Spitzer} surveys of nearby low-mass star forming regions have resulted in more than 4000 Young Stellar Objects (YSOs). Using color criteria we have selected a sample of $\sim$150 candidates, and an additional 40 candidates and known transition disks from the literature. The \emph{Spitzer} data were complemented by new observations at longer wavelengths, including new JCMT and APEX submillimeter photometry, and \emph{WISE} and \emph{Herschel}-PACS mid and far-infrared photometry. Furthermore, optical  spectroscopy was obtained and stellar types were  derived for 85\% of the sample, including information from the literature. The SEDs were fit to a grid of RADMC-3D disk models with a limited number of parameters: disk mass, inner disk mass, scale height and flaring, and disk cavity radius, where the latter is the main parameter of interest.
A large fraction of the targets possibly have dust cavities based on the SED. The derived cavity sizes are consistent with imaging/modeling results in the literature, where available. Trends are found with $L_{\rm disk}/L_*$ and stellar mass and a possible connection with exoplanet orbital radii. A comparison with a previous study where color observables are used (Cieza et al. 2010) reveals large overlap between their category of planet-forming disks and our transition disks with cavities.
A large number of the new transition disk candidates are suitable for follow-up observations with ALMA.}


\maketitle

\section{Introduction}
A central question in planet formation is how the optically thick protoplanetary disks around classical T Tauri stars evolve into the optically thin debris disks around older systems \citep{WilliamsCieza2011}. An important part of the evolution occurs in the transitional phase between these two regimes. Transitional disks, disks with inner dust cavities, are considered to form the evolutionary link, although it remains uncertain whether all disks go through this phase at some point during their lifetime \citep[e.g.][]{Cieza2007,Currie2009}. One of the most exciting explanations for transition disks is the presence of a young planet that has cleared out its orbit \citep{LinPapaloizou1979}. This scenario has been confirmed through the (tentative) detection of planets embedded in transition disks through direct imaging for a handful of disks \citep{KrausIreland2012,Quanz2013,Reggiani2014,Quanz2015,Sallum2015}. As it remains unclear how and at what stage planets are formed in a disk, finding them at the earliest stage and study of their environment can provide important clues on the planet formation process. For a better understanding of the role of transition disks in the disk evolution and planet formation process, a large unbiased sample of transition disks with large holes should be studied.

The transition disk fraction is thought to be 5\%-25\% depending on the definition, and with the fraction varying with stellar age \citep{Currie2011}, implying that the evolutionary path through a transition disk is either rapid or uncommon. Transitional disk candidates are traditionally identified through a deficit of infrared flux in the mid-IR spectral energy distribution (SED) \citep[e.g.][for review]{Strom1989,Calvet2002,Espaillat2014}. The deficit arises from the absence of hot small dust particles close to the star, which can be caused by either grain growth  \citep[e.g.][]{DullemondDominik2005}, photoevaporative clearing \citep[e.g.][]{Alexander2006a} or interaction with a stellar companion or recently formed planet \citep[e.g.][]{Artymowicz1994}, all processes closely linked to disk evolution. Thanks to \emph{Spitzer} mid-infrared spectroscopy surveys, a large number of transitional disks has been discovered through a minimum in the infrared part of their SED \citep[e.g.][]{Brown2007,Najita2007,Kim2009,Merin2010}. Submillimeter observations of about two dozen of the brightest disks have directly resolved large holes with pioneering interferometers, confirming their transition disk status  \citep[e.g.][]{Pietu2005,Brown2008,Brown2009,Isella2010rytau,Isella2010mwc758,Andrews2011}. The hole sizes generally match well with estimates from SED modeling, suggesting that the current interpretation and modeling of SEDs can correctly infer this parameter provided that the mid-infrared part of the SED is well covered observationally. The Atacama Large Millimeter/submillimeter Array (ALMA) has produced even sharper dust images of a small sample of transition disks with evidence for dust trapping \citep{vanderMarel2013,Casassus2013,Perez2014,Zhang2014}. ALMA has also revealed the gas distribution through CO observations, showing that substantial amounts of gas are present inside the dust cavities  \citep{Bruderer2014,vanderMarel2015-12co,SPerez2015,vanderMarel2015-isot} indicating the presence of planets. However, ALMA has so far focused on the most well-studied and brightest transition disks. A larger sample is required to derive a more general picture.

Transition disk candidates have historically been identified through a range of different criteria \citep{Brown2007,Muzerolle2010,Oliveira2010,Merin2010,Cieza2010,Cieza2012,Romero2012}, usually involving the Spitzer colors in the (mid) infrared. The availability of \emph{Spitzer} IRS spectra between 5-35 $\mu$m was crucial for classification and determination of the hole size in these studies especially in covering the 8-20 $\mu$m region where the SEDs reach their minimum but which is not well covered by the 8 and 24 micron photometry points. In recent years, far infrared \emph{Herschel} PACS and SPIRE photometry has been used to identify and characterize (transition) disks \citep[e.g][]{Ribas2013,Bustamante2015,Rebollido2015}. Other studies identified candidates by comparing the infrared part of their SEDs with the 'median' T Tauri disk SED \citep[e.g.][]{Harvey2007,Merin2008}. These studies define a separate class of transition disks as `anemic' disks: disks with homologous depletion of dust due to grain growth or settling at all radii, exhibiting a low infrared excess at all wavelengths. Furthermore, some studies distinguish between pre-transitional and transitional disks: disks with a gap (inner disk present inside the cavity) and disks with a hole \citep{Espaillat2007} although there is no obvious evolutionary connection. A `cold disk' \citep{Brown2007} refers to a transition disk with a strong deficit in the mid infrared, implying a cavity with a steep inner wall. Note that a few transition disks have been found in millimeter imaging without evidence for mid infrared dips in their SED, e.g. MWC~758 \citep{Isella2010mwc758}

Selection of candidates is sometimes followed up by radiative transfer modeling of the radial disk structure, to constrain the dust cavity size and disk mass \citep{Kim2009,Merin2010} to determine the origin of the cavity besides clearing by a companion. Increased grain growth in the inner part of disk would result in the appearance of a dust deficit in the SED \citep{DullemondDominik2005}, although this would not be visible in millimeter imaging \citep{Birnstiel2010}. Furthermore, multiplicity studies can define the origin of the cavity as circumbinary disk whereas measuring the accretion through optical H$\alpha$ can determine photoevaporative clearing  \citep{Najita2007,Espaillat2007,Cieza2010}. Theoretical work has also shown that photoevaporative clearing cannot explain the largest observed cavities and a combination of processes may be responsible \citep{OwenClarke2012,Rosotti2013}.  

Overall, the definition of a transition disk candidate remains loose and has been used in various contexts in different studies. Due to lack of a large sample of transition disks, general properties remain uncertain and it is still unclear whether the origin for all transition disk cavities is the same, or whether disks follow different evolutionary paths \citep{Cieza2007}. Also, the distribution of cavity radii is not known, while this could constrain the birth sites of giant planets before migration. The analysis of a large unbiased sample of transition disks and candidates can provide firm constraints on their general properties. \emph{Spitzer} surveys in all nearby ($<$500 pc) star-forming regions (Cores to Disks (c2d), Gould-Belt (GB) and Taurus) have provided identification and SEDs of several thousands of Young Stellar Objects (YSOs) \citep[e.g.][and references therein]{Evans2009,Rebull2010,Dunham2015}, out of which many transition disk candidates. In addition, in recent years the AllWISE catalog with mid infrared targets has become available \citep{Wright2010}, and the \emph{Herschel Space Observatory} \citep{Pilbratt2010} has observed large parts of nearby star forming regions in the far infrared. Due to the availability of \emph{Spitzer} data combined with WISE and \emph{Herschel} data, the timing is perfect for a large transition disk SED survey.

In this work, we analyze transition disk candidates selected from the \emph{Spitzer} catalogs using robust color criteria developed by \citet{Merin2010}. These criteria were developed after deep analysis of the SEDs including IRS spectra. Our sample is complemented by additional candidates and known transition disks from the literature. The SEDs are complemented with optical, new archival far infrared \emph{Herschel}, \emph{Spitzer} IRS spectra (where available) and new submillimeter observations and are modeled using the dust radiative transfer code RADMC-3D with a generic disk structure with a cavity. The main parameter of interest is the cavity size $r_{\rm cav}$. In Section \ref{sct:obs} we discuss the selection criteria of the sample and the additional observations, Section \ref{sct:results} presents the results of the observations, Section \ref{sct:modeling} discusses the modeling procedure and limitations and the resulting disk parameters and in Section \ref{sct:discussion} we discuss the robustness of the sample and comparison with previous studies. One of the aims of this study is to define a large sample of transition disk candidates with dust cavities that are large enough to be imaged in the future by ALMA ($\geq$10 AU or $\sim$0.03", for the largest distances). The resolved images of gas and dust will provide more clues on the origin of the dust cavities and the place of transition disks in disk evolution.

\section{Observations}
\label{sct:obs}

\subsection{Target selection}
\begin{table*}[!ht]
\begin{center}
\caption{Overview Spitzer papers of YSOs in star forming regions}
\label{tbl:spitzerpapers}
\begin{tabular}{llll}
\hline
\hline
\multicolumn{2}{c}{Cores to Disks (c2d)}&d (pc)&Paper\\
\hline
Ophiuchus (MIPS)&\citet{Padgett2008}&120&VII\\
Serpens&\citet{Harvey2007}&250-400\tablefootmark{a}&IX\\
Cham II&\citet{Alcala2008}&180&X\\
Lupus I,III,IV&\citet{Merin2008}&150-200&XI\\
Perseus&\citet{Young2015}&250&XII\\
WTTS (c2d)&\citet{Padgett2006,Cieza2007,Wahhaj2010}&-&\\
Disks with holes (c2d)&\citet{Merin2010}&-&\\
\hline
\multicolumn{2}{c}{Gould Belt (GB)}&&\\
\hline
IC5146&\citet{Harvey2008}&950&I\\
Cepheus&\citet{Kirk2009}&300&II\\
CrA&\citet{Peterson2011}&150&III\\
Lupus V \& VI (full) &\citet{Spezzi2011}&150&IV\\ 
Ophiuchus North&\citet{Hatchell2012}&120&V\\
Auriga&\citet{Broekhoven-Fiene2014}&450&VI\\
\hline
\multicolumn{2}{c}{Others}&&\\
\hline
$\eta$ Cham (IRAC)&\citet{Megeath2005}&97&\\
$\eta$ Cham (MIPS)&\citet{Sicilia2009}&97&\\
Cham I&\citet{Luhman2008}&160&\\
Taurus&\citet{Rebull2010,Luhman2010}&140&\\
$\lambda$ Orionis&\citet{Hernandez2010}&450&\\
Orion&\citet{Megeath2012}&450&\\
FEPS&\citet{Carpenter2008}&-&\\
\hline
\end{tabular}
\tablefoot{
\tablefoottext{a}{The distance to Serpens is uncertain, but recent VLBA observations put it at 415 pc \citep{Dzib2011}, which has been used in this study.}
}
\end{center}
\end{table*}

The c2d, GB and Taurus \emph{Spitzer} Legacy programs completed full infrared surveys using the Infrared Array Camera (IRAC; 3.6-8.0 $\mu$m) and Multiband Imaging Photometer (MIPS; 24-160 $\mu$m) in the nearby star-forming regions ($\leq$450 pc), resulting in more than 3000 identified YSOs (see Table \ref{tbl:spitzerpapers} for an overview of papers presenting the data). Several bright YSOs from the c2d survey were targeted for additional observation with the \emph{Spitzer} InfraRed Spectrograph (IRS; 5-35 $\mu$m). \citet[][hereafter M10]{Merin2010} analyzed 35 possible transition disk candidates for which IRS spectra were available in detail through SED modeling, and defined two sets of color criteria: 
\begin{eqnarray}
{\rm [A]}: 0.0 < [3.6]-[8.0] < 1.1; \nonumber \\
3.2 < [8.0]-[24.0] < 5.3;\\
{\rm [B]}: 1.1 < [3.6]-[8.0] < 1.8; \nonumber \\
3.2 < [8.0]-[24.0] < 5.3;
\end{eqnarray}
where the bracketed numbers refer to the magnitudes at the \emph{Spitzer} wavelengths. The Region A criteria select 'clean' inner holes (disks for which there is no substantial excess in any IRAC band and there is a clear signature of an inner dust hole) and the Region B criteria select disks with a clear signature of an inner dust hole, but some excess in the IRAC bands, possibly resulting from an inner disk. The latter criterion includes several of the confirmed imaged transition disks \citep{Brown2009,Andrews2009}, but may also include some disks without holes (M10). 

M10 finds one transition disk with a particularly large hole (Sz~84, object 17), which falls outside of the color criteria mentioned above. Inspection of its SED reveals a steep slope between the 24 $\mu$m and 70 $\mu$m flux. Therefore we set an additional color criterium:
\begin{eqnarray}
{\rm [L]}: 0.0 < [3.6]-[8.0] < 1.1; \nonumber \\
10.0 > [24.0]-[70.0] > 3.8;
\end{eqnarray}
In this case the MIPS-2 flux at 70 $\mu$m has to be detected rather than an upper limit. Due to the large beam size of \emph{Spitzer} at 70 $\mu$m of 18", this flux can be confused with nearby sources. The long wavelength flux thus has to be taken with extra care for the Region L criteria. The Region L targets are not mutually exclusive with the Region A criteria: some targets follow in both.

The color criteria were applied to the three main \emph{Spitzer} catalogs, listed in Table \ref{tbl:subsamples}, resulting in 153 candidates. In addition to the catalogs, we searched the literature for additional transition disk candidates, by using the color criteria on \emph{Spitzer} targets that were not included in the catalogs (row 'Other' in Table \ref{tbl:subsamples}), finding an additional 12 disks. Targets in Orion, Cepheus \citep{Kirk2009} and IC 5146 \citep{Harvey2008} are not included due to their large distances (450, 500 and 950 pc respectively). Finally, we added 7 confirmed transition disks known from resolved millimeter imaging and 21 targets that were marked as transition disk candidate by various authors, but were not yet included by the color criteria. The number of targets from various selections are listed in Table \ref{tbl:subsamples} with corresponding references. All targets in the sample are listed in Table \ref{tbl:fullsample}. Several of the color-selected targets have been identified as transition disk candidates or confirmed by millimeter imaging, as indicated in the last column of Table \ref{tbl:fullsample}.

The distance to Serpens is uncertain, with values between 250 and 400 pc \citep[discussion in e.g.][]{Oliveira2009}. However, VLBA observations have set a distance of 415 pc for the Main Cloud \citep{Dzib2010}, which has been used in more recent work \citep{Erickson2015,Ortiz-Leon2015}, and has also been used in this study.

\begin{table}[!ht]
\begin{center}
\caption{Target selection in each catalog}
\label{tbl:subsamples}
\begin{tabular}{l|lll}
\hline
\hline
Catalog/Criterion&[A]&[B]&[L]\\
\hline
c2d \citep{Evans2009}&30&34&9\\
GB \citep{Dunham2015}&25&15&31\\
Taurus \citep{Rebull2010}&7&12&6\\
Other samples\tablefootmark{a}&7&4&1\\
\hline
Additional targets\tablefootmark{b}&\multicolumn{3}{c}{7 imaging}\\
&\multicolumn{3}{c}{21 SED}\\
\hline
\end{tabular}
\end{center}
\tablefoot{
\tablefoottext{a}{Targets were selected using our color criteria in the following papers, for targets not included in the c2d/GB/Taurus catalogs: \citet{Padgett2006,Silverstone2006,Carpenter2008,Luhman2008,Kim2009,Sicilia2009,Cieza2010,Luhman2010}}
\tablefoottext{b}{Some targets were added from the literature that did not follow the color criteria. Imaging targets were taken from \citet{Pietu2006,Ohashi2008,Brown2009,Isella2010rytau,Andrews2010,Andrews2011,Rosenfeld2013,vanderMarel2013}. The other targets were identified as transition disk candidate by \citet{Megeath2005,Hernandez2007,Merin2008,Monnier2008,Hughes2008,Sicilia2008,Ireland2008,Kim2009, McClure2010,Najita2010,Espaillat2011,Furlan2011}
}
}
\end{table}

\subsection{Additional photometry}
For each target, an SED was constructed using the \emph{Spitzer} IRAC and MIPS photometry, complemented with optical B, V and R data from the NOMAD catalog \citep{Zacharias2005} and near infrared J, H and K photometry from 2MASS \citep{Cutri2003}. Reduced \emph{Spitzer} IRS low-res spectra of 5-35 $\mu$m were taken from the Cornell Atlas of Spitzer/IRS Sources (CASSIS) \citep{Lebouteiller2011} when available. For ID63 (DoAr28), the IRS spectrum in CASSIS included extended emission, a properly reduced spectrum was kindly provided by Melissa McClure \citep{McClure2010}. Unfortunately IRS spectra are not available for the entire sample, while colors only provide limited constraints on the derived cavity size. Bright isolated targets could be complemented with \emph{IRAS} photometry, especially when \emph{Spitzer} data were saturated. The \emph{Wide-field Infrared Survey Explorer (WISE)} performed an all sky survey in four wavelength bands: 3.4, 4.6, 12.0 and 22 $\mu$m leading to the AllWISE Source catalog \citep{Wright2010}. The coordinates of the targets in our sample were matched with the WISE targets (within 2") and the fluxes were added to the SEDs. Although 3 of the 4 bands overlap with \emph{Spitzer}, the 12 $\mu$m flux provides an important data point in between IRAC and MIPS wavelengths when no IRS spectra are available. Furthermore, the diffraction limited beam size of the \emph{WISE} satellite is twice as large as the \emph{Spitzer} beam (see Table \ref{tbl:beamsize}). The comparison between the \emph{WISE} 22 $\mu$m flux with the MIPS-1 24 $\mu$m flux gives an independent check of confusion at longer wavelengths: if the 22 $\mu$m flux is much larger, there is likely a nearby source that will confuse 70 $\mu$m MIPS-2 flux as well. Although the \emph{Spitzer} c2d and GB catalogs provide a quality flag on the MIPS-2 flux (MP2{\_}Q{\_}det{\_}c) for possible confusion, this independent alternative check showed more directly which targets were confused at longer wavelengths. A difference between the 22 and 24 $\mu$m flux could also originate from infrared variability, for example due to  scale height changes in the inner disk \citep[e.g.][]{Flaherty2010,Espaillat2011}. However, such variability is typically on the order of 20-40\%. Therefore, we only consider confusion if the difference in flux is  more than 50\%. The fluxes of different telescopes are taken with years in between, so without infrared monitoring there is no possibility to quantify this effect for the targets in our sample, but the effect on our SED modeling is expected to be minor. The following targets were removed from the sample due to possible confusion and their SEDs were not further analyzed: IDs 30, 32, 82, 85, 86, 88, 90, 92, 93, 95, 97, 98, 116, 123, 126, 202, 346 and 347. 

\begin{table}[!ht]
\begin{center}
\caption{Beam sizes and apertures for photometry}
\label{tbl:beamsize}
\begin{tabular}{l|lll}
\hline
\hline
Telescope&Instrument&Wavelength&Beam size/\\
&&range ($\mu$m)&Aperture(")\\
\hline
\emph{Spitzer}&IRAC&3.6,4.5,5.8,8.0&1.7--1.9\\
&MIPS&24.0,70.0&6.0,18\\
\emph{WISE}&&3.4,4.6,12,22&6.1,6.4,6.5,12\\
\emph{Herschel}&PACS&70,100,160&5.5,6.5,11\\
APEX&SABOCA&350&7.8\\
&LABOCA&870&19\\
JCMT&SCUBA-2&850&15\\
\hline
\end{tabular}
\end{center}
\end{table}

At longer wavelengths, the SEDs were complemented with (sub)millimeter data from the literature where available (see refs in Table \ref{tbl:submmphotometry}). A subsample of the remaining targets were observed with the James Clerk Maxwell Telescope (JCMT) \footnote{The James Clerk Maxwell Telescope has historically been operated by the Joint Astronomy Centre on behalf of the Science and Technology Facilities Council of the United Kingdom, the National Research Council of Canada and the Netherlands Organisation for Scientific Research. Additional funds for the construction of SCUBA-2 were provided by the Canada Foundation for Innovation. } and the Atacama Pathfinder Experiment (APEX) \footnote{This publication is based on data acquired with the Atacama Pathfinder Experiment (APEX). APEX is a collaboration between the Max-Planck-Institut fur Radioastronomie, the European Southern Observatory, and the Onsala Space Observatory.}. Targets were selected on their expected submillimeter brightness considering their 70 $\mu$m flux (brighter than $\sim$140 mJy). The details of these observations are discussed in Section \ref{sct:submmobs}. 

The SEDs were further complemented with far infrared fluxes from the \emph{Herschel} Space Observatory \citep{Pilbratt2010}. The data reduction is discussed in Section \ref{sct:pacs}.

\subsection{Submillimeter observations}
\label{sct:submmobs}
Observations of 32 of our targets were taken with the SABOCA and/or LABOCA instruments at the APEX telescope at the Chajnantor plateau in Chile. Observations were taken in service mode in 2012 and 2013 in ESO programs 089.C-0940, 090.C-0820 and 091.C-0822 and Max Planck programs M0010{\_}88 and M0003{\_}90. SABOCA is a 39-channel bolometer array operating at 350 $\mu$m \citep{Siringo2010}, LABOCA is a 295-channel bolometer array at 870 $\mu$m \citep{Siringo2009}. Imaging was performed in wobbler on-off mode. For a few sources, imaging was also performed in mapping mode (map size 1.5') to check the pointing and to check for extended emission. One source (MP Mus, ID20) was observed with the new ArTeMiS camera in mapping during its commissioning phase, operating at 350 $\mu$m \citep{Reveret2014}. 
Integration times were 5-40 minutes on source. The data were reduced using the CRUSH software \citep{Kovacs2008} and (for the wobbler observations) verified using the BoA software \citep{Schuller2012}. The results from both reduction techniques were found to agree within error bars and the CRUSH results are reported in Table \ref{tbl:apexfluxes}. Flux calibration uncertainties (not included in Table \ref{tbl:apexfluxes}) are typically 10\% for LABOCA and 25-30\% for SABOCA.

Observations of 41 of our targets were taken with the SCUBA-2 instrument at the JCMT telescope at Mauna Kea, Hawaii. Observations were taken in service mode in 2012 and 2013 in programs M12AN07, M12BN13 and M13AN01. SCUBA-2 is a 10,000 pixel bolometer camera operating simultaneously at 450 and 850 $\mu$m \citep{Holland2013}. Imaging was performed in the smallest possible map size (Daisy 3' pattern). Observations were taken in grade 3-5 weather, which is generally insufficient for observing at 450 $\mu$m, so only the 850 $\mu$m data are considered. Integration times were 5-50 minutes on source. The data were reduced using the default online pipeline. The resulting FITS images were inspected by eye for extended emission and fluxes and noise levels were derived. The noise levels were estimated by measuring the standard deviation in the map, after subtraction of point sources. The results are reported in Table \ref{tbl:jcmtfluxes}. The flux calibration uncertainty (not included in Table \ref{tbl:jcmtfluxes}) is typically 10\% for SCUBA-2.

\subsection{Herschel observations}
\label{sct:pacs}
We have searched the \emph{Herschel} Science Archive for observations with the PACS broadband photometer \citep{Poglitsch2010} at the coordinates of all targets in the sample. In photometry mode, PACS observes simultaneously at either 70 (PACS blue) and 160 $\mu$m (PACS red) or 100 (PACS green) and 160 $\mu$m. Therefore, targets are recovered in either two or three of these wavelength bands. Only data products of reduction level higher than 2.0 were used, using the high pass filter.

Photometry of the PACS data was performed using the \emph {annularSkyAperturePhotometry}-task in the Herschel Interactive Processing Environment (HIPE), version 12.1.0. This task derives background-corrected fluxes from point sources by comparing the flux inside a region centered on the point source and an annulus around it. We used the values for the aperture and annulus radii as used by \citet{Ribas2013}. The background was estimated using the DAOPhot algorithm. Errors were estimated manually at several positions near the source position, to avoid including nearby extended emission originating from clouds. The presence of nearby clouds is indicated in Table \ref{tbl:herschelphotometry}. The flux calibration uncertainty (not included in Table \ref{tbl:herschelphotometry}) is typically 5\% for PACS photometry.

\subsection{Optical spectroscopy}
Stellar properties such as the spectral type must be determined to correct for the extinction and deredden the SED flux points. The stellar luminosity is required to understand and interpret the SEDs properly. For about half of the targets in the sample, spectral types are available from the literature. The targets without known spectral type were observed with optical spectroscopy.

Optical spectra were taken for 90 targets, including reobservation of 24 targets for which the literature spectral type was still uncertain. We obtained low resolution optical spectra with the Intermediate dispersion Spectrograph and Imaging System (ISIS) on the William Herschel Telescope \footnote{The William Herschel Telescope is operated on the island of La Palma by the Isaac Newton Group in the Spanish Observatorio del Roque de los Muchachos of the Instituto de Astrof{\'i}sica de Canarias.} from 4-8 August 2012.  The D5700 dichroic splits the light at 5700 \AA\ into red and blue channels.  The red emission then passes through the GG495 filter and is dispersed by the R158R grating to generate spectra from 5600--10000 \AA\ at $R\sim1200$.  The blue emission is dispersed by the R300B grating to generate spectra from 3200--5800 \AA\ at $R\sim1800$.

R magnitudes ranged between 9 and 19 mags, requiring integration times between 1 and 60 minutes. The slit width was set each night depending on the seeing.

The spectra were reduced with custom written codes in IDL, including flatfield and cosmic ray corrections.  The wavelength calibration was obtained from arc lamp spectra.  An initial flux calibration was performed with observations of spectrophotometric standards G191 B2B, EG 274, G93-48, and LTT 6248, repeated several times each night \citep{Oke1990,Hamuy1992}

The spectral types of stars in our sample were estimated following the approximate method described by \citet{Herczeg2014}, based on the spectral compilation by \citet{Pickles1998} at early spectral types and \citet[e.g.][]{Luhman2004} for late K and M dwarfs. For K and M stars with accretion, the spectral types estimates include a rough correction for veiling from the accretion continuum. The spectral types are estimated to be accurate to a few subclasses at spectral types earlier than K5 and 0.5-1 subclass for late K and M stars. The H$\alpha$ line equivalent width was calculated by fitting a Gaussian profile to the line.

\section{Results}
\label{sct:results}
\subsection{Stellar parameters}
Spectral types as derived from our observations and taken from the literature are given in Table \ref{tbl:stellarparams}. The observations of previously characterized stars resulted generally in the same spectral types as derived before. Some of the WHT-ISIS targets did not show any lines and no spectral type could be determined: 19002346-3712242 (ID32), ISO-Oph43 (ID47), 18294721-0148301 (ID101), J182821.6+000016 (ID112), 18392594+0006382 (ID114), serp22 (124), 18401205+0029276 (ID125), Serp111 (ID131), J034219.3+314327 (ID164), J034345.17+320358.6 (ID202) and J162715.89-243843.2 (ID204), these SEDs were fit assuming a K7 star. For these stars, the temperature of 4060 K is marked as (4060) in Table \ref{tbl:stellarparams}. J18272873-0406248 (ID68), 18304127-0242335 (ID80), 18291450-0220575 (ID84), 18314110-0128035 (ID104), 18385989-0008097 (ID110), 18374209+0016519 (ID119), 18381580+0024218 (ID122) and 18381447+0035099 (ID129) turned out to be giants, these SEDs were not further analyzed. There may be additional contamination in the sample by giants, especially in Serpens. For a handful of targets, the spectral type could not be determined to subtype accuracy. This paper presents new spectral types for 85 targets. For our final sample, spectral types are known for $\sim$85\% of our targets. The uncertainty in the spectral type of a few subclasses results in less than 0.2 dex uncertainty in the bolometric luminosity, which is sufficient for our purposes of modeling the SED with a simple disk structure.

Spectral types are converted to the effective temperature $T_{\rm eff}$ using the scales in \citet{KenyonHartmann1995}. The extinction $A_V$ and stellar luminosity $L_*$ (or stellar radius $R_*$, as $L_*=4\pi R_*^2\sigma T^4$) are fit simultaneously to the SEDs, assuming the distances listed at the bottom of Table \ref{tbl:fullsample}. Kurucz models of stellar photospheres \citep{CastelliKurucz2004} are used as templates for the broadband emission. The 2MASS J-band and optical V and R band fluxes are taken as reference to constrain the fit, assuming no excess in these bands, and assuming no significant veiling or variability through accretion or extinction \citep{Cody2014,Stauffer2014}. When both V and R were missing, the extinction was estimated adopting $A_J$ = 1.53$\times E(J-K)$, where $E(J-K)$ is the observed color excess with respected to the expected photospheric color \citep{KenyonHartmann1995}, depending on its spectral type. The extinction law is parametrized as a function of wavelength assuming $R_V$=5.5 \citep{Indebetouw2005} and scaled to the visual extinction $A_V$. The resulting values are listed in Table \ref{tbl:stellarparams}. 
Stellar masses are derived by interpolation of evolutionary models of \citet{Baraffe1998} in the position of the target on the HR diagram, although these are only approximations due to the uncertainties in the spectral type. For targets that could not be fit by the Baraffe models (which only include stars up to 1 $M_{\odot}$), masses were derived using the evolutionary models by \citet{Siess2000}. Since uncertainties in stellar age are large, they are not tabulated here. We note that for the Serpens targets an alternative distance of 250 pc as used in previous work would often result in very high age estimates ($>$10 Myr), confirming that the 415 pc used here is likely more accurate \citep[also demonstrated in][]{Oliveira2009,Oliveira2013}. For 10  targets no stellar mass could be derived, suggesting that their derived stellar properties are uncertain. Most of these are targets without known spectral type or late M stars. 

The presence or absence of accretion can be assessed from the strength and shape of emission of the H$\alpha$ and other optical lines \citep[e.g.][]{WhiteBasri2003,Natta2006}. Although a proper treatment of the accretion requires simultaneous fitting of extinction, luminosity and accretion through broadband spectroscopy \citep[e.g. with X-shooter,][]{Manara2014}, as accretion also results in broadband UV/blue excess, the analysis in this study is limited to a simple designation of accretion by the width of the H$\alpha$ line. We do not aim to quantify the accretion in terms of $M_{\odot}$ yr$^{-1}$ due to the large uncertainties when deriving accretion from the line width only. Both the equivalent width EW[H$\alpha$] and the H$\alpha$ 10\% width have been used to distinguish between accretors and non-accretors, where the EW[H$\alpha$] cut-off depends on the spectral type \citep{WhiteBasri2003}. Typically, a star is classified as an accretor if the H$\alpha$ 10\% width is $>$300 km s$^{-1}$ \citep{Natta2004}, or if EW[H$\alpha$] $>3\AA$ for an early-K star, $>10\AA$ for a late-K star and $>20\AA$ for an M star. Since other studies often only list the EW[H$\alpha$] values, our accretion designation is largely based on that. 

In recent years, several YSOs have been analyzed with broadband high resolution spectroscopy, including some of the targets in our sample \citep[e.g.][]{Alcala2014,Manara2014}. This accretion information is preferred to that derived from the equivalent width as this method is more reliable, and those targets have been marked explicitly in Table \ref{tbl:stellarparams}. Accretion properties are known for 84\% of our sample: about 64\% of these targets are accreting, the remaining targets show little or no signs of accretion.

\subsection{Long wavelength photometry}
The submillimeter photometry resulted in a total of 34 detections and 39 upper limits, listed in Table \ref{tbl:apexfluxes} and \ref{tbl:jcmtfluxes}. In addition, we have taken (sub)millimeter photometry from the literature (see Table \ref{tbl:submmphotometry}). With 57 (sub)mm detections and 47 upper limits, about 50\% of the targets in our sample have constraints at longer wavelengths.

\begin{table}[!ht]
\begin{center}
\caption{APEX photometry at 350 and 870 $\mu$m for our sample.}
\label{tbl:apexfluxes}
\small
\begin{tabular}{lll|lll}
\hline
\hline
ID	&		F$_{350\mu{m}}$			&		F$_{870\mu{m}}$			&	ID	& F$_{350\mu{m}}$			&		F$_{870\mu{m}}$			\\
&(Jy)&(mJy)&&(Jy)&(mJy)\\
\hline
1	&		2.4	$\pm$	0.2		&		210	$\pm$	20		&	40	&	$<$	0.7			&	$<$	20				\\
2	&		-				&	$<$	40				&	43	&		-			&	$<$	18				\\
6	&		0.69	$\pm$	0.18		&						&	46	&		0.8	$\pm$	0.2	&		164	$\pm$	14		\\
9	&		-				&	$<$	15				&	55	&	$<$	0.9			&		92	$\pm$	6		\\
10	&		-				&	$<$	20				&	58	&		-			&	$<$	18				\\
11	&		2.4	$\pm$	0.5		&						&	62	&	$<$	0.6			&		62	$\pm$	9		\\
14	&		-				&		24	$\pm$	6		&	95	&		-			&		49	$\pm$	11		\\
15	&		4	$\pm$	0.2		&		420	$\pm$	50		&	98	&		-			&		63	$\pm$	8		\\
16	&		9.8	$\pm$	2.4		&						&	99	&		-			&		360	$\pm$	30	\tablefootmark{a}	\\
18	&		-				&		20	$\pm$	6		&	132	&		-			&		109	$\pm$	11	\tablefootmark{a}	\\
20	&		-				&		390	$\pm$	10		&	200	&		-			&	$<$	40				\\
22	&	$<$	0.3				&						&	203	&		-			&		55	$\pm$	10	\tablefootmark{a}	\\
25	&		-				&	$<$	30				&	303	&		-			&		136		7		\\
27	&		0.19	$\pm$	0.04		&						&	307	&		-			&		123	$\pm$	14		\\
35	&		-				&		28	$\pm$	4	\tablefootmark{a}	&	316	&		-			&		98	$\pm$	13		\\
36	&		0.22	$\pm$	0.06	\tablefootmark{a}	&	$<$	30				&	321	&		-			&	$<$	20				\\
	\hline
\end{tabular}
\end{center}
\tablefoot{
\tablefoottext{a}{The flux is contaminated by extended emission near the source position.}
}
\end{table}

\begin{table}[!ht]
\begin{center}
\caption{JCMT photometry at 850 $\mu$m for our sample}
\label{tbl:jcmtfluxes}
\small
\begin{tabular}{ll|ll}
\hline
\hline
ID	&		$F_{850\mu{m}}$&ID	&		$F_{850\mu{m}}$ 	\\
& (mJy)&& (mJy)\\
\hline
22	&	$<$	31			&	124	&	$<$	19			\\
23	&		153			&	127	&	$<$	29			\\
26	&	$<$	30			&	137	&		63	$\pm$	18	\\
29	&		31	$\pm$	7	&	154	&	$<$	56			\\
32	&	$<$	47			&	155	&		55	$\pm$	18	\\
36	&		73	$\pm$	18	&	156	&	$<$	31	\tablefootmark{a}		\\
44	&		38	$\pm$	11	&	161	&	$<$	115			\\
47	&	$<$	50			&	163	&	$<$	117	\tablefootmark{a}		\\
56	&	$<$	59			&	166	&		167	$\pm$	14	\\
60	&	$<$	24			&	171	&		17	$\pm$	6	\\
63	&		95	$\pm$	16	&	174	&	$<$	27			\\
70	&		32	$\pm$	11	&	178	&		222	$\pm$	16	\\
82	&	$<$	57			&	187	&		126	$\pm$	18	\\
86	&	$<$	107	\tablefootmark{a}		&	189	&		93	$\pm$	16	\\
88	&	$<$	137	\tablefootmark{a}		&	192	&	$<$	29			\\
89	&	$<$	35			&	193	&		32	$\pm$	9	\\
92	&	$<$	79			&	196	&		69	$\pm$	19	\\
100	&	$<$	37			&	202	&	$<$	56			\\
107	&	$<$	17	\tablefootmark{a}		&	333	&	$<$	56			\\
113	&	$<$	18			&		&					\\
120	&	$<$	21			&		&					\\
\hline
\end{tabular}
\end{center}
\tablefoot{
\tablefoottext{a}{The flux is contaminated by extended emission near the source position.}
}
\end{table}

\begin{table*}[!ht]
\begin{center}
\caption{Comparison PACS photometry with previous estimates}
\label{tbl:herschelcomparison}
\small
\begin{tabular}{llllllll}
\hline
\hline
ID&\multicolumn{2}{c}{F$_{70\mu{m}}$ (Jy)}&\multicolumn{2}{c}{F$_{100\mu{m}}$ (Jy)}&\multicolumn{2}{c}{F$_{160\mu{m}}$ (Jy)}&Ref\\
&This study&Previous&This study&Previous&This study&Previous&\\
\hline
4 & $<0.1$ & $<0.08$ & $<0.07$ & $<0.14$ & $<0.41$ & $<1.10$&1\\
5 & $0.18 \pm 0.05$ & $0.15 \pm 0.02$ & $0.17 \pm 0.03$ & $0.17 \pm 0.04$ & $<0.33$ & $<1.07$&1\\
6 & $3.11 \pm 0.31$ & $3.08 \pm 0.46$ & $2.90 \pm 0.29$ & $2.82 \pm 0.42$ & $2.15 \pm 0.25$ & $2.32 \pm 0.35$&1\\
7 & $0.21 \pm 0.04$ & $<0.28$ & $0.21 \pm 0.03$ & $0.21 \pm 0.01$ & $<0.31$ & $<0.32$&2\\
9 & $<0.65$ & $0.60 \pm 0.09$ & $<0.71$ & $0.77 \pm 0.12$ & $<1.06$ & $0.98 \pm 0.15$&1\\
11 & $3.86 \pm 0.39$ & $3.88 \pm 0.58$ & $3.80 \pm 0.38$ & $3.63 \pm 0.54$ & $3.65 \pm 0.37$ & $3.86 \pm 0.58$&1\\
12 & $0.44 \pm 0.05$ & $0.38 \pm 0.06$ & $0.40 \pm 0.05$ & $0.36 \pm 0.06$ & $<0.39$ & $0.20 \pm 0.03$&1\\
13 & $<0.11$ & $<0.04$ & $0.14 \pm 0.03$ & $<0.07$ & $<0.55$ & $<0.85$&1\\
14 & $0.69 \pm 0.08$ & $0.68 \pm 0.10$ & $0.55 \pm 0.06$ & $0.57 \pm 0.09$ & $0.41 \pm 0.07$ & $<0.30 \pm 0.05$&1\\
15 & $1.58 \pm 0.16$ & $1.61 \pm 0.24$ & $2.31 \pm 0.23$ & $2.19 \pm 0.33$ & $2.80 \pm 0.28$ & $2.74 \pm 0.41$&1\\
16 & $26.06 \pm 2.92$ & $25.91 \pm 3.88$ & $36.06 \pm 3.9$ & $32.32 \pm 4.85$ & $38.45 \pm 6.0$ & $27.3 \pm 4.10$&1\\
17 & $0.21 \pm 0.05$ & $<0.25$ & $0.25 \pm 0.03$ & $0.23 \pm 0.01$ & $0.30 \pm 0.09$ & $0.28 \pm 0.05$&2\\
24 & $0.17 \pm 0.04$ & $0.07 \pm 0.02$ & $0.11 \pm 0.03$ & $0.10 \pm 0.02$ & $<0.09$ & $<0.13$&3\\
25 & $<0.34$ & $0.11 \pm 0.03$ & $<0.32$ & $0.16 \pm 0.04$ & $<0.38$ & $<0.23$&3\\
26 & $<0.26$ & $0.10 \pm 0.02$ & $<0.12$ & $0.18 \pm 0.04$ & $<0.02$ & $<0.19$&3\\
27 & $0.61 \pm 0.07$ & $0.51 \pm 0.13$ & $0.80 \pm 0.08$ & $0.68 \pm 0.17$ & $0.96 \pm 0.17$ & $0.72 \pm 0.18$&3\\
179 & $1.23 \pm 0.13$ & $1.04 \pm 0.26$ & $1.41 \pm 0.14$ & $1.26 \pm 0.31$ & $1.69 \pm 0.19$ & $1.57 \pm 0.39$&3\\
185 & $0.21 \pm 0.06$ & $0.17 \pm 0.04$ & $0.24 \pm 0.05$ & $0.23 \pm 0.06$ & $<0.91$ & $0.29 \pm 0.07$&3\\
200 & $0.48 \pm 0.06$ & $0.36 \pm 0.09$ & $0.37 \pm 0.05$ & $0.37 \pm 0.09$ & $0.47 \pm 0.15$ & $0.26 \pm 0.07$&3\\
\hline
\end{tabular}
\end{center}
{\bf Refs.} 1) \citet{Ribas2013}, 2) \citet{Olofsson2013}, 3) \citet{Bustamante2015}
\end{table*}

\emph{Herschel} PACS surveys cover 92\% of our targets. The derived fluxes and upper limits are listed in Table \ref{tbl:herschelphotometry} and images of the cut out maps are given in Figure. For 152 targets at least one of the three wavelengths results in a detection. For 18 targets the emission is confused by cloud emission at all three wavelengths, for 27 only at 100 and 160 $\mu$m and for 62 targets only at 160 $\mu$m. For 25 of the targets without cloud confusion no flux is detected at any of the wavelengths.

The PACS 70 $\mu$m fluxes and upper limits are consistent with the MIPS-2 fluxes and upper limits. The PACS sensitivity is sometimes shallower than the MIPS-2, resulting in a higher upper limit. For some targets, a more thorough data reduction of the PACS data was performed in other work \citep{Ribas2013,Olofsson2013,Bustamante2015}. In Table \ref{tbl:herschelcomparison} the derived fluxes and upper limits are compared. Our values are similar within  errors with previous estimates, confirming the validity of our data reduction.

\subsection{Disk parameters}
Millimeter fluxes can be used to obtain a rough estimate of the disk mass (gas+dust) assuming optically thin dust emission and a gas-to-dust ratio of 100. Disk masses $M_{\rm disk,mm}$ in our sample are calculated following the relations presented in \citet{Cieza2008} with standard assumptions and parameters:
\begin{equation}
M_{\rm disk} = 0.17 \left(F_{1.3mm} ({\rm mJy})\times\frac{d}{\rm 140 pc}^2\right) M_{\rm Jup}
\end{equation}
\begin{equation}
M_{\rm disk} = 0.08 \left(F_{0.85mm} ({\rm mJy})\times\frac{d}{\rm 140 pc}^2\right) M_{\rm Jup}
\end{equation}
with $F_{\lambda}$ the flux at wavelength $\lambda$ and $d$ the distance to the star. Using this relation, disk masses of our sample range between $<$0.4 and 168 M$_{\rm Jup}$, and an average disk mass of 14 Jupiter masses, similar to large millimeter studies of disks \citep[e.g.][]{AndrewsWilliams2007oph}. However, these disk masses remain highly uncertain as the vertical structure, cavities and the stellar radiation field are not taken into account and the dust opacities and gas-to-dust ratio are uncertain.

Furthermore, we derive $L_{\rm disk}$ for each target by integrating over all data points after subtraction of the fitted stellar photosphere. The ratio $L_{\rm disk}/L_*$ is a measure of disk processing, as it traces the total amount of dust that is reprocessing stellar light. As disks become more tenuous, settle and eventually disappear, $L_{\rm disk}$ is expected to decline. The majority of the disks have $0.001<L_{\rm disk}/L_*<0.4$, as expected for flared disks. Disks with $L_{\rm disk}/L_*<10^{-3}$ are generally considered as debris disks \citep[e.g.][]{Wahhaj2010}. On the other hand, targets with $L_{\rm disk}/L_*>>1$ are either embedded Class I objects or edge-on disks which are more difficult to analyze \citep{Merin2010}. ID178 has $L_{\rm disk}/L_*\sim17$ and is thus removed from the analyzed sample. 

Both $L_{\rm disk}/L_*$ and $M_{\rm disk,mm}$ are listed in Table \ref{tbl:diskparams}. The final sample consists of 184 targets for which the SEDs will be analyzed.

\section{Modeling}
\label{sct:modeling}
In order to determine the presence of a dust cavity and measure its size, the SEDs are modeled using the dust radiative transfer code RADMC-3D \footnote{\url{www.ita.uni-heidelberg.de/~dullemond/software/radmc-3d/}} \citep{DullemondDominik2004}. This code performs a Monte Carlo continuum radiative transfer calculation based on the input dust density profile and stellar photosphere, followed by raytracing of the SED. The model has a large number of input parameters and we have fixed as many as possible that are not important for our science goals. The model assumes a passive disk which reprocesses the stellar radiation field. 

The modeling procedure consists of two steps: first using a rough grid with a broad range of parameters, followed by a finer grid for the specific stellar type. The modeling was performed blindly, without taking any results from previous SED modeling or imaging studies, for an uniform approach for each disk in this sample. 
In Section \ref{sct:discussion} the derived parameters are compared with previously found results.

\subsection{Approach}

\begin{figure}[!ht]\begin{center}
\includegraphics[scale=0.5]{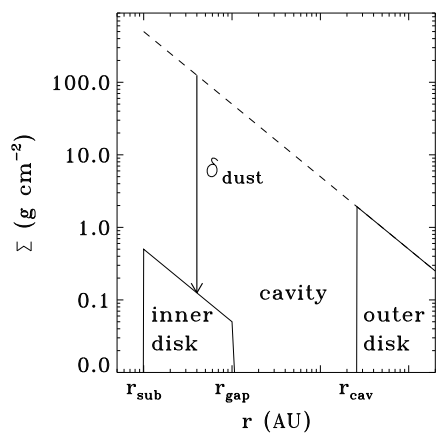}\end{center}
\caption{Gas surface density profile used for the modeling, assuming a gas-to-dust ratio of 100.}
\label{fig:duststruct}
\end{figure}

The disks are modeled using a large grid of models, computed by RADMC-3D. The model assumes an axisymmetric gas surface density profile, following a radial power-law
\begin{equation}
\Sigma_g(r)={\rm GDR}\cdot\Sigma_c\left(\frac{r}{r_c}\right)^{-1}
\end{equation}
with $r_c$ the characteristic radius and GDR the gas-to-dust ratio set to 100. The outer radius is set to 200 AU and the inner radius to the sublimation radius $r_{\rm sub}$ with $r_{\rm sub}=0.07(L_*/L_{\odot})^{1/2}$, assuming a sublimation temperature of 1500 K \citep{Dullemond2001}. The characteristic radius $r_c$ is set to 25 AU. The dust density inside the cavity is parametrized by setting the density equal to zero between $r_{\rm gap}$ and $r_{\rm cav}$. The inner disk (between $r_{\rm sub}$ and $r_{\rm gap}$) is set by varying $\delta_{\rm dust}$ to fit the near infrared excess (see Figure \ref{fig:duststruct}). The $r_{\rm gap}$ is fixed to 1 AU as it can not be constrained well by the SED. A full disk without a cavity is simulated by setting $r_{\rm cav}=r_{\rm gap}$.

\begin{figure*}[!ht]
\begin{center}
\includegraphics[width=0.95\textwidth]{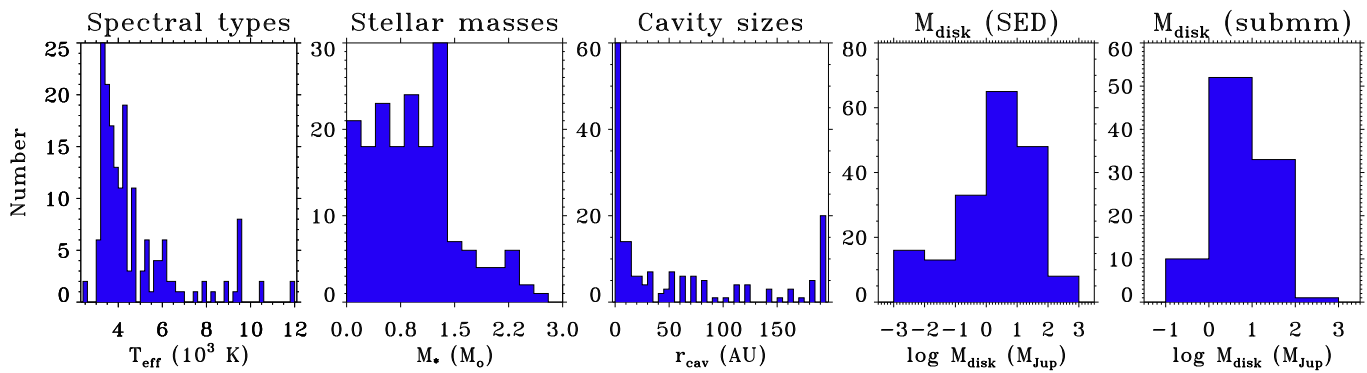}
\end{center}
\caption{Distribution of spectral types, stellar masses, cavity sizes and disk masses (derived from SED fitting and from submillimeter flux) in this study. The disk mass is calculated assuming a gas-to-dust ratio of 100.}
\label{fig:histograms}
\end{figure*}

The stellar photosphere in the model is described by its temperature and stellar luminosity, which has been fit independently together with the extinction. The disk is assumed to be flared, so that the vertical structure of the disk is described by
\begin{equation}
h(r) = h_c\left(\frac{r}{r_c}\right)^{\psi}
\end{equation}
with $h_c$ the scale height at $r_c$ and $\psi$ the flaring angle, which are both varied to fit the near and mid infrared part of the SED. As the scale height is degenerate with the cavity radius, the flaring angle is taken as a conservative value of either 1/7 or 2/7, following \citet{ChiangGoldreich1997}. The derived cavity radius is thus likely a lower limit if the disk is flatter. With the inclusion of the \emph{Herschel} fluxes, the scale height is better constrained than in previous SED modeling studies. Dust composition and settling is prescribed following \citet{Andrews2011}, with a large and small dust grain population where the large grains have a lower scale height than the small grains. The inclination of the disks is taken as a constant of 30$^{\circ}$ and was not varied in the modeling, as only very high inclination angles (edge-on disks) result in a significant difference in the near infrared emission (and in addition, obscuration of the star). With our color criteria, edge-on disks are not expected to be included \citep{Merin2010} and also the computation of the stellar masses from the stellar luminosities implies that most of the targets are not edge-on (although higher inclinations than 30$^{\circ}$ are still possible). 

The five free parameters are thus $r_{\rm cav}$, $\delta_{\rm dust}$, $\Sigma_c$ or disk mass, $h_c$ and $\psi$, where $r_{\rm cav}$ is the main parameter of interest. Note that $\Sigma_c$ represent the dust surface density. The fitting was performed in two steps. First, a large grid of models with a broad range of disk parameters and a limited number of stellar parameters was fit to each SED. Second, each SED was fine-tuned individually, using the exact stellar photosphere and starting from the best fit from the broad grid. The disk grids per object have a large range of cavity radii (our main parameter of interest), in combination with a small range of scale heights and disk masses. Although this approach is rather simple, results of SED modeling are known to be highly uncertain, especially for those targets without known spectral type, and the fitting results should only be considered as a first approximation of the structure. More detailed analysis and imaging data are required to fit individual targets more accurately.

In the fitting procedure, a $\chi^2$ minimization was performed between the dereddened SED data points and the model SEDs. In the grid fitting, the data points were weighted by their excess above the stellar photosphere at each wavelength: fluxes at longer wavelengths got a larger weight than those in the optical and near infrared since the stellar photosphere is largely known from the extinction fitting. Uncertainties on the cavity radii are given in Table \ref{tbl:diskparams}, based on fits with up to 10\% variation in $\chi^2$.

\subsection{Results}
Each SED can be fit to a disk model, with the majority of disks containing a cavity. Table \ref{tbl:diskparams} presents the results of the fitting procedure. Figure \ref{fig:histograms} presents the distribution of hole sizes and disk masses (assuming a gas-to-dust ratio of 100) of the full sample, showing a broad distribution of both parameters. The disk masses obtained from the fit generally agree within a factor of 2-3 with the mass estimate from the millimeter flux. During the fit procedure it became clear that certain disks have really large cavities ($>$100 AU) but very low scale heights, which can not be well reproduced by our flared models. These disks are likely debris disks, as also suggested by their low $L_{\rm disk}/L_*$ values. Recently, a sample of similar young WTTS disks were found to be gas-poor debris disks by ALMA observations \citep{Hardy2015}.

Figures \ref{fig:SEDs1} to \ref{fig:SEDs6} present the SEDs with the best fitting models overlaid. The SEDs are grouped into 5 different classifications:
\begin{itemize}
\item NH: Disks without holes ($r_{\rm cav}$=1 AU)
\item LS: Low-mass disks with small holes \\ ($r_{\rm cav}<$10 AU, $M_{\rm disk}<5 M_{\rm Jup}$)
\item LL: Low-mass disks with large holes \\ ($r_{\rm cav}>$10 AU, $M_{\rm disk}<5 M_{\rm Jup}$, $h_c>0.01$)
\item MS: Massive disks with small holes \\ ($r_{\rm cav}<$10 AU, $M_{\rm disk}>5 M_{\rm Jup}$)
\item ML: Massive disks with large holes \\ ($r_{\rm cav}>$10 AU, $M_{\rm disk}>5 M_{\rm Jup}$)
\item DD: Low-mass disks with large holes and very low scale heights \\ ($r_{\rm cav}>$100 AU, $M_{\rm disk}<5 M_{\rm Jup}$, $h_c\sim0.01$)
\end{itemize}

\begin{table*}[!ht]
\begin{center}
\caption{Comparison cavity radii with literature values.}
\label{tbl:comparercav}
\small
\begin{tabular}{llllll|llllll}
\hline
\hline
ID&Name&$r_{\rm cav}^{\rm here}$&$r_{\rm cav}^{\rm lit}$&Type\tablefootmark{a}&Ref&ID&Name&$r_{\rm cav}^{\rm here}$&$r_{\rm cav}^{\rm lit}$&Type\tablefootmark{a}&Ref\\
\hline
       1&                     TCha&     140$^{+    10}_{   -10}$ &   19&I&1			&	      51&                    IRS48&     120$^{+    10}_{  -100}$ &   60&I&13	\\
       3&                    RECX5&      10$^{+     2}_{    -2}$ &   33&M&2			&	      52&                   DoAr44&      80$^{+    10}_{   -20}$ &   30&I&10	\\
       4&                  CHXR22E&      45$^{+    15}_{    -5}$ &    7&M&3			&	      54&                     SR21&      60$^{+    20}_{   -15}$ &   36&I&8	\\
       6&                    CSCha&      60$^{+    10}_{   -10}$ &   38&M&4			&	      60&                    oph62&       2$^{+     2}$ &    3&M&6	\\
       9&                      T54&     120$^{+    20}_{   -10}$ &   37&M&3			&	      64&         J160421&      70$^{+    20}_{   -30}$ &   80&I&14	\\
      10&                      T21&     190$_{   -10}$ &  146&M&3			&	     120&         J182911&      10$^{+     5}_{    -8}$ &    8&M&6	\\
      11&                    SZCha&      30$^{+    10}_{   -10}$ &   29&M&3			&	     127&                  Serp127&      80$^{+    20}_{   -10}$ &   25&M&6\\	
      12&                      T35&      15$^{+     5}_{    -5}$ &   15&M&3			&	     128&         J182935&      25$^{+    45}_{   -23}$ &    7&M&6\\	
      14&                      T56&      10$^{+     4}_{    -4}$ &   18&M&3			&	     135&                    DMTau&       4$^{+     2}_{    -2}$ &   18&I&10\\	
      15&                    CRCha&       1$^{+     1}$ &   10&M&4			&	     136&                   UXTauA&      50$^{+    40}_{   -10}$ &   25&I&10\\	
      18&                      T25&      30$^{+     5}_{    -5}$ &    8&M&3			&	     142&                   MWC758&      25$^{+    15}_{    -5}$ &   73&I&10\\	
      21&                 HD142527&     110$^{+    10}_{   -20}$ &  100&I&5			&	     148&                    IPTau&     100$^{+    10}_{   -30}$ &    2&M&4\\	
      24&                    Lup60&      16$^{+     2}_{    -4}$ &    3&M&6			&	     153&                    RYTau&       2$^{+     2}$ &   26$^{(b)}$&I&15\\	
      27&                     Sz91&     120$^{+    30}_{   -20}$ &   97&I&7			&	     159&                    ABAur&       1$^{+     4}$ &  115&I&16\\	
      29&                     Sz84&      70$^{+    40}_{   -10}$ &   55&M&6			&	     161&                   ASR118&       2$^{+     4}$ &    1&M&6\\	
      33&                 HD135344&      80$^{+    10}_{   -10}$ &   46&I&8			&	     165&         J034227&       8$^{+     4}_{    -4}$ &    5&M&6\\	
      35&                     Sz76&       2$^{+     2}$ &    1&M&9			&	     168&         J034434&       5$^{+     20}_{    -3}$ &    3&M&6\\	
      38&           RXJ1615&      10$^{+    10}_{    -2}$ &   30&I&10			&	     169&              IC348LRL190&       2$^{+     2}$ &    5&M&6\\	
      39&                 V4046Sgr&      16$^{+     4}_{    -6}$ &   29&I&11			&	     173&             LkH-alpha330&     120$^{+    10}_{   -20}$ &   68&I&10\\	
      45&                    SR24S&      50$^{+    40}_{   -20}$ &   30&I&10			&	     309&                    TWHya&      10$^{+     2}_{    -2}$ &    4&M&17\\	
      46&           RXJ1633&      20$^{+    10}_{    -5}$ &   27&I&12			&	     325&                   LkCa15&      80$^{+    40}_{   -35}$ &   50&I&10\\	
      48&                    WSB60&       8$^{+     6}_{    -6}$ &   15&I&10			&	     326&               CoKu-Tau-4&       6$^{+     2}_{    -2}$ &   10&M&18\\	
      50&         J162245&       2$^{+     2}$ &    1&M&6			&	     329&                    GMAur&      30$^{+    10}_{    -5}$ &   20&I&10\\	
\hline
\end{tabular}
\end{center}
 {\bf Refs.}
       1) \citet{         Huelamo2015}, 
       2) \citet{         Bouwman2010}, 
       3) \citet{             Kim2009}, 
       4) \citet{       Espaillat2011}, 
       5) \citet{        Fukagawa2013}, 
       6) \citet{           Merin2010}, 
       7) \citet{         Canovas2015}, 
       8) \citet{           Brown2009}, 
       9) \citet{         Padgett2006}, 
      10) \citet{         Andrews2011}, 
      11) \citet{       Rosenfeld2013}, 
      12) \citet{           Cieza2012rxj}, 
      13) \citet{     vanderMarel2013}, 
      14) \citet{         Mathews2012}, 
      15) \citet{     Isella2010rytau}, 
      16) \citet{           Pietu2005}, 
      17) \citet{         Andrews2012}, 
      18) \citet{         Alessio2005}  \\ 
      {\bf Notes.}$^{(a)}$ M = derived from SED modeling, I = derived from millimeter imaging. $^{(b)}$ This value is not a real cavity size, but a transition radius: the disk was fit with a surface density profile that radially increases and decreases, peaking at 26 AU.
\end{table*}

For the disks classified as NH (no hole), we have excluded the targets that could be fit with a cavity $>$1 AU within the 10\% $\chi^2$ limit. 

A large fraction of the disks ($\sim$23\%) falls in the ML category of large holes in massive disks. It turns out that several of these disks are indeed the famous, bright disks with large inner holes known from imaging surveys \citep{Andrews2011,WilliamsCieza2011}, confirming the strength of our SED modeling, even if rather simple. The new targets in the ML, MS and some in the LL groups are promising disks for follow up observations with ALMA. Excluding the DD and NH disks, a total of 133 targets (72\% of our analyzed sample) can be labeled as disks with cavities, transition disks. More than half of these ($\sim70$ targets) are new transition disks; about 40 had been imaged or modeled before and another 20 had been recognized as a possible transition disk. Of the new transition disks, two thirds have a known spectral type.

\section{Discussion}
\label{sct:discussion}
The SED modeling has confirmed the presence of cavities in a large sample of transition disks. At least 72\% of the sample could be modeled as a disk with a cavity, including about 110 new transition disks that had not been identified as transition disk before.

\subsection{Comparison of cavity radii with literature values}

\begin{figure}[!ht]
\includegraphics[width=0.45\textwidth]{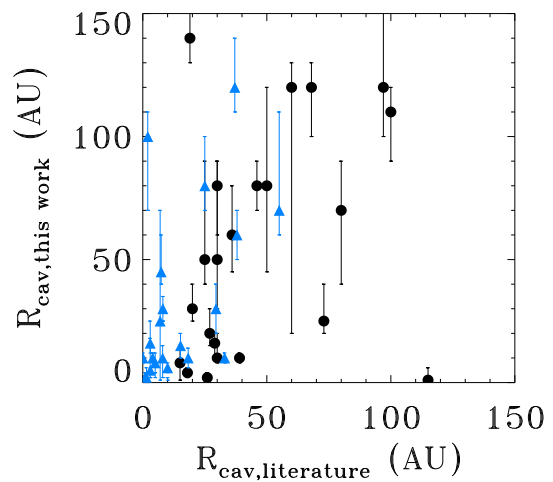}
\caption{Comparison derived cavity radii with literature values, based on Table \ref{tbl:comparercav}. The black circles indicate the literature results from imaging, the blue diamonds from modeling.}
\label{fig:comparercavs}
\end{figure}

\begin{table*}[!ht]
\caption{Multiplicity properties}
\label{tbl:binarity}
\begin{center}\small
\begin{tabular}{llllll|llllll}
\hline
\hline
   ID&  Name&Sep.& Sep.&$r_{\rm cav}$&Ref&ID&Name&  Sep.& Sep.&$r_{\rm cav}$&Ref\\
   &&(arcsec)&(AU)&(AU)&&&&(arcsec)&(AU)&(AU)&\\
\hline
1&TCha&0.062&6.7&     140$^{+    10}_{   -10}$ &1			&	   60&oph62&        <0.1&         <12&       2$^{+     2}$ &9\\
4&CHXR22E&        <0.1&         <16&      45$^{+    15}_{    -5}$ &2			&	   62&J162218.5-232148&        <0.1&         <12&1&9\\
5&ISO52&        <0.1&         <16&      30$^{+    50}_{   -18}$ &2			&	   63&DoAr28&       <0.13&         <16&      20$^{+     5}_{    -5}$ &10\\
6&CSCha&           -&4&      60$^{+    10}_{   -10}$ &3			&	   64&J160421.7-213028&  <0.01&    <1.5&      70$^{+    20}_{   -30}$ &11\\
7&11094742-7726290&       <0.75&        <120&       8$^{+     2}_{    -2}$ &4			&	  134&RXJ0432.8+1735&       <0.13&         <18&     190$_{   -10}$ &12\\
9&T54&0.25&40&     120$^{+    20}_{   -10}$ &2			&	  135&DMTau&       >0.03&          >4&       4$^{+     2}_{    -2}$ &13\\
10&T21&0.14&22&     190$_{   -10}$ &2			&	  136&UXTauA&       >0.03&          >4&      50$^{+    40}_{   -10}$ &13\\
11&SZCha& <0.07&     <11&      30$^{+    10}_{   -10}$ &2			&	  140&043649.1+241258&        <0.1&         <14&     190$_{   -10}$ &14\\
12&T35&       <0.07&         <11&      15$^{+     5}_{    -5}$ &2			&	  149&V892Tau&    0.06&       8&      10$^{+     8}_{    -8}$ &15\\
14&T56&       <0.07&         <11&      10$^{+     4}_{    -4}$ &2			&	  150&V410X-ray6&        <0.1&         <14&      15$^{+     5}_{    -5}$ &14\\
15&CRCha&       <0.08&         <13&       1$^{+     1}$ &2			&	  152&V819Tau&        <0.1&         <14&     150$^{+    20}_{   -20}$ &14\\
16&WWCha&       <0.12&         <19&      50$^{+    30}_{   -49}$ &2			&	  153&RYTau&       >0.03&          >4&       2$^{+     2}$ &13\\
17&11062554-7633418&       <0.75&        <120&      15$^{+    15}_{    -5}$ &4			&	  162&MBO22&        <0.1&         <25&       2$^{+     2}$ &14\\
18&T25&       <0.08&         <13&      30$^{+     5}_{    -5}$ &2			&	  172&IC348-67&        <0.1&         <25&       2$^{+     1}$ &14\\
21&HD142527&0.088&13&     110$^{+    10}_{   -20}$ &5			&	  174&J04300424+3522238&        <0.1&         <45&      18$^{+    10}_{    -6}$ &14\\
23&Sz111&        <0.7&        <140&      60$^{+    10}_{   -10}$ &6			&	  175&J04303235+3536133&0.83&116&       4$^{+     8}_{    -2}$ &14\\
24&Lup60&        <0.1&         <20&      16$^{+     2}_{    -4}$ &7			&	  177&J04304004+3542101&1.2&168&      25$^{+    10}_{   -10}$ &14\\
25&J160830.3-390611&        <0.8&        <160&       4$^{+     4}_{    -2}$ &6			&	  179&J160044.5-415531&        <0.1&         <15&       1$^{+    59}$ &7\\
27&Sz91&        <0.1&         <20&     120$^{+    30}_{   -20}$ &7			&	  180&J190058.1-364505&0.5&72&      14$^{+     4}_{    -4}$ &7\\
28&J160855.5-390234&        <0.1&         <20&       2$^{+     2}$ &7			&	  181&03445614+3209152&        <0.1&         <25&       6$^{+    12}_{    -5}$ &14\\
33&HD135344&        <0.1&         <14&      80$^{+    10}_{   -10}$ &8			&	  182&03442156+3215098&        <0.1&         <25&       2$^{+    23}$ &14\\
40&J163154.7-250324&        <0.1&         <12&1&9			&	  183&03442257+3201536&        <0.1&         <25&       4$^{+     1}_{    -2}$ &14\\
41&J163205.5-250236&        <0.1&         <12&1&9			&	  184&04330422+2921499&        <0.1&         <25&     160$^{+    20}_{   -10}$ &14\\
43&J163023.4-245416&       <0.13&         <16&      45$^{+     5}_{   -20}$ &10			&	  191&042921.6+270125&0.22&30&       2$^{+     2}$ &16\\
44&WSB63&        <0.1&         <12&       4$^{+     2}_{    -2}$ &9			&	  200&J160710.08-391103.5&       <0.06&         <12&1&6\\
46&RXJ1633.9-2442&        <0.1&         <12&      20$^{+    10}_{    -5}$ &9			&	  309&TWHya&        <0.1&          <5&      10$^{+     2}_{    -2}$ &8\\
47&ISO-Oph43&       <0.13&         <16&       1$^{+     1}$ &10			&	  318&DoAr21&           -&1.5&      70$^{+    30}_{   -10}$ &17\\
48&WSB60&        <0.1&         <12&       8$^{+     6}_{    -7}$ &8			&	  319&J162740.3-242204&0.638&80&       1$^{+    17}$ &10\\
49&J163115.7-243402&0.33&41&      20$^{+    30}_{   -10}$ &9			&	  325&LkCa15&       >0.03&          >4&      80$^{+    40}_{   -35}$ &13\\
50&J162245.4-243124&0.54&68&       2$^{+     2}$ &9			&	  326&CoKu-Tau-4&0.053&8&       6$^{+     2}_{    -2}$ &18\\
58&J162648.6-235634&       <0.13&         <16&1&10			&	  329&GMAur&       >0.03&          >4&      30$^{+    10}_{    -5}$ &13\\
59&J162802.6-235504&        <0.1&         <12&       2$^{+     2}$ &9			&	&&&&&\\
    \hline
\end{tabular}
\end{center}
{\bf Refs.}
       1) \citet{         Huelamo2011}, 
       2) \citet{      Lafreniere2008}, 
       3) \citet{        Guenther2007}, 
       4) \citet{         Comeron2012}, 
       5) \citet{          Biller2012}, 
       6) \citet{            Ghez1997}, 
       7) \citet{          Romero2012}, 
       8) \citet{         Vicente2011}, 
       9) \citet{           Cieza2010}, 
      10) \citet{          Ratzka2005}, 
      11) \citet{           Kraus2008}, 
      12) \citet{   KohlerLeinert1998}, 
      13) \citet{            Pott2010}, 
      14) \citet{           Cieza2012}, 
      15) \citet{         Leinert1997}, 
      16) \citet{          Biller2011}, 
      17) \citet{         Loinard2008}, 
      18) \citet{    Ireland2008}
\end{table*}

In order to quantify the quality of our models, the fit results for the cavity size are compared with values from the literature from both SED modeling and resolved millimeter imaging in Table \ref{tbl:comparercav} and Figure \ref{fig:comparercavs}. The cavity radii generally agree well within a factor of two with previously derived parameters. Especially the similarity to the imaging results is encouraging: this implies that a large number of our new targets are suitable for resolved imaging. Exceptions for the imaging targets are T Cha, RY Tau and AB Aur. For T Cha the overestimate of the cavity size could be caused by the assumed low inclination in our models: imaging has shown that the inclination is in reality $\sim67^{\circ}$ so close to edge-on \citep{Huelamo2015}, affecting the near infrared emission from the inner disk. Also their flaring angle is lower than ours. For RY Tau, the cavity radius found by imaging is not defined in the same way as here: it is the peak of the mm dust surface density, assuming a surface density that first increases and then decreases with radius \citep{Isella2010rytau}. Therefore, the values can not be compared directly. For AB Aur, only a very small hole of at most 2 AU can be fit with our models, while millimeter imaging has revealed a large cavity of 115 AU at 1.4mm, with a complex, possibly spiral-arm structure \citep{Pietu2005,Tang2012}. As the AB Aur disk is still embedded in a envelope, the mid infrared emission is likely confused by cloud emission, which can explain this discrepancy between the SED and the millimeter image. The comparison with SED modeling shows large discrepancies for CHXR22E, T54, CR Cha and IP Tau. These targets were not modeled with a full radiative transfer code but a parametrized temperature profile and optically thin dust emission inside the cavity rather than an inner disk \citep{Kim2009,Espaillat2011}, so the results can not be compared directly.

\subsection{Binaries}
Some transition disks can be explained as circumbinary disks due to the dynamical interaction between the disk and a stellar companion. The cavity size is expected to be $\sim$2 times as large as the binary separation \citep{Artymowicz1996}. The fraction of the transition disks in our sample for which binarity has been studied is limited, but for those targets where spatially resolved information is available from the literature (either detections or upper limits), the properties are listed in Table \ref{tbl:binarity}, together with the cavity sizes found in this study. 

Although for a handful of targets the cavity can indeed be explained by a binary companion, for the bulk of the disks the limits are not sufficient to exclude circumbinary disks. Previous binary studies of transition disks also revealed that most of the sharp cavities are not due to binary systems \citep{Pott2010,Vicente2011}.  

\subsection{Accretion}
\begin{figure}[!ht]
\begin{center}
\includegraphics[width=0.48\textwidth]{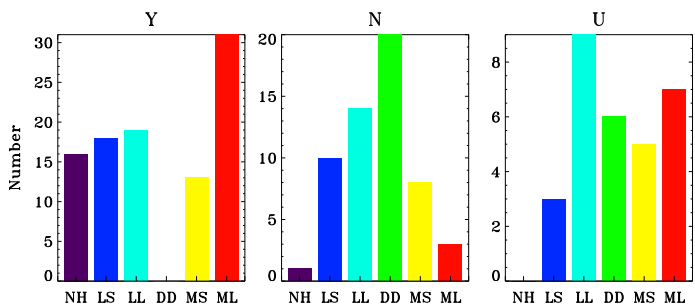}
\end{center}
\caption{Comparison of accretion properties from Table \ref{tbl:stellarparams} with disk hole parameters: Y means accreting, N means non-accreting, U means unknown.}
\label{fig:accretion}
\end{figure}

By combining the outcome of the SED modeling with our information on accretion, the possibility of photoevaporation as origin of the cavities can be checked. According to photoevaporation models \citep[e.g.][]{Alexander2006a}, UV photons from the star heat and ionize the gas in the disk; beyond a critical radius, the thermal velocity of the ionized gas exceeds its escape velocity and the material is dissipated as a wind. During the lifetime of the disk, the accretion rate is  expected to gradually decrease: when the rate drops below the photoevaporation rate, the outer disk can no longer resupply the inner disk with material and an inner hole is formed. This process is called photoevaporative clearing, and transition disks created by this mechanism are expected to have no or very low accretion, although only disks with small inner cavities can be explained by this mechanism \citep{Owen2016}. Clearing of a gap by a planet and photoevaporation could also happen simultaneously \citep[e.g.][]{Rosotti2013}, making the distinction not purely measurable by accretion alone. Figure \ref{fig:accretion} shows the number of objects in each class that are accreting/non-accreting. The accretors are dominated by disks without holes and massive disks with large holes, which are likely transition disks with a cavity due to clearing by a companion. The non-accretors are dominated by the low-scale height low-mass disks (DD), confirming that they are likely debris disks. The non-accreting low-mass disks are possibly disks where the hole is caused by photoevaporative clearing, consistent with the so-called mm-faint disks \citep{OwenClarke2012,Owen2016} in contrast with the mm-bright transitional disks. On the other hand, there are several low-mass accreting disks as well, so there is no general trend for the low-mass disks. The non-accreting disk without a hole (J034520.5+320634, ID171) is an outlier, but the equivalent width of this target is on the edge of accreting/non-accreting, probably due to the ubiquitous variable accretion \citep{Mendigutia2012,Venuti2014}, and should thus have been classified as an accretor.

\begin{figure}[!ht]
\begin{center}
\includegraphics[width=0.45\textwidth]{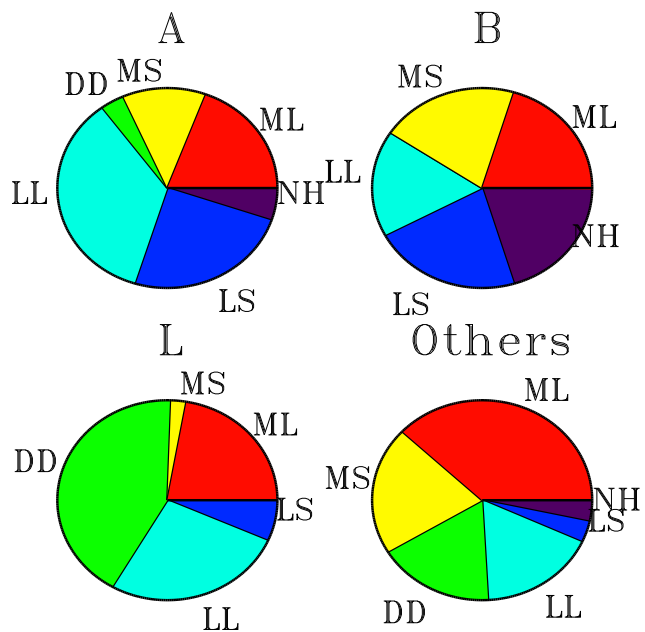}
\end{center}
\caption{Evaluation of the color criteria (Region A, B and L colors) vs the outcome of the SED modeling.}
\label{fig:criteria}
\end{figure}

\begin{figure}[!ht]
\begin{center}
\includegraphics[width=0.45\textwidth]{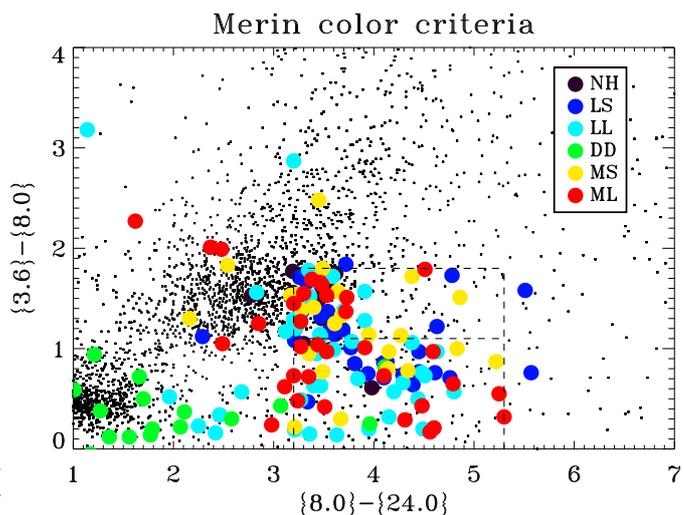}
\end{center}
\caption{Evaluation of the color criteria (A and B) vs the outcome of the SED modeling. The filled symbols are the targets analyzed in this study, the dots are all YSOs in the c2d, GB and Taurus catalogs. The dashed lines mark the Region B (top) and Region A (bottom) criteria.}
\label{fig:merincolorplot}
\end{figure}

\begin{figure}[!ht]
\begin{center}
\includegraphics[width=0.45\textwidth]{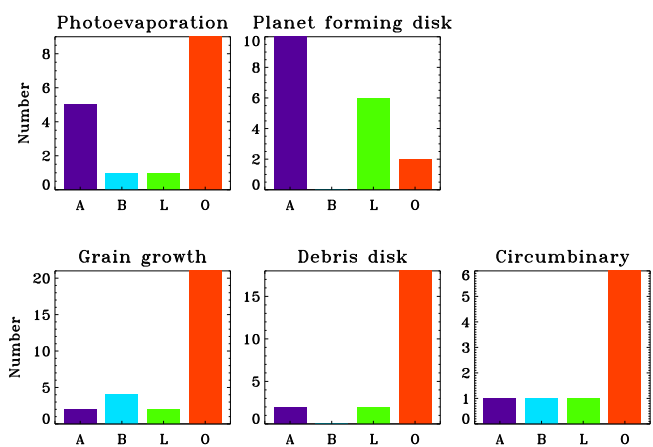}
\end{center}
\caption{Comparison of the categorization by the Cieza et al. studies with the color criteria used in this study applied to the 92 targets in the Cieza sample.}
\label{fig:ciezacolors}
\end{figure}

\begin{figure}[!ht]
\begin{center}
\includegraphics[width=0.4\textwidth,trim=0 -40 10 0]{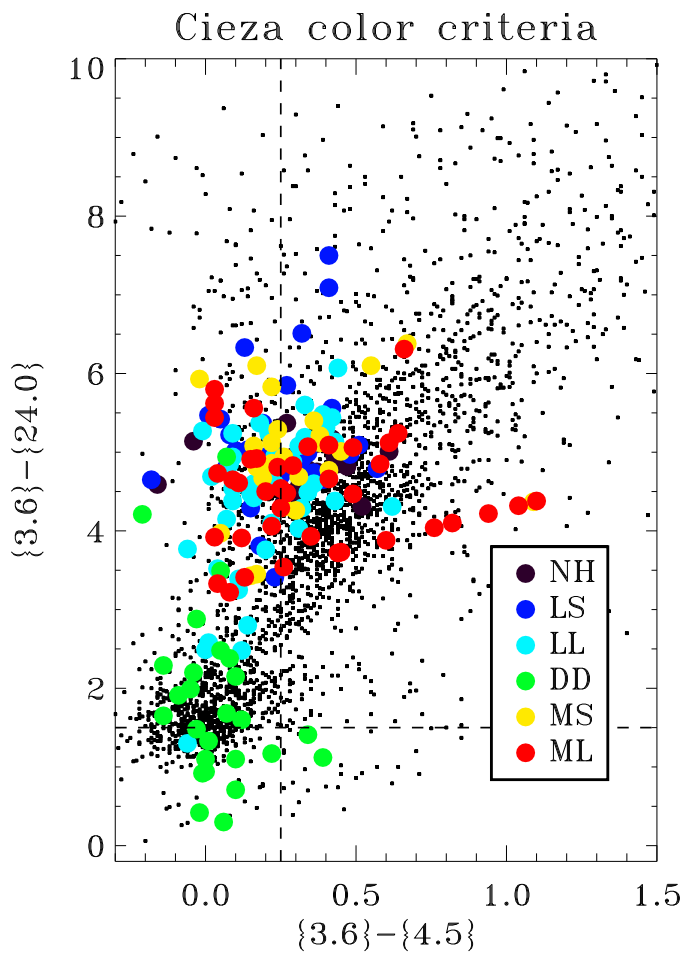}
\end{center}
\caption{Evaluation of the initial color criteria by the Cieza survey vs the outcome of our SED modeling. The filled symbols are the targets analyzed in our study, the dots are all YSOs in the c2d, GB and Taurus catalogs. The dashed lines mark the regions: the upper left quartile are the transition disks according to Cieza, the lower left quartile are diskless stars.}
\label{fig:ciezacolorplot}
\end{figure}

\begin{figure}[!ht]
\begin{center}
\includegraphics[width=0.55\textwidth,trim=0 0 50 0]{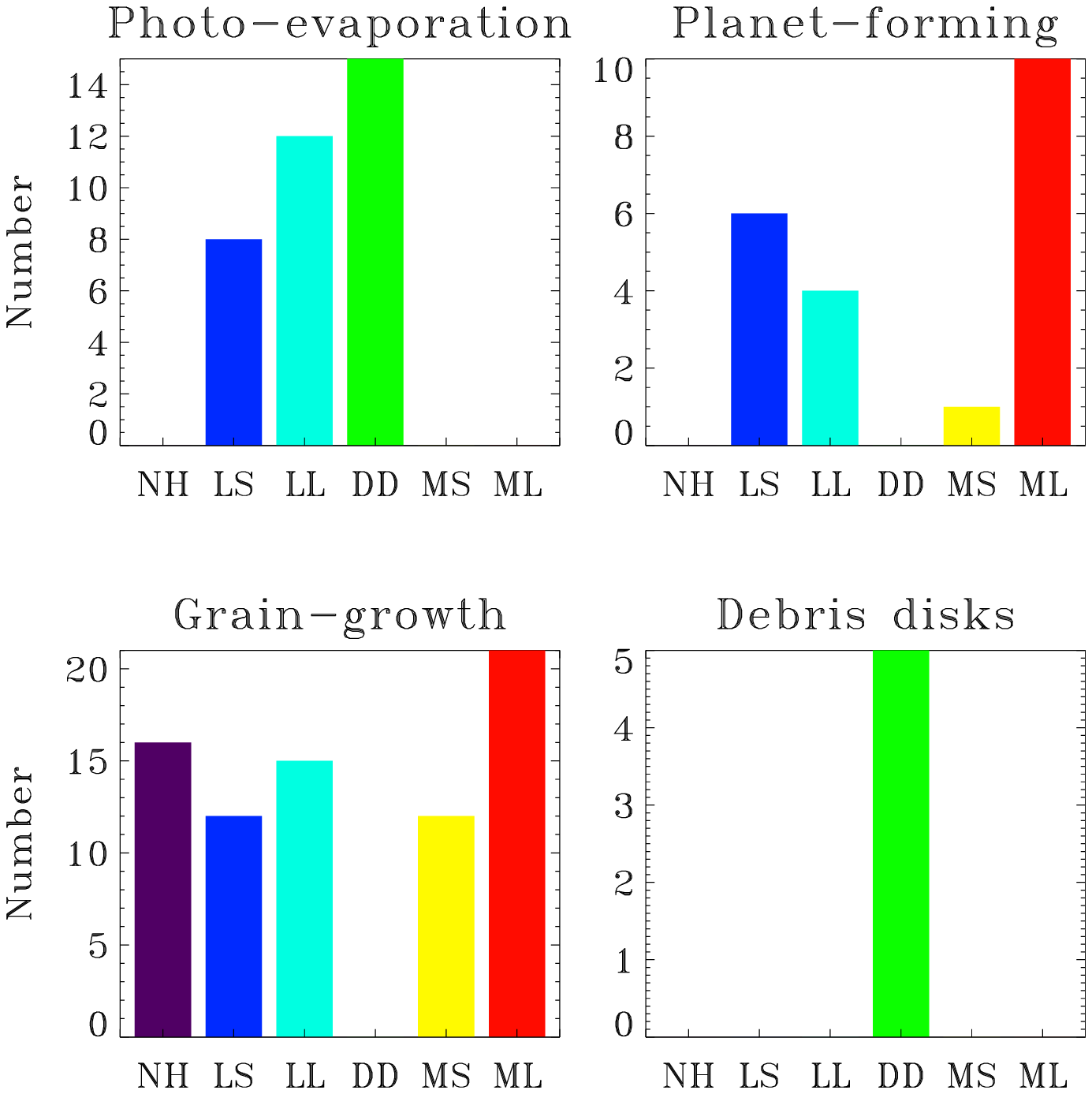}
\end{center}
\caption{Comparison of the categorization by the Cieza et al. studies with the classification of this study for our targets.}
\label{fig:ciezaclass}
\end{figure}

\subsection{Evaluation of color criteria}
Considering the high success rate of new transition disks found in the sample in this study, it is now possible to re-evaluate the criteria used to select the targets. Figure \ref{fig:criteria} presents the the resulting classifications from our SED modeling for each of the four color selection criteria.

From Figure \ref{fig:criteria} it is clear that the disks without holes are most dominant in the B criteria, but overall at a low fraction. The L criteria are particularly biased towards the low scale height disks and low-mass disks with large holes. Massive disks with large holes are found in all colors. Evaluating the M10 criteria directly in the color-color plot, the DD disks fall outside the A and B range. A small amount of disks with holes falls outside the range of the A, B criteria (these are targets from the literature), generally with a shallower 8-24 $\mu$m slope.

\subsection{Comparison with Cieza survey}

A previous large survey of transition disk candidates was performed by \citet{Cieza2010,Cieza2012} and \citet{Romero2012}, who selected a sample of candidates using their own color criteria. Rather than SED modeling, they apply criteria based on a range of observables (disk mass, $L_{\rm disk}/L_*$, accretion, multiplicity, infrared spectral slope $\alpha_{\rm excess}$ and the wavelength where the disk emission starts to dominate, $\lambda_{\rm turnoff}$) to determine the origin of the dust deficit in their disks: circumbinary disk, photoevaporative clearing, debris disk, grain growth or planetary clearing.

\begin{figure*}[!ht]
\begin{center}
\subfigure{\includegraphics[width=0.35\textwidth]{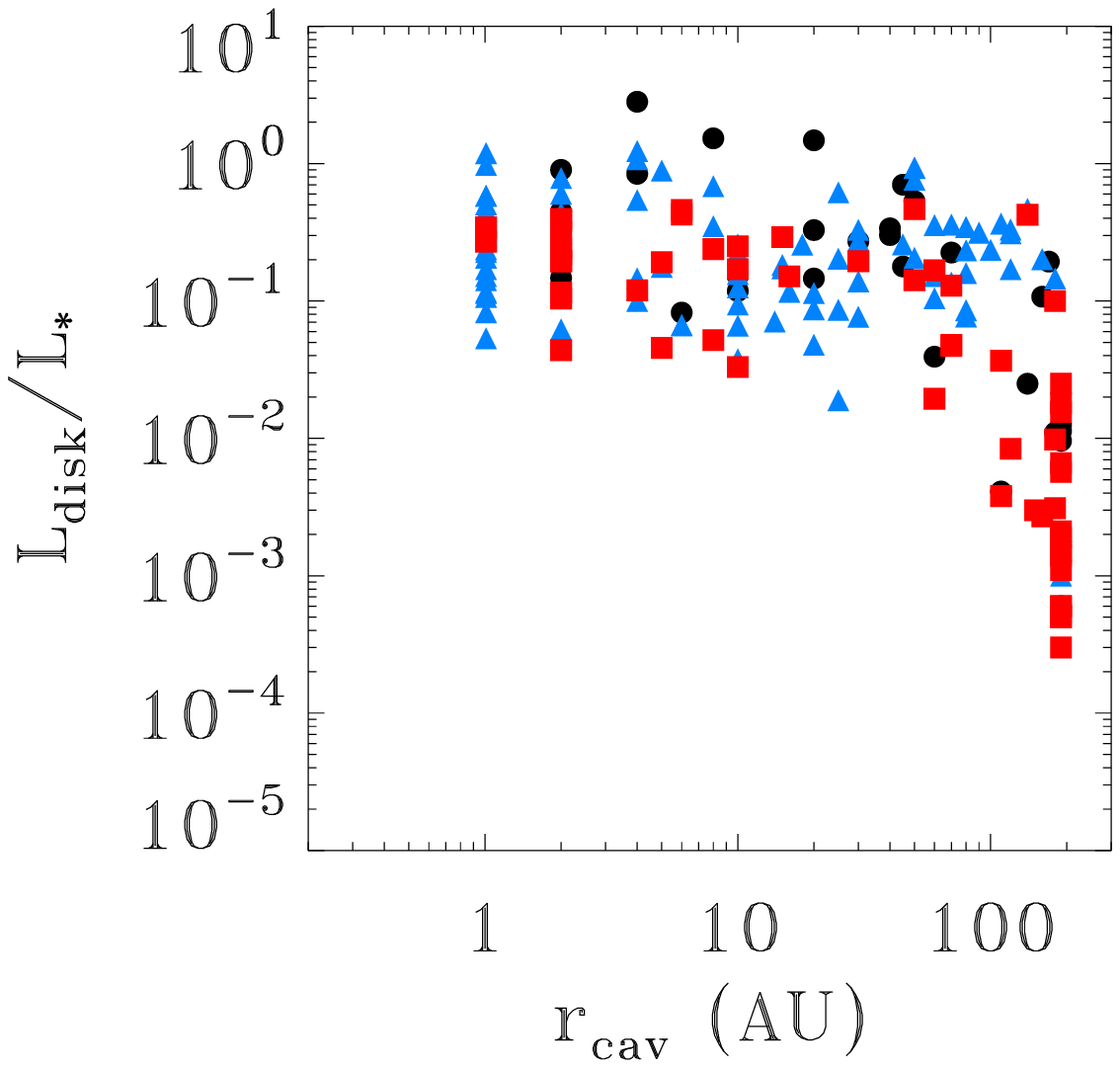}}
\subfigure{\includegraphics[width=0.35\textwidth]{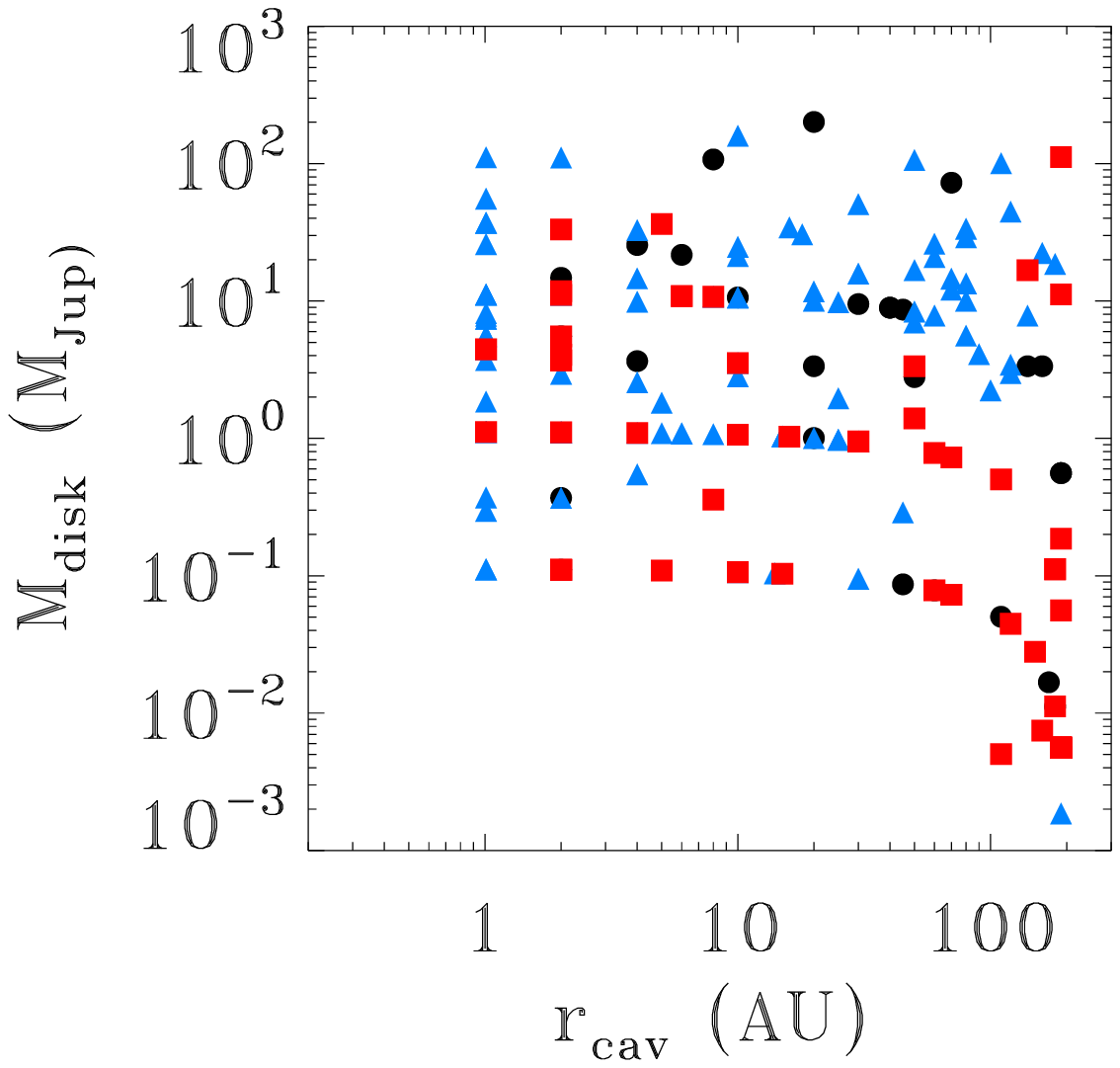}}
\end{center}
\caption{The relation between cavity radius and $L_{\rm disk}/L_*$ and disk mass. Blue triangles indicate accretors, red squares non-accretors, black circles unknown.}
\label{fig:diskevolution}
\end{figure*}

\begin{figure}[!ht]
\begin{center}
\includegraphics[width=0.45\textwidth]{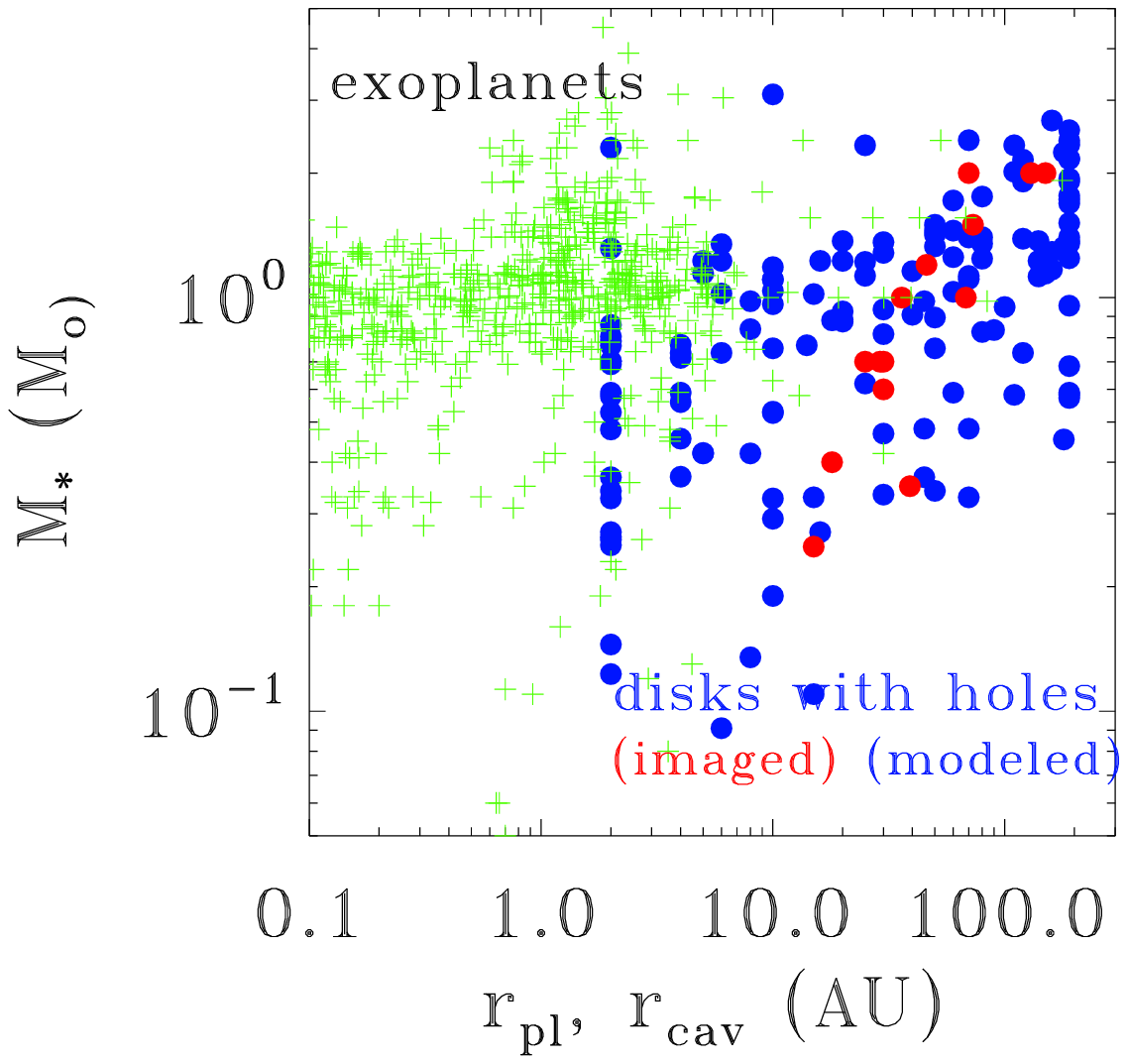}
\end{center}
\caption{Comparison of exoplanet orbital radii with transition disk cavity radii. Only cavity radii $>$2 AU are considered. Exoplanet orbital radii are indicated as green crosses, disk cavities with circles in blue (SED-modeling of this study) and red (millimeter imaging).}
\label{fig:msplot}
\end{figure}

Of particular interest are their planet-forming disks, which are massive, accreting disks with sharp cavities ($\alpha_{\rm excess}>0$). Their final target list is compared with our sample, and the colors used for our color criteria have been derived for all their targets. Note that 27 of their 92 targets are already present in our sample, either selected by the color criteria or by their classification in the literature. Figure \ref{fig:ciezacolors} shows how their categorization overlaps with our color criteria. Our A and L color criteria are clearly favored in the planet forming disks, while many of the less interesting disks from our perspective (grain growth, debris disks) fall in category, 'O', outside our color criteria. Several of the photoevaporation disks also follow the A criteria. This comparison confirms that our color criteria are good at selecting disks with sharp inner cavities.

Figure \ref{fig:ciezacolorplot} shows the inital color selection of Cieza et al. ([3.6]-[4.5]$<$0.25, [3.6]-[24.0]$>$1.5) in comparison with the outcome of our classification. This Figure shows that only 50\% of our disks with holes fall within these criteria. One of the reasons is that Cieza et al. have stricter constraints for their near infrared emission, which excludes the transition disks with strong near infrared excess (indicating an inner disk). The DD targets fall in the same quartile as the diskless stars, as expected. 

Finally, Figure \ref{fig:ciezaclass} shows how the classification of Cieza et al. compares with our classification for our targets. Note that we have included all our targets in the comparison, computing the values of $\alpha_{\rm excess}$ and $\lambda_{\rm turnoff}$ ourselves, in order to put them in the Cieza classification. Planet forming disks fall mostly within the class of massive disks with large cavities, while photoevaporation and debris disks are mostly consistent with low-mass disks. Neither of this is a surprise, considering the categorization of \citet{Cieza2010}. Disks without holes all fall within the category of grain-growth disks. On the other hand, several other disks in the grain-growth category could be fit with a disk including a cavity. Note that circumbinarity is not well-constrained for most of our sample and therefore not included.

\subsection{Evolution}
The $L_{\rm disk}/L_*$ ratio is generally taken as a measure of disk evolution. For the targets in this study, there is a hint of a trend, with larger cavity radii and generally more non-accretors for lower $L_{\rm disk}/L_*$ (Figure \ref{fig:diskevolution}a). This implies a general growth of cavity sizes with time while accretion decreases. 

A relation between disk mass and cavity radius has been noted in previous studies and interpreted as a gravitational process, where larger disk masses produce more massive planets, clearing larger cavities \citep{Merin2010}. This study shows no trend between disk mass and cavity size. However, accretors are generally more massive than non-accretors (Figure \ref{fig:diskevolution}b). 

Both trends are susceptible due to the uncertainties in $r_{\rm cav}$ and biased due to the presence of many low-mass disks with large cavities in our sample, which are more likely to be debris disks than transition disks. When the DD disks are removed from the sample, there is barely a visible trend.

\subsection{Cavity radii and exoplanets}
If the origin of the dust cavity radii lies in forming planets, a connection should exist between the orbital radii of exoplanets and disk cavity radii according to planet-disk interaction models \citep{Pinilla2012b}, unless planets migrate from their birth location. Planets are expected to clear cavities up to twice their orbital radius. We investigate this relation in Figure \ref{fig:msplot}. Only cavity radii $>$2 AU are considered. Both samples are heavily biased, especially there is a lack of planets on orbits larger than 5 AU (radial velocity limit), whereas disk cavity radii can not be detected below 2 AU. The exoplanet data is only fairly complete for $r<$1 AU (from Kepler) whereas there is no information about transition disk cavity radii at this limit. Therefore, we cannot test any connection with the current data but as both disk and exoplanet will start to fill the critical 2-10 AU range in the coming years it will be important to search for such a relation. A trend is seen between cavity radius and stellar mass (blue and red points in Figure \ref{fig:msplot}) as seen in previous work \citep{Kim2009,Merin2010}.

\subsection{Transition disks within the full YSO sample}
A total number of 133 transition disk candidates (ML, MS, LL or LS category) is confirmed through the SED modeling. 108 of these are from one of the main \emph{Spitzer} catalogs (c2d, GB or Taurus) of YSOs in nearby star forming regions. These three catalogs add up to 3331 YSOs, out of which 1387 are classified as disks (Class II objects). This means that approximately 8\% of the disks in star forming regions are expected to be transition disks, considering only the targets from the \emph{Spitzer} catalogs. Assuming a Class II lifetime of 3 Myr \citep{Dunham2015,Ribas2015}, this implies a transition disk lifetime of 0.24 Myr, assuming that disk evolution is continuous over the timespan of the Class II lifetime. More specifically, assuming that transition disks evolve by dissipating mass, the first stage of massive disks (MS+ML, 45 targets) is about 90 kyr, followed by a longer period of low-mass disks (LS+LL, 63 targets) of 150 kyr. Although it is tempting to conclude that disk dissipation is rapid \citep{WilliamsCieza2011,Owen2016}, it should be noted that about half of the low-mass disks in the sample is still accreting (see Figure \ref{fig:accretion}) and disk mass alone cannot be used as a robust measurement of transition disk evolution.

\section{Summary}
The main results of this work are summarized as follows.
\begin{enumerate}
\item A large sample of $\sim$200 transition disk candidates is presented and analyzed. Candidates are primarily selected from the \emph{Spitzer} catalogs using the color criteria from \citet{Merin2010}, with some additional targets from the literature.
\item The \emph{Spitzer} data of the targets have been complemented with new submillimeter fluxes, \emph{Herschel}-PACS archival data and optical/near infrared spectroscopy for spectral typing and accretion properties.
\item All targets are analyzed using RADMC-3D modeling with a limited number of parameters: disk mass, inner disk mass, scale height and flaring, and disk cavity radius, where the latter is the main parameter of interest.
\item The derived cavity sizes are consistent with imaging/modeling results in the literature where available.
\item Using the derived properties, the disks are categorized in disks without holes, large/small holes and massive/low-mass disks and very low scale height disks. The latter are likely debris disks. A large fraction of the targets falls in the category of disks with large holes, including several previously imaged disks.
\item Based on this classification, 133 targets (72\% of the sample) are disks with cavities, including about 70 new transition disk candidates that had not been identified before.
\item The color criteria are evaluated and compared for the targets in the Cieza studies \citep{Cieza2010,Cieza2012,Romero2012}. In general, our color criteria are a proper tool to select transition disk candidates.
\item The cavity radius increases with stellar mass.
\item The sample list provides a large number of transition disk candidates that are suitable for follow-up observations with ALMA.
\end{enumerate}

  \begin{acknowledgements}
  The authors would like to thank C. Manara for useful discussions and M. McClure for providing the IRS spectrum on DoAr28.
  N.M. is supported by the Netherlands Research School for Astronomy
  (NOVA). Astrochemistry in Leiden is supported by the Netherlands
  Research School for Astronomy (NOVA), by a Royal Netherlands Academy
  of Arts and Sciences (KNAW) professor prize, and by the European
  Union A-ERC grant 291141 CHEMPLAN.  This publication makes use of data products from the Wide-field Infrared Survey Explorer, which is a joint project of the University of California, Los Angeles, and the Jet Propulsion Laboratory/California Institute of Technology, funded by the National Aeronautics and Space Administration. 
  \end{acknowledgements}

\bibliographystyle{aa}

\appendix

\section{Tables}
\onecolumn
\begin{longtable}{llllllllll}
\caption{Sample of transition disk candidates}
\label{tbl:fullsample}\\
\hline
\hline
ID&Target&RA&Dec&Reg\tablefootmark{a}&d\tablefootmark{b}&Color\tablefootmark{c}&Origin&Prev.\tablefootmark{d}&Ref\\
&&&&&(pc)&&data&&\\
\hline
\endfirsthead

\caption{Sample continued.}\\
\hline
\hline
ID&Target&RA&Dec&Reg\tablefootmark{a}&d\tablefootmark{b}&Color\tablefootmark{c}&Origin&Prev.\tablefootmark{d}&Ref\\
&&&&&(pc)&&data&&\\
\hline
\endhead

\hline
\endfoot

\hline
\endlastfoot

1	&	TCha	&	 11 57 13.53	&	-79 21 31.5	&	$\epsilon$ Cha	&	108	&	[L]		&	Other	&	Imag.	&	1,2\\
2	&	RECX11	&	 08 47 01.80	&	-78 59 35.2	&	$\eta$ Cha	&	97	&	-		&	Other	&	TD	&	3,4\\
3	&	RECX5	&	 08 42 27.09	&	-78 57 47.9	&	$\eta$ Cha	&	97	&	[A]		&	Other	&	PF	&	5,4\\
4	&	CHXR22E	&	 11 07 13.30	&	-77 43 49.9	&	ChaI	&	160	&	-		&	Other	&	TD	&	6,7\\
5	&	ISO52	&	 11 04 42.58	&	-77 41 57.1	&	ChaI	&	160	&	[B]		&	GB	&	V	&	8\\
6	&	CSCha	&	 11 02 25.11	&	-77 33 35.9	&	ChaI	&	160	&	[L]		&	c2d	&	V	&	8\\
7	&	11094742-7726290	&	 11 09 47.27	&	 -77 26 29.5	&	ChaI	&	160	&	[B]		&	GB	&		&	\\
9	&	T54	&	 11 12 42.69	&	-77 22 23.1	&	ChaI	&	160	&	[L]		&	GB	&	CB	&	9\\
10	&	T21	&	 11 06 15.41	&	-77 21 56.9	&	ChaI	&	160	&	-		&	Other	&	TD	&	6,7\\
11	&	SZCha	&	 10 58 16.77	&	-77 17 17.1	&	ChaI	&	160	&	-		&	-	&	TD	&	9\\
12	&	T35	&	 11 08 39.05	&	-77 16 04.2	&	ChaI	&	160	&	[A]		&	GB	&	TD/V	&	6,10\\
13	&	ISO-ChaII29	&	 12 59 10.19	&	-77 12 13.7	&	ChaII	&	180	&	[L]		&	c2d	&		&	\\
14	&	T56	&	 11 17 37.01	&	-77 04 38.1	&	ChaI	&	160	&	-		&	-	&	TD/V	&	6,10\\
15	&	CRCha	&	 10 59 06.97	&	-77 01 40.3	&	ChaI	&	160	&	-		&	-	&	V	&	8\\
16	&	WWCha	&	 11 10 00.11	&	-76 34 57.9	&	ChaI	&	160	&	[B]		&	-	&	V	&	8\\
17	&	11062554-7633418	&	 11 06 25.47	&	 -76 33 42.2	&	ChaI	&	160	&	[B]		&	GB	&		&	\\
18	&	T25	&	 11 07 19.15	&	-76 03 04.9	&	ChaI	&	160	&	[A][L]		&	Other	&	V	&	11,7\\
20	&	MPMus	&	13 22 07.55	&	-69 38 12.2	&	Isol.	&	100	&	[B]		&	Other	&	TD	&	12,13\\
21	&	HD142527	&	15 56 41.89	&	 -42 19 23.3 	&	Isol.	&	140	&	-		&	-	&	Imag.	&	14\\
22	&	J16232807-4015368	&	 16 23 28.09	&	-40 15 36.9	&	LupVI	&	150	&	[A][L]		&	GB	&		&	\\
23	&	Sz111	&	16 08 54.69	&	 -39 37 43.1	&	LupIII	&	200	&	[A][L]		&	c2d	&	LU	&	15\\
24	&	Lup60	&	 16 10 29.56	&	-39 22 14.7	&	LupIII	&	200	&	[A]		&	c2d	&	GG	&	16\\
25	&	J160830.3-390611	&	 16 08 30.26	&	 -39 06 11.1	&	LupIII	&	200	&	[B]		&	c2d	&	L	&	15\\
26	&	Sz104	&	 16 08 30.80	&	-39 05 48.8	&	LupIII	&	200	&	[B]		&	c2d	&	H	&	15\\
27	&	Sz91	&	 16 07 11.60	&	-39 03 47.7	&	LupIII	&	200	&	[L]		&	c2d	&	PF	&	16\\
28	&	J160855.5-390234	&	 16 08 55.52	&	-39 02 33.9	&	LupIII	&	200	&	[A]		&	c2d	&	L/PE	&	17,18\\
29	&	Sz84	&	 15 58 02.50	&	-37 36 02.8	&	LupI	&	150	&	[L]	*	&	c2d	&	TD	&	19\\
30	&	16182186-3730298	&	 16 18 21.88	&	-37 30 29.9	&	LupV	&	150	&	[L]		&	GB	&		&	\\
31	&	16225309-3724373	&	 16 22 53.10	&	-37 24 37.4	&	LupV	&	150	&	[L]		&	GB	&		&	\\
32	&	J19002346-3712242	&	 19 00 23.47	&	-37 12 24.2	&	CrA	&	150	&	[A]		&	GB	&		&	\\
33	&	HD135344	&	 15 15 48.44	&	-37 09 16.0	&	Isol.	&	140	&	[B]		&	-	&	Imag.	&	20\\
34	&	CrA-466	&	 19 01 18.95	&	-36 58 28.3	&	CrA	&	150	&	-		&	Other	&	TD	&	21\\
35	&	Sz76	&	 15 49 30.80	&	-35 49 52.0	&	LupI	&	150	&	[B]		&	Other	&		&	22\\
36	&	J154508.9-341734	&	 15 45 08.88	&	-34 17 33.7	&	LupI	&	150	&	-		&	Other	&	LU	&	15\\
38	&	RXJ1615.3-3255	&	 16 15 20.23	&	-32 55 04.9	&	Lup	&	185	&	[B]	*	&	c2d	&	Imag.	&	23\\
39	&	V4046Sgr	&	18 14 10.47	&	 -32 47 34.5	&	Isol.	&	73	&	-		&	-	&	Imag.	&	24\\
40	&	J163154.7-250324	&	 16 31 54.73	&	 -25 03 24.0	&	Oph	&	125	&	[B]		&	c2d	&	GG	&	25\\
41	&	J163205.5-250236	&	16 32 05.52	&	-25 02 36.2	&	Oph	&	125	&	[A]		&	c2d	&	PF	&	25\\
43	&	J163023.4-245416	&	 16 30 23.39	&	 -24 54 16.1	&	Oph	&	125	&	[B]		&	c2d	&		&	\\
44	&	WSB63	&	 16 28 54.06	&	-24 47 44.3	&	Oph	&	125	&	[A]		&	c2d	&	PF	&	25\\
45	&	SR24S	&	 16 26 58.51	&	-24 45 37.0	&	Oph	&	125	&	-		&	c2d	&	Imag.	&	26\\
46	&	RXJ1633.9-2442	&	 16 33 55.60	&	-24 42 05.0	&	Oph	&	125	&	[A]		&	c2d	&	Imag.	&	25,50\\
47	&	ISO-Oph43	&	 16 26 27.53	&	-24 41 53.6	&	Oph	&	125	&	[B]		&	c2d	&		&	\\
48	&	WSB60	&	 16 28 16.51	&	-24 36 58.3	&	Oph	&	125	&	[B]		&	c2d	&	Imag.	&	23\\
49	&	J163115.7-243402	&	 16 31 15.74	&	 -24 34 02.0	&	Oph	&	125	&	[B]		&	c2d	&	GG	&	25\\
50	&	J162245.4-243124	&	 16 22 45.39	&	 -24 31 23.8	&	Oph	&	125	&	[A]	*	&	c2d	&	PE	&	25\\
51	&	IRS48	&	16 27 37.19	&	-24 30 34.8	&	Oph	&	125	&	-		&	Other	&	Imag.	&	27,28\\
52	&	DoAr44	&	 16 31 33.46	&	-24 27 37.4	&	Oph	&	125	&	[B]		&	c2d	&	Imag.	&	29\\
53	&	J162435.2-242620	&	16 24 35.20	&	-24 26 20.0	&	Oph	&	125	&	[A]		&	c2d	&		&	\\
54	&	SR21	&	 16 27 10.28	&	-24 19 12.5	&	Oph	&	125	&	-		&	-	&	Imag.	&	20\\
55	&	J162309.2-241705	&	 16 23 09.22	&	-24 17 04.6	&	Oph	&	125	&	[A]		&	c2d	&	PTD	&	30\\
56	&	J163136.8-240420	&	 16 31 36.77	&	-24 04 19.8	&	Oph	&	125	&	[L]		&	c2d	&		&	\\
58	&	J162648.6-235634	&	 16 26 48.64	&	 -23 56 34.1	&	Oph	&	125	&	[B]		&	c2d	&		&	\\
59	&	J162802.6-235504	&	16 28 02.60	&	-23 55 04.0	&	Oph	&	125	&	[A]		&	c2d	&	PE	&	25\\
60	&	oph62	&	 16 25 06.92	&	-23 50 50.4	&	Oph	&	125	&	[A]	*	&	c2d	&	PF	&	25\\
61	&	J162532.5-232626	&	16 25 32.50	&	-23 26 26.0	&	Oph	&	125	&	[A]		&	c2d	&		&	\\
62	&	J162218.5-232148	&	 16 22 18.52	&	 -23 21 48.1	&	Oph	&	125	&	[B]		&	c2d	&	GG	&	25\\
63	&	DoAr28	&	16 26 47.42	&	 -23 14 52.2	&	Oph	&	125	&	-		&	HREL	&	TD	&	30\\
64	&	J160421.7-213028	&	16 04 21.70	&	 -21 30 28.4	&	UppS	&	145	&	[L]		&	-	&	Imag.	&	31\\
65	&	18015423-0437531	&	 18 01 54.24	&	-04 37 53.1	&	Ser	&	415	&	[L]		&	GB	&		&	\\
66	&	18044921-0436413	&	 18 04 49.20	&	 -04 36 41.5	&	Ser	&	415	&	[B]		&	GB	&		&	\\
67	&	18270980-0414297	&	 18 27 09.79	&	-04 14 29.8	&	Ser	&	415	&	[L]		&	GB	&		&	\\
68	&	18272873-0406248	&	 18 27 28.73	&	-04 06 24.8	&	Ser	&	415	&	[L]		&	GB	&		&	\\
69	&	18273408-0403247	&	 18 27 34.08	&	-04 03 24.8	&	Ser	&	415	&	[L]		&	GB	&		&	\\
70	&	18273858-0402289	&	 18 27 38.57	&	-04 02 28.9	&	Ser	&	415	&	[A][L]		&	GB	&		&	\\
71	&	18255765-0357040	&	 18 25 57.66	&	-03 57 04.0	&	Ser	&	415	&	[L]		&	GB	&		&	\\
73	&	18291383-0342355	&	 18 29 13.84	&	-03 42 35.5	&	Ser	&	415	&	[L]		&	GB	&		&	\\
74	&	18284156-0341507	&	 18 28 41.56	&	-03 41 50.7	&	Ser	&	415	&	[L]		&	GB	&		&	\\
75	&	18283439-0339371	&	 18 28 34.40	&	-03 39 37.2	&	Ser	&	415	&	[L]		&	GB	&		&	\\
76	&	18272161-0314158	&	 18 27 21.62	&	-03 14 15.9	&	Ser	&	415	&	[L]		&	GB	&		&	\\
77	&	18222604-0304383	&	 18 22 26.04	&	-03 04 38.3	&	Ser	&	415	&	[L]		&	GB	&		&	\\
78	&	18330328-0244021	&	 18 33 03.30	&	 -02 44 02.2	&	Ser	&	415	&	[B]		&	GB	&		&	\\
79	&	18324685-0243273	&	 18 32 46.86	&	-02 43 27.4	&	Ser	&	415	&	[L]		&	GB	&		&	\\
80	&	18304127-0242335	&	 18 30 41.26	&	-02 42 33.7	&	Ser	&	415	&	[L]		&	GB	&		&	\\
81	&	18324783-0239401	&	 18 32 47.83	&	 -02 39 40.1	&	Ser	&	415	&	[B]		&	GB	&		&	\\
82	&	J18321275-0222377	&	 18 32 12.75	&	-02 22 37.8	&	Ser	&	415	&	[A][L]		&	GB	&		&	\\
83	&	18292883-0221157	&	 18 29 28.84	&	-02 21 15.7	&	Ser	&	415	&	[A][L]		&	GB	&		&	\\
84	&	18291450-0220575	&	 18 29 14.50	&	-02 20 57.5	&	Ser	&	415	&	[L]		&	GB	&		&	\\
85	&	18304121-0220189	&	 18 30 41.20	&	 -02 20 19.1	&	Ser	&	415	&	[B]		&	GB	&		&	\\
86	&	J18314556-0218408	&	 18 31 45.57	&	-02 18 40.9	&	Ser	&	415	&	[A]		&	GB	&		&	\\
88	&	18311986-0208161	&	18 31 19.86 	&	-02 08 16.1	&	Ser	&	415	&	[A]		&	GB	&		&	\\
89	&	18323005-0204130	&	 18 32 30.06	&	-02 04 13.0	&	Ser	&	415	&	[A][L]		&	GB	&		&	\\
90	&	18292804-0204042	&	 18 29 28.07	&	-02 04 04.7	&	Ser	&	415	&	[A]		&	GB	&		&	\\
91	&	18293961-0202414	&	 18 29 39.60	&	 -02 02 41.4	&	Ser	&	415	&	[B]		&	GB	&		&	\\
92	&	J18303289-0200514	&	 18 30 32.89	&	-02 00 51.3	&	Ser	&	415	&	[A][L]		&	GB	&		&	\\
93	&	18311732-0200461	&	18 31 17.32 	&	-02 00 46.1	&	Ser	&	415	&	[A]		&	GB	&		&	\\
94	&	18312875-0159125	&	18 31 28.75 	&	-01 59 12.5	&	Ser	&	415	&	[A]		&	GB	&		&	\\
95	&	18315497-0157330	&	 18 31 54.98	&	-01 57 33.1	&	Ser	&	415	&	[L]		&	GB	&		&	\\
96	&	18313657-0157320	&	18 31 36.57 	&	-01 57 32.0	&	Ser	&	415	&	[A]		&	GB	&		&	\\
97	&	18313343-0155182	&	18 31 33.43 	&	-01 55 18.2	&	Ser	&	415	&	[A]		&	GB	&		&	\\
98	&	18315077-0153393	&	 18 31 50.77	&	 -01 53 39.3	&	Ser	&	415	&	[B]		&	GB	&		&	\\
99	&	J18303321-0152563	&	 18 30 33.22	&	-01 52 56.2	&	Ser	&	415	&	[A][L]		&	GB	&		&	\\
100	&	18295741-0151541	&	 18 29 57.41	&	-01 51 54.1	&	Ser	&	415	&	[A][L]		&	GB	&		&	\\
101	&	18294721-0148301	&	 18 29 47.21	&	-01 48 30.2	&	Ser	&	415	&	[A][L]		&	GB	&		&	\\
102	&	18293368-0145103	&	 18 29 33.69	&	 -01 45 10.3	&	Ser	&	415	&	[B]		&	GB	&		&	\\
103	&	18290819-0139215	&	 18 29 08.19	&	 -01 39 21.5	&	Ser	&	415	&	[B]		&	GB	&		&	\\
104	&	18314110-0128035	&	 18 31 41.10	&	-01 28 03.6	&	Ser	&	415	&	[L]		&	GB	&		&	\\
105	&	18290391-0115357	&	 18 29 03.92	&	-01 15 35.8	&	Ser	&	415	&	[L]		&	GB	&		&	\\
106	&	18371575-0026561	&	 18 37 15.75	&	-00 26 56.1	&	Ser	&	415	&	[L]		&	GB	&		&	\\
107	&	18381010-0023452	&	 18 38 10.10	&	-00 23 45.2	&	Ser	&	415	&	[L]		&	GB	&		&	\\
108	&	18371444-0023261	&	 18 37 14.45	&	-00 23 26.2	&	Ser	&	415	&	[L]		&	GB	&		&	\\
110	&	18385989-0008097	&	 18 38 59.90	&	-00 08 09.9	&	Ser	&	415	&	[L]		&	GB	&		&	\\
111	&	J182813.5+000-249	&	 18 28 13.51	&	 -00 02 49.1	&	Ser	&	415	&	[B]		&	c2d	&	TT	&	32\\
112	&	J182821.6+000016	&	 18 28 21.58	&	 +00 00 16.4	&	Ser	&	415	&	[B]		&	c2d	&	L	&	32\\
113	&	18384257+0001324	&	 18 38 42.59	&	+00 01 32.5	&	Ser	&	415	&	[A][L]		&	GB	&		&	\\
114	&	18392594+0006382	&	 18 39 25.96	&	+00 06 38.4	&	Ser	&	415	&	[A]		&	GB	&		&	\\
115	&	J182850.2+000950	&	 18 28 50.21	&	 +00 09 49.7	&	Ser	&	415	&	[B]		&	c2d	&	F	&	32\\
116	&	183549.4+001002	&	 18 35 49.38	&	+00 10 01.7	&	Ser	&	415	&	[L]		&	GB	&		&	\\
117	&	18385571+0014431	&	 18 38 55.72	&	+00 14 43.1	&	Ser	&	415	&	[A][L]		&	GB	&		&	\\
118	&	18394048+0014497	&	 18 39 40.50	&	 +00 14 49.7	&	Ser	&	415	&	[B]		&	GB	&		&	\\
119	&	18374209+0016519	&	 18 37 42.09	&	+00 16 52.0	&	Ser	&	415	&	[L]		&	GB	&		&	\\
120	&	J182911.5+002039	&	 18 29 11.49	&	 +00 20 38.8	&	Ser	&	415	&	[A]	*	&	c2d	&		&	\\
121	&	18375663-0023253	&	18 37 56.63 	&	-00 23 25.3	&	Ser	&	415	&	[A]		&	GB	&		&	\\
122	&	18381580+0024218	&	 18 38 15.81	&	+00 24 21.9	&	Ser	&	415	&	[L]		&	GB	&		&	\\
123	&	J18295130+0027477	&	 18 29 51.30	&	+00 27 47.9	&	Ser	&	415	&	[L]		&	c2d	&	LU	&	32\\
124	&	serp22	&	 18 28 29.06	&	+00 27 56.0	&	Ser	&	415	&	[A]	*	&	c2d	&		&	\\
125	&	18401205+0029276	&	 18 40 12.06	&	 +00 29 27.7	&	Ser	&	415	&	[B]		&	GB	&		&	\\
126	&	J182901.2+002933	&	 18 29 01.22	&	 +00 29 33.0	&	Ser	&	415	&	[B]		&	c2d	&	L	&	32\\
127	&	Serp127	&	 18 29 44.10	&	+00 33 56.0	&	Ser	&	415	&	[A][L]	*	&	c2d	&	LU	&	32\\
128	&	J182935.6+003504	&	 18 29 35.62	&	+00 35 03.9	&	Ser	&	415	&	[A]	*	&	c2d	&	LU	&	32\\
129	&	18381447+0035099	&	 18 38 14.48	&	+00 35 09.8	&	Ser	&	415	&	[L]		&	GB	&		&	\\
130	&	18401486+0037042	&	 18 40 14.88	&	+00 37 04.2	&	Ser	&	415	&	[A][L]		&	GB	&		&	\\
131	&	Serp111	&	 18 29 36.19	&	+00 42 16.7	&	Ser	&	415	&	[A]	*	&	c2d	&	LU	&	32\\
132	&	J182955.3+004939	&	 18 29 55.32	&	+00 49 39.3	&	Ser	&	415	&	[A]		&	c2d	&	TT	&	32\\
133	&	J183008.6+005847	&	 18 30 08.62	&	 +00 58 46.7	&	Ser	&	415	&	[B]		&	c2d	&	TT	&	32\\
134	&	RXJ0432.8+1735	&	04 32 53.24	&	+17 35 33.7	&	Tau	&	140	&	-		&	Tau	&	TD	&	33\\
135	&	DMTau	&	 04 33 48.73	&	+18 10 10.0	&	Tau	&	140	&	-		&	Tau	&	Imag./V	&	34,10\\
136	&	UXTauA	&	 04 30 04.00	&	+18 13 49.3	&	Tau	&	140	&	[A]		&	Other	&	Imag.	&	34,35\\
137	&	043339.0+222720	&	 04 33 39.00	&	+22 27 20.0	&	Tau	&	140	&	[A]		&	Tau	&		&	33\\
138	&	043326.2+224529	&	 04 33 26.20	&	+22 45 29.0	&	Tau	&	140	&	[L]		&	Tau	&		&	36\\
139	&	J04390525+2337450	&	04 39 05.25	&	+23 37 45.0	&	Tau	&	140	&	[B]		&	GB	&	PTD	&	33\\
140	&	043649.1+241258	&	 04 36 49.10	&	+24 12 58.0	&	Tau	&	140	&	[L]		&	Tau	&	DD	&	37\\
142	&	MWC758	&	 05 30 27.53	&	+25 19 56.9	&	Isol.	&	200	&	-		&	Tau	&	Imag.	&	38\\
144	&	044555.7+261858	&	 04 45 55.70	&	+26 18 58.0	&	Tau	&	140	&	[L]		&	Tau	&		&	\\
145	&	DHTauAB	&	 04 29 41.50	&	+26 32 58.0	&	Tau	&	140	&	[A]		&	Tau	&		&	\\
146	&	043044.7+263308	&	 04 30 44.70	&	+26 33 08.0	&	Tau	&	140	&	[L]		&	Tau	&		&	\\
147	&	J04214631+2659296	&	04 21 46.32	&	+26 59 29.6	&	Tau	&	140	&	[A]		&	Tau	&		&	\\
148	&	IPTau	&	 04 24 57.08	&	+27 11 56.5	&	Tau	&	140	&	-		&	Tau	&	V	&	8\\
149	&	V892Tau	&	04 18 40.62	&	+28 19 15.5	&	Tau	&	140	&	-		&	Tau	&	CB-disk	&	39\\
150	&	V410X-ray6	&	 04 19 01.10	&	+28 19 42.0	&	Tau	&	140	&	[A]		&	Tau	&	PF/TD	&	40,41\\
151	&	042254.6+282354	&	 04 22 54.60	&	+28 23 54.0	&	Tau	&	140	&	[L]		&	Tau	&		&	\\
152	&	V819Tau	&	04 19 26.26 	&	+28 26 14.3	&	Tau	&	140	&	-		&	Tau	&	TD	&	33\\
153	&	RYTau	&	 04 21 57.41	&	+28 26 35.6	&	Tau	&	140	&	-		&	Tau	&	Imag./V	&	42,10\\
154	&	V410X-ray2	&	 04 18 34.40	&	+28 30 30.0	&	Tau	&	140	&	[A]		&	Tau	&	TD	&	33\\
155	&	041542.7+290959	&	 04 15 42.78	&	+29 09 59.0	&	Tau	&	140	&	[A][L]		&	Tau	&	TD	&	33\\
156	&	041332.3+291726	&	 04 13 32.30	&	+29 17 26.0	&	Tau	&	140	&	[A]		&	Tau	&		&	\\
157	&	J032800.1+300847	&	 03 28 00.09	&	 +30 08 47.0	&	Per	&	250	&	[B]		&	c2d	&		&	\\
158	&	LkCa19	&	04 55 36.96	&	+30 17 55.2	&	Tau	&	140	&	-		&	Other	&	TD	&	43,35\\
159	&	ABAur	&	 04 55 45.85	&	+30 33 04.3	&	Tau	&	140	&	-		&	-	&	Imag.	&	44\\
160	&	J033341.3+311341	&	 03 33 41.29	&	 +31 13 41.0	&	Per	&	250	&	[B]		&	c2d	&		&	\\
161	&	ASR118	&	 03 28 56.97	&	+31 16 22.3	&	Per	&	250	&	[B]	*	&	c2d	&		&	\\
162	&	MBO22	&	 03 29 29.27	&	+31 18 34.7	&	Per	&	250	&	[A]	*	&	c2d	&	PF	&	25\\
163	&	J032856.6+311836	&	 03 28 56.65	&	 +31 18 35.5	&	Per	&	250	&	[B]		&	c2d	&		&	\\
164	&	J034219.3+314327	&	 03 42 19.27	&	 +31 43 27.0	&	Per	&	250	&	[B]		&	c2d	&		&	\\
165	&	J034227.1+314433	&	 03 42 27.12	&	 +31 44 32.9	&	Per	&	250	&	[A]	*	&	c2d	&		&	\\
166	&	J034109.1+314438	&	 03 41 09.13	&	 +31 44 37.9	&	Per	&	250	&	[B]		&	c2d	&		&	\\
167	&	J034355.2+315532	&	03 43 55.20	&	+31 55 32.0	&	Per	&	250	&	[A]		&	c2d	&		&	\\
168	&	J034434.8+315655	&	 03 44 34.81	&	 +31 56 55.2	&	Per	&	250	&	[A]	*	&	c2d	&		&	\\
169	&	IC348LRL190	&	 03 44 29.23	&	+32 01 15.7	&	Per	&	250	&	[A]	*	&	c2d	&		&	\\
171	&	J034520.5+320634	&	 03 45 20.46	&	 +32 06 34.5	&	Per	&	250	&	[B]		&	c2d	&		&	\\
172	&	IC348-67	&	 03 43 44.63	&	+32 08 17.8	&	Per	&	250	&	[A]		&	c2d	&	PF	&	37\\
173	&	LkH-alpha330	&	 03 45 48.29	&	+32 24 11.8	&	Per	&	250	&	[B]		&	c2d	&	Imag.	&	20\\
174	&	J04300424+3522238	&	 04 30 04.25	&	+35 22 23.8	&	Aur	&	450	&	[A][L]		&	GB	&	PF	&	37\\
175	&	J04303235+3536133	&	 04 30 32.35	&	+35 36 13.4	&	Tau	&	140	&	[L]		&	GB	&	PF	&	37\\
176	&	04300980+3540355	&	 04 30 09.80	&	 +35 40 35.6	&	Tau	&	140	&	[B]		&	GB	&		&	\\
177	&	J04304004+3542101	&	 04 30 40.05	&	+35 42 10.3	&	Tau	&	140	&	[L]		&	GB	&	GG	&	37\\
178	&	J04303831+3549591	&	 04 30 38.27	&	+35 49 59.3	&	Aur	&	450	&	[L]		&	GB	&		&	\\
179	&	J160044.5-415531	&	 16 00 44.53	&	 -41 55 31.2	&	LupIV	&	150	&	[B]		&	c2d	&	PF	&	16\\
180	&	J190058.1-364505	&	 19 00 58.05	&	 -36 45 05.0	&	UppS	&	145	&	[A]		&	GB	&	PF	&	16\\
181	&	03445614+3209152	&	 03 44 56.14	&	 +32 09 15.1	&	Per	&	250	&	[A]		&	c2d	&	PF	&	37\\
182	&	03442156+3215098	&	 03 44 21.58	&	 +32 15 09.7	&	Per	&	250	&	[A]		&	c2d	&	PE	&	37\\
183	&	03442257+3201536	&	 03 44 22.58	&	 +32 01 53.8	&	Per	&	250	&	[A]		&	c2d	&	PE	&	37\\
184	&	04330422+2921499	&	 04 33 04.22	&	 +29 21 50.0	&	Per	&	250	&	[A]		&	Tau	&	DD	&	37\\
185	&	J160825.76-390601.1	&	16 08 25.76	&	 -39 06 01.1	&	LupIII	&	200	&	[B]		&	c2d	&	TT	&	15\\
186	&	RXJ1556.1-3655	&	 15 56 02.10	&	-36 55 28.2	&	LupIII	&	150	&	[B]		&	Other	&		&	22\\
187	&	043150.5+242418	&	 04 31 50.50	&	+24 24 18.0	&	Tau	&	140	&	[B]		&	Tau	&		&	\\
188	&	041413.5+281249	&	 04 14 13.50	&	+28 12 49.0	&	Tau	&	140	&	[B]		&	Tau	&		&	\\
189	&	041841.3+282725	&	 04 18 41.30	&	+28 27 25.0	&	Tau	&	140	&	[B]		&	Tau	&		&	\\
190	&	042025.5+270035	&	 04 20 25.50	&	+27 00 35.0	&	Tau	&	140	&	[B]		&	Tau	&		&	\\
191	&	042921.6+270125	&	 04 29 21.60	&	+27 01 25.0	&	Tau	&	140	&	[B]		&	Tau	&		&	\\
192	&	043249.1+225302	&	 04 32 49.10	&	+22 53 02.0	&	Tau	&	140	&	[B]		&	Tau	&		&	\\
193	&	044221.0+252034	&	 04 42 21.00	&	+25 20 34.0	&	Tau	&	140	&	[B]		&	Tau	&		&	\\
194	&	041539.1+281858	&	 04 15 39.10	&	+28 18 58.0	&	Tau	&	140	&	[B]		&	Tau	&		&	\\
195	&	042318.2+264115	&	 04 23 18.20	&	+26 41 15.0	&	Tau	&	140	&	[B]		&	Tau	&		&	\\
196	&	041414.5+282758	&	 04 14 14.50	&	+28 27 58.0	&	Tau	&	140	&	[B]		&	Tau	&		&	\\
197	&	041915.8+290626	&	 04 19 15.80	&	+29 06 26.0	&	Tau	&	140	&	[B]		&	Tau	&		&	\\
198	&	042155.6+275506	&	 04 21 55.60	&	+27 55 06.0	&	Tau	&	140	&	[B]		&	Tau	&		&	\\
200	&	J160710.08-391103.5	&	 16 07 10.08	&	-39 11 03.5	&	LupIII	&	200	&	[B]		&	c2d	&	L	&	15\\
201	&	J032741.47+302016.8	&	 03 27 41.47	&	+30 20 16.8	&	Per	&	250	&	[B]		&	c2d	&		&	\\
202	&	J034345.17+320358.6	&	 03 43 45.17	&	+32 03 58.6	&	Per	&	250	&	[A]		&	c2d	&		&	\\
203	&	J182815.26-000243.3	&	 18 28 15.26	&	-00 02 43.3	&	Ser	&	415	&	[B]		&	c2d	&		&	\\
204	&	J162715.89-243843.2	&	 16 27 15.89	&	-24 38 43.2	&	Oph	&	125	&	[B]		&	c2d	&		&	30\\
301	&	J130455.7-773949	&	 13 04 55.74	&	-77 39 49.5	&	ChaII	&	180	&	[B]	*	&	c2d	&		&	\\
303	&	J160115.5-415235	&	 16 01 15.55	&	-41 52 35.3	&	LupIV	&	150	&	[B]		&	c2d	&	F	&	15\\
307	&	16083070-3828268	&	16 08 30.70	&	-38 28 26.8	&	LupIII	&	200	&	[B]		&	HREL	&	L	&	15\\
309	&	TWHya	&	 11 01 51.91	&	-34 42 17.0	&	TWH	&	50	&	-		&	-	&	TD	&	45\\
310	&	15395742-3414567	&	15 39 57.42	&	-34 14 56.7	&	LupI	&	150	&	-		&	HREL	&		&	\\
314	&	16281385-2456113	&	16 28 13.85	&	-24 56 11.3	&	Oph	&	125	&	[B]		&	HREL	&		&	\\
316	&	16271587-2438433	&	16 27 15.87	&	-24 38 43.3	&	Oph	&	125	&	[B]		&	HREL	&		&	\\
317	&	16312019-2430009	&	16 31 20.19	&	-24 30 00.9	&	Oph	&	125	&	[B]		&	HREL	&		&	\\
318	&	DoAr21	&	 16 26 03.03	&	-24 23 36.4	&	Oph	&	125	&	[A]	*	&	c2d	&		&	\\
319	&	J162740.3-242204	&	 16 27 40.27	&	-24 22 04.0	&	Oph	&	125	&	[A]		&	c2d	&	CB	&	25\\
321	&	Serp48	&	 18 28 58.08	&	+00 17 24.5	&	Ser	&	415	&	[LL]	*	&	c2d	&		&	\\
322	&	18302986+0035004	&	18 30 29.86	&	+00 35 00.4	&	Ser	&	415	&	[B]		&	HREL	&		&	\\
325	&	LkCa15	&	 04 39 17.78	&	+22 21 03.5	&	Tau	&	140	&	[A]		&	Tau	&	Imag./V	&	46,10\\
326	&	CoKu-Tau-4	&	 04 41 16.79	&	+28 40 00.5	&	Tau	&	140	&	-		&	-	&	CB-disk	&	47\\
329	&	GMAur	&	 04 55 10.98	&	+30 21 59.4	&	Tau	&	140	&	[A]		&	-	&	Imag./V	&	48,10\\
333	&	03370363+3039291	&	03 37 03.63	&	+30 39 29.1	&	Per	&	250	&	-		&	HREL	&		&	\\
334	&	03401579+3055047	&	03 40 15.79	&	+30 55 04.7	&	Per	&	250	&	[B]		&	HREL	&		&	\\
335	&	J033234.0+310056	&	03 32 34.00	&	+31 00 56.0	&	Per	&	250	&	[B]	*	&	c2d	&		&	\\
348	&	UScoJ155837.1-225724	&	15 58 36.91	&	-22 57 15.3	&	UppS	&	145	&	[B]		&	Other	&		&	49\\
349	&	RXJ1842.9-3532	&	18 42 57.95	&	-35 32 42.7	&	UppS	&	145	&	[A]		&	Other	&		&	49\\
350	&	RXJ1852.3-3700	&	18 52 17.30	&	-37 00 11.9	&	UppS	&	145	&	[A]		&	Other	&		&	49\\
\end{longtable}
{\bf Refs.}
       1) \citet{Wahhaj2010}, 
       2) \citet{Huelamo2015}, 
       3) \citet{Megeath2005}, 
       4) \citet{Sicilia2009}, 
       5) \citet{Bouwman2010}, 
       6) \citet{Kim2009}, 
       7) \citet{Luhman2008}, 
       8) \citet{Espaillat2011}, 
       9) \citet{Kim2009}, 
      10) \citet{Espaillat2011}, 
      11) \citet{Espaillat2011}, 
      12) \citet{Cortes2009}, 
      13) \citet{Silverstone2006}, 
      14) \citet{Ohashi2008}, 
      15) \citet{Merin2008}, 
      16) \citet{Romero2012}, 
      17) \citet{Merin2008}, 
      18) \citet{Romero2012}, 
      19) \citet{Merin2010}, 
      20) \citet{Brown2009}, 
      21) \citet{Sicilia2008}, 
      22) \citet{Padgett2006}, 
      23) \citet{Andrews2011}, 
      24) \citet{Rosenfeld2013}, 
      25) \citet{Cieza2010}, 
      26) \citet{Andrews2010}, 
      27) \citet{vanderMarel2013}, 
      28) \citet{vanKempen2009}, 
      29) \citet{Andrews2009}, 
      30) \citet{McClure2010}, 
      31) \citet{Mathews2012}, 
      32) \citet{Harvey2007}, 
      33) \citet{Furlan2011}, 
      34) \citet{Andrews2011}, 
      35) \citet{Wahhaj2010}, 
      36) \citet{Rebull2010}, 
      37) \citet{Cieza2012}, 
      38) \citet{Isella2010mwc758}, 
      39) \citet{Monnier2008}, 
      40) \citet{Cieza2012}, 
      41) \citet{Furlan2011}, 
      42) \citet{Isella2010rytau}, 
      43) \citet{Furlan2011}, 
      44) \citet{Pietu2005}, 
      45) \citet{Najita2010}, 
      46) \citet{Pietu2006}, 
      47) \citet{Ireland2008}, 
      48) \citet{Hughes2009}, 
      49) \citet{Carpenter2008},
      50) \citet{Cieza2012rxj}
\tablefoot{
\tablefoottext{a}{Full names of the regions: Cha = Chamaeleon, Lup = Lupus, CrA = Corona Australis, Oph = Ophiuchus, UppS = Upper Sco, Oph = Ophiuchus, Tau = Taurus, Aur = Auriga, Per = Perseus, TWH = TW Hydrae, Isol. = Isolated.}
\tablefoottext{b}{Distances were adopted from the literature as follows: 120 pc for Oph \citep{Loinard2008}; 150 pc for Lup I, IV, V and VI and 200 pc for LupIII \citep{Comeron2008,Comeron2009}; 250 pc for Per \citep{Jorgensen2006}; 140 pc for Tau \citep{Kenyon2008}; 450 pc for Aur \citep{Broekhoven-Fiene2014}; 145 pc for Upp Sco \citep{Carpenter2008}; 150 pc for Corona Australis \citep{Sicilia2008}; 109 pc for $\epsilon$ Cha\citep{Torres2008}; 160 pc for ChaI \citep{Kim2009}; 180 pc for ChaII \citep{Alcala2008}; 97 pc for $\eta$ Cha\citep{Mamajek1999}; and 430 pc for Ser \citep{Dzib2010}.}
\tablefoottext{c}{An asterix (*) indicates this target was part of the M10 sample.}
\tablefoottext{d}{The full explanation previous classifications is as follows, according to their recording papers. 1) \citet{Cieza2010}: "PF"=Planet-forming disk, "GG"=Grain-growth dominated disk, "PE"=Photoevaportive disk. 2) \citet{Merin2008,Harvey2007}: "L"=low infrared excess or anemic disk, "H"=high infrared excess, "T"=T Tauri-like infrared excess. 3) Others: "PTD" = pre-transitional disk, "V" = sea-saw variability.}
}

\newpage

{\small
\begin{longtable}{lllllllllll}
\caption{Stellar parameters}
\label{tbl:stellarparams}\\
\hline
\hline
ID&Name&SpT&T$_{\rm eff}$&A$_V$&L$_*$&M$_*$&EW[H$\alpha$]&FW10\%[H$\alpha$]\tablefootmark{a}&Accretion\tablefootmark{d}&Ref\\
&&&(K)&(mag)&(L$_{\odot}$)&(M$_{\odot}$)&($\AA$)&(km s$^{-1}$&(Y/N)&\\
\hline
\endfirsthead

\caption{Stellar parameters continued.}\\
\hline
\hline
ID&Name&SpT&T$_{\rm eff}$&A$_V$&L$_*$&M$_*$&EW[H$\alpha$]&FW10\%[H$\alpha$]\tablefootmark{a}&Accretion\tablefootmark{d}&Ref\\
&&&(K)&(mag)&(L$_{\odot}$)&(M$_{\odot}$)&($\AA$)&(km s$^{-1}$&(Y/N)&\\
\hline
\endhead

\hline
\endfoot

\hline
\endlastfoot

1&TCha&K0&5250&2&1.34&1.1&-&400&Y&1,2\\
2&RECX11&K5&4350&0.4&0.73&1.2&4.4&330&Y&3\\
3&RECX5&M4&3370&2.4&0.14&0.3&35&330&Y&3\\
4&CHXR22E&M3.5&3370&4.9&0.26&0.4&-&-&U&4\\
5&ISO52&M4&3370&3&0.17&0.3&\tablefootmark{b}&-&N&5\\
6&CSCha&K2&4780&1.3&1.88&1.5&\tablefootmark{b}&-&Y&6\\
7&11094742-7726290&M3.25&3470&5.7&0.16&0.4&200&-&Y&7\\
9&T54&K0&5250&0&2.47&1.4&\tablefootmark{b}&-&N&5\\
10&T21&G5&5770&3&15.19&2.3&-&-&U&4\\
11&SZCha&K2&4780&0.7&1.36&1.4&\tablefootmark{b}&-&Y&6\\
12&T35&K7&4060&3.4&0.41&1&\tablefootmark{b}&-&Y&6\\
13&ISO-ChaII29&M0&3850&4.3&0.57&1&1&249&N&8\\
14&T56&M0.5&3720&0.3&0.34&0.8&\tablefootmark{b}&-&Y&6\\
15&CRCha&K0&5250&2.4&4.99&2&\tablefootmark{b}&-&Y&5\\
16&WWCha&K5&4350&3.4&4.41&0.9&65&-&Y&9,4\\
17&11062554-7633418&M6&3050&6.9&0.17&0.1&43.6&-&Y&10\\
18&T25&M2.5&3470&2.3&0.33&0.5&\tablefootmark{b}&-&Y&6\\
20&MPMus&K1&5080&0.9&1.35&1.2&-47&-&Y&11\\
21&HD142527&F6&6360&1.4&23.58&2.3&\tablefootmark{c}&-&Y&12\\
22&J16232807-4015368&M3.5&3370&4.1&0.2&0.3&-0.4&-&N&13\\
23&Sz111&M1.5&3580&0.1&0.38&0.6&-&375&Y&14\\
24&Lup60&M4.5&3240&2.6&0.22&0.3&12.9&-&N&13\\
25&J160830.3-390611&M4&3370&3.7&0.47&0.5&-&426&Y&14\\
26&Sz104&M5&3240&1&0.19&0.3&-&201&Y&14\\
27&Sz91&M0.5&3720&0.8&0.29&0.7&-&374&Y&14\\
28&J160855.5-390234&M6&3050&0&0.22&0.1&-&189&Y&14\\
29&Sz84&M5&3240&2.7&0.32&0.3&\tablefootmark{b}&-&Y&6\\
30&16182186-3730298&K5&4350&0.8&0.05&1&0.1&-&N&13\\
31&16225309-3724373&M5.5&3050&4.9&1.51&-&-&-&U&13\\
32&J19002346-3712242&-&(4060)&14&0.79&1&0.3&-&N&13\\
33&HD135344&F4&6590&0.8&11.48&1.8&\tablefootmark{c}&-&Y&12\\
34&CrA-466&M2&3580&5.7&0.12&0.5&-14.5&-&N&15\\
35&Sz76&M1&3720&1.3&0.21&0.7&10.3&227&N&16\\
36&J154508.9-341734&M6&3050&8.4&0.49&0.1&174.5&-&Y&13\\
38&RXJ1615.3-3255&K7&4060&0&1.08&1.2&\tablefootmark{b}&-&Y&6\\
39&V4046Sgr&K5&4350&0.5&0.73&1.2&26.7&-&Y&13\\
40&J163154.7-250324&K7&4060&4.6&0.88&1.2&-&470&Y&13,17\\
41&J163205.5-250236&M2&3580&2.9&0.12&0.5&-&567&Y&17\\
43&J163023.4-245416&M3&3470&4.2&0.38&0.5&88.7&-&Y&13\\
44&WSB63&M2&3580&3.9&0.38&0.6&-&365&Y&17\\
45&SR24S&K2&4780&5.8&2.07&1.5&\tablefootmark{c}&-&Y&18\\
46&RXJ1633.9-2442&M0&3850&3.3&0.44&0.9&-&301&Y&13,17\\
47&ISO-Oph43&-&(4060)&15&0.41&1&-0.6&-&N&13\\
48&WSB60&M5.5&3050&5.8&0.38&0.1&\tablefootmark{b}&-&Y&6\\
49&J163115.7-243402&K5&4350&1.9&2.53&0.9&-&450&Y&13,17\\
50&J162245.4-243124&M2&3580&2.9&0.38&0.6&\tablefootmark{b}&-&Y&6\\
51&IRS48&A0&9520&10.6&14.5&1.9&\tablefootmark{c}&-&Y&19\\
52&DoAr44&K3&4730&2.9&1.46&1.4&\tablefootmark{b}&-&Y&6\\
53&J162435.2-242620&M4&3370&6.7&0.12&0.3&60.2&-&Y&13\\
54&SR21&G3&5830&5.5&6.5&1.7&\tablefootmark{b}&-&Y&6\\
55&J162309.2-241705&G&5830&6&1.76&1.2&8.4&-&Y&13\\
56&J163136.8-240420&-&(4060)&11.2&0.04&-&-&-&U&\\
58&J162648.6-235634&K8&3960&3.3&0.43&1&28.4&-&Y&13,20\\
59&J162802.6-235504&M4&3370&5.2&0.2&0.3&-&159&N&13,17\\
60&oph62&M0&3850&3.7&0.28&0.9&\tablefootmark{b}&-&Y&6\\
61&J162532.5-232626&M2.5&3470&1.5&0.19&0.4&7.4&-&N&13\\
62&J162218.5-232148&K5&4350&1.8&0.83&1.3&-&493&Y&13,17\\
63&DoAr28&K5&4350&2.6&0.73&1.2&44.5&-&Y&13\\
64&J160421.7-213028&K5&4350&0&0.55&1.1&5.3&-&Y&13\\
65&18015423-0437531&F3&6740&0.8&10.71&1.7&-4.3&-&N&13\\
66&18044921-0436413&M5&3240&6.6&3.45&-&32&-&Y&13\\
67&18270980-0414297&-&(4060)&6.7&3.92&0.6&-&-&U&\\
68&18272873-0406248&M-GIANT&-&-&-&-&-&-&N&13\\
69&18273408-0403247&B9&10500&4.4&39.52&2.4&-9.8&-&N&13\\
70&18273858-0402289&G5&5770&5.2&3.23&1.4&5.8&-&Y&13\\
71&18255765-0357040&FG&6030&7.4&3.05&1.3&1.2&-&N&13\\
73&18291383-0342355&-&4060&14.4&7.14&0.7&-&-&U&\\
74&18284156-0341507&-&(4060)&10.6&1.53&1.2&-&-&U&\\
75&18283439-0339371&-&(4060)&13.2&3.92&0.6&-&-&U&\\
76&18272161-0314158&A0&9520&3.4&29.59&2.2&-9.2&-&N&13\\
77&18222604-0304383&F5&6440&4.5&4.48&1.4&-3.2&-&N&13\\
78&18330328-0244021&M5.5&3050&7&1.25&-&21.6&-&Y&13\\
79&18324685-0243273&A0&9520&1.8&23.97&2.2&-10.3&-&N&13\\
80&18304127-0242335&M-GIANT&-&-&-&-&-&-&N&13\\
81&18324783-0239401&FG&6030&10.8&4.3&1.4&-15.8&-&N&13\\
82&J18321275-0222377&-&(4060)&9.3&0.02&1&-&-&U&\\
83&18292883-0221157&M4.5&3240&6&0.22&0.3&68.5&-&Y&13\\
84&18291450-0220575&M-GIANT&-&-&-&-&-&-&N&13\\
85&18304121-0220189&-&(4060)&14&0.12&1&-&-&U&\\
86&J18314556-0218408&-&(4060)&14.9&0.12&1&-&-&U&\\
88&18311986-0208161&-&(4060)&11&2.83&1&-&-&U&\\
89&18323005-0204130&A6&8350&5.7&25.22&2&-5.7&-&N&13\\
90&18292804-0204042&-&(4060)&10.6&0.2&1&-&-&U&\\
91&18293961-0202414&-&(4060)&12&0.3&0.9&-&-&U&\\
92&J18303289-0200514&-&(4060)&14.9&0.12&1&-&-&U&\\
93&18311732-0200461&A0&9520&12.4&32.63&1&-&-&N&13\\
94&18312875-0159125&-&(4060)&14.9&2.83&0.6&-&-&U&\\
95&18315497-0157330&-&(4060)&8.7&0.71&1&-&-&U&\\
96&18313657-0157320&-&(4060)&7&1.92&1.3&-&-&U&\\
97&18313343-0155182&-&(4060)&9.3&0.06&1&-&-&U&\\
98&18315077-0153393&-&(4060)&11.4&0.12&1&-&-&U&\\
99&J18303321-0152563&A2&8970&2.7&16.85&1.9&-9.6&-&N&13\\
100&18295741-0151541&A7&7850&8.8&13.68&1.8&-6.6&-&N&13\\
101&18294721-0148301&-&(4060)&7.2&0.79&1.1&0.3&-&N&13\\
102&18293368-0145103&-&(4060)&19.8&0.35&1&-&-&U&\\
103&18290819-0139215&-&(4060)&15&0.88&1.2&-&-&U&\\
104&18314110-0128035&M-GIANT&-&-&-&-&-&-&N&13\\
105&18290391-0115357&FG&6030&8.7&3.86&1.4&1.7&-&N&13\\
106&18371575-0026561&A7&7850&1.9&46.82&2.5&-6.6&-&N&13\\
107&18381010-0023452&M3&3470&3.4&0.38&0.5&4.8&-&N&13\\
108&18371444-0023261&-&(4060)&7&8.52&-&-&-&U&\\
110&18385989-0008097&M-GIANT&-&-&-&-&-&-&N&13\\
111&J182813.5+000-249&K2&4780&3.8&4.23&1.6&13.4&-&Y&13,21\\
112&J182821.6+000016&-&(4060)&4&1.53&1.2&17.5&-&Y&13\\
113&18384257+0001324&FG&6030&5.2&0.19&-&-1.3&-&N&13\\
114&18392594+0006382&-&(4060)&7.1&8.52&-&-5.1&-&N&13\\
115&J182850.2+000950&K7&4060&7.5&1.53&1.2&131.3&-&Y&13,21\\
116&183549.4+001002&FG&6030&6.6&2.68&1&-1.5&-&N&13\\
117&18385571+0014431&GK&5250&4.7&0.99&1&-5.3&-&N&13\\
118&18394048+0014497&M1&3720&1&0.21&0.7&9.3&-&N&13\\
119&18374209+0016519&M-GIANT&-&-&-&-&-&-&N&13\\
120&J182911.5+002039&M2&3580&2.9&0.15&0.5&\tablefootmark{b}&-&N&6\\
121&18375663-0023253&-&(4060)&3.8&0.12&1&-&-&U&\\
122&18381580+0024218&M-GIANT&-&-&-&-&-&-&N&13\\
123&J18295130+0027477&B8&11900&4&52.2&1&-9.1&-&N&13\\
124&serp22&-&(4060)&11&0.63&1.1&-&-&U&\\
125&18401205+0029276&F4&6590&8.9&5.5&1.4&42.7&-&Y&13\\
126&J182901.2+002933&M2&3580&1&0.21&1&88.6&-&Y&13,21\\
127&Serp127&M1&3720&2.8&0.76&0.8&\tablefootmark{b}&-&Y&6\\
128&J182935.6+003504&K7&4060&3.8&1.41&1.2&10.9&273&Y&21\\
129&18381447+0035099&M-GIANT&-&-&-&-&-&-&N&13\\
130&18401486+0037042&K7&4060&4.3&0.2&0.8&38&-&Y&13\\
131&Serp111&-&(4060)&5.1&0.63&1.1&9.8&-&Y&13\\
132&J182955.3+004939&A2&8970&5.6&39.42&2.2&9.9&-&Y&13,21\\
133&J183008.6+005847&K5&4350&0.7&1.04&1.4&7.4&-&Y&13,21\\
134&RXJ0432.8+1735&M2&3580&0.7&0.38&0.6&1.4&-&N&13\\
135&DMTau&M1&3720&0.7&0.25&0.7&\tablefootmark{b}&-&Y&6\\
136&UXTauA&K2&4780&0.8&1.88&1.5&9.5&-&Y&22\\
137&043339.0+222720&M2.5&3470&2&0.03&0.4&28&-&Y&13,23\\
138&043326.2+224529&M3&3470&5.6&0.29&0.5&4.2&-&N&13,23\\
139&J04390525+2337450&K5&4350&5.1&0.08&-&18.2&-&Y&13,23\\
140&043649.1+241258&F2&6890&1.5&3.98&1.4&-5.5&-&N&23\\
142&MWC758&A8&7580&0.7&34.37&2.3&\tablefootmark{c}&-&Y&24\\
144&044555.7+261858&K1&5080&3&0.15&-&-1.1&-&N&13,23\\
145&DHTauAB&M1&3720&0.5&0.44&0.8&-&348&Y&23,25\\
146&043044.7+263308&K3&4730&2.6&3.54&1.5&-1.1&-&N&23\\
147&J04214631+2659296&M6&3050&3&0.02&0.1&17.3&-&N&23\\
148&IPTau&M0&3850&0.6&0.51&0.9&11&-&Y&22\\
149&V892Tau&B8&11900&9.5&131.67&3.1&-&-&U&26\\
150&V410X-ray6&M4.5&3240&1.7&0.32&0.3&-&210&N&23\\
151&042254.6+282354&A0&9520&0.9&16.65&1.9&10&-&Y&23\\
152&V819Tau&K7&4060&1.1&0.79&1.1&-&180&N&23\\
153&RYTau&G1&5945&2.8&17.11&2.3&\tablefootmark{c}&-&Y&27\\
154&V410X-ray2&M0&3850&17&1.05&1&-&-&U&23\\
155&041542.7+290959&M1.25&3720&2.8&0.34&0.8&2.3&-&N&28\\
156&041332.3+291726&G5&5770&2.4&0.04&-&-2.4&-&N&13\\
157&J032800.1+300847&M5&3240&3.7&0.44&0.2&21.8&-&Y&13\\
158&LkCa19&K0&5250&1&1.75&1.2&-1.2&110&N&16\\
159&ABAur&A0&9520&1.6&112.53&3.1&\tablefootmark{c}&-&Y&29,24\\
160&J033341.3+311341&K4&4560&5.7&0.19&0.8&205&-&Y&13\\
161&ASR118&K4&4560&5.7&0.19&0.8&17.6&-&Y&13,30\\
162&MBO22&M0&3850&2.5&0.2&0.8&4.8&-&N&30\\
163&J032856.6+311836&K6&4205&3.3&0.34&1&104.7&-&Y&13\\
164&J034219.3+314327&-&(4060)&7&0.3&0.9&64.7&-&Y&13\\
165&J034227.1+314433&K7&4060&7.1&0.35&1&4.3&-&N&30\\
166&J034109.1+314438&FG&6030&5.4&3.86&1.4&-1.2&-&N&13\\
167&J034355.2+315532&K&4730&9.3&0.37&0.8&-2.9&-&N&13\\
168&J034434.8+315655&M3&3470&3.1&0.16&0.4&130&504&Y&30\\
169&IC348LRL190&M4&3370&7.9&0.14&0.3&-5.7&-&N&13,30\\
171&J034520.5+320634&M1&3720&2.6&0.69&0.8&11.5&-&N&31\\
172&IC348-67&M0.75&3720&0.7&0.17&0.7&-&280&N&32\\
173&LkH-alpha330&G3&5830&3&12.75&2.2&\tablefootmark{b}&-&Y&6\\
174&J04300424+3522238&M0&3850&2.1&0.33&0.9&-&370&Y&13,32\\
175&J04303235+3536133&M0&3850&3.2&0.16&0.8&-&350&Y&13,32\\
176&04300980+3540355&M4&3370&4.1&0.02&0.3&10.3&-&N&13\\
177&J04304004+3542101&K7&4060&2.7&0.09&0.6&-&310&Y&32\\
178&J04303831+3549591&-&(4060)&8.8&0.09&1&-&-&U&\\
179&J160044.5-415531&K0&5250&1.3&1.34&1.1&-&532&Y&33\\
180&J190058.1-364505&M0.75&3720&1.9&0.39&0.8&-&440&Y&33\\
181&03445614+3209152&K0&5250&4.8&2.22&1.3&-&360&Y&32\\
182&03442156+3215098&M2&3580&1.1&0.09&0.5&-&130&N&32\\
183&03442257+3201536&M2&3580&2.1&0.25&0.6&-&140&N&32\\
184&04330422+2921499&B9&10500&2.8&79.81&2.7&-&-&N&32\\
185&J160825.76-390601.1&M5&3240&0&0.22&0.3&\tablefootmark{b}&-&Y&14\\
186&RXJ1556.1-3655&M1&3720&0&0.25&0.7&82.6&416&Y&16\\
187&043150.5+242418&M0.5&3720&1.5&0.29&0.7&29.3&-&Y&23,34\\
188&041413.5+281249&M0&3850&1&0.33&0.9&99&-&Y&23\\
189&041841.3+282725&K5&4350&20.4&0.32&0.9&-&-&U&23\\
190&042025.5+270035&M2.25&3580&1.5&0.04&0.5&-&-&U&23\\
191&042921.6+270125&M6&3050&4.1&0.49&0.1&-&-&U&23\\
192&043249.1+225302&K6&4205&2.5&0.55&1.1&22&-&Y&23,35\\
193&044221.0+252034&M4.5&3240&4.3&0.29&0.3&114.6&-&Y&13\\
194&041539.1+281858&M5&3240&4&0.57&0.2&997.2&-&Y&23\\
195&042318.2+264115&M3.5&3370&8.6&0.26&0.4&-&-&U&23\\
196&041414.5+282758&M3.5&3370&3.7&1.34&0.3&-&-&N&36\\
197&041915.8+290626&K7&4060&2.5&1.65&1.2&-&458&Y&23,25\\
198&042155.6+275506&M1&3720&0.9&0.83&0.8&-&453&Y&25\\
200&J160710.08-391103.5&M0&3850&0.5&0.57&1&24.2&-&Y&13\\
201&J032741.47+302016.8&M3&3470&1.7&0.52&0.5&51.3&-&Y&13\\
202&J034345.17+320358.6&-&(4060)&7.1&0.25&1&150.2&-&Y&13\\
203&J182815.26-000243.3&-&(4060)&5.4&3.17&0.6&-&-&U&\\
204&J162715.89-243843.2&-&(4060)&16.1&0.63&1&-&-&U&\\
301&J130455.7-773949&M0.5&3720&0.7&0.5&0.8&0.2&311&N&30\\
303&J160115.5-415235&-&(4060)&12.8&0.04&-&-&0&U&\\
307&16083070-3828268&-&(4060)&0&2.2&1.3&-&0&U&\\
309&TWHya&K6&4205&1.3&0.34&1&\tablefootmark{b}&-&Y&6\\
310&15395742-3414567&-&(4060)&0.5&0.02&-&-&0&U&\\
314&16281385-2456113&-&(4060)&4.6&0.25&0.9&-&0&U&\\
316&16271587-2438433&-&(4060)&16.1&0.63&1&-&0&U&\\
317&16312019-2430009&-&(4060)&2.2&2.35&1&-&0&U&\\
318&DoAr21&K1&5080&5.5&8.66&2.4&1.5&450&N&30\\
319&J162740.3-242204&K7&4060&1.9&2.06&1.3&29.2&-&Y&13\\
321&Serp48&F6&6360&4.4&9.96&1.7&-2.3&-&N&13,21\\
322&18302986+0035004&-&(4060)&2.3&0.71&1.1&-&0&U&\\
325&LkCa15&K3&4730&1.2&1.01&1.2&\tablefootmark{b}&-&Y&6\\
326&CoKu-Tau-4&M1&3720&0.1&0.29&0.7&-&185&N&23,25\\
329&GMAur&K5&4350&0.3&0.83&1.3&\tablefootmark{b}&-&Y&6\\
333&03370363+3039291&-&(4060)&0&2.35&1.3&-&0&U&\\
334&03401579+3055047&-&(4060)&2.7&0.01&-&-&0&U&\\
335&J033234.0+310056&K6&4205&4&1.02&1.3&71.4&-&Y&13,30\\
348&UScoJ155837.1-225724&G7&5630&0.6&3.27&1.4&-&-&U&37\\
349&RXJ1842.9-3532&K2&4780&0.7&0.92&1.2&\tablefootmark{b}&-&Y&6\\
350&RXJ1852.3-3700&K2&4780&0.7&0.68&1&\tablefootmark{b}&-&Y&6\\
\end{longtable}
}
{\bf Refs.}
       1) \citet{Alcala1995}, 
       2) \citet{Schisano2009}, 
       3) \citet{Lawson2004}, 
       4) \citet{Luhman2007}, 
       5) \citet{Manara2015cham}, 
       6) \citet{Manara2014}, 
       7) \citet{Luhman2008new}, 
       8) \citet{Spezzi2008}, 
       9) \citet{Luhman2004}, 
      10) \citet{Comeron2004}, 
      11) \citet{Silverstone2006}, 
      12) \citet{GarciaLopez2006}, 
      13) This work, 
      14) \citet{Alcala2014}, 
      15) \citet{Sicilia2008}, 
      16) \citet{Wahhaj2010}, 
      17) \citet{Cieza2010}, 
      18) \citet{Natta2006}, 
      19) \citet{Brown2012a}, 
      20) \citet{Wilking2005}, 
      21) \citet{Oliveira2009}, 
      22) \citet{WhiteGhez2001}, 
      23) \citet{Rebull2010}, 
      24) \citet{Salyk2013}, 
      25) \citet{Nguyen2012}, 
      26) \citet{Furlan2006}, 
      27) \citet{Calvet2004}, 
      28) \citet{Furlan2011}, 
      29) \citet{Mooley2013}, 
      30) \citet{Merin2010}, 
      31) \citet{Cieza2007}, 
      32) \citet{Cieza2012}, 
      33) \citet{Romero2012}, 
      34) \citet{Rigliaco2015}, 
      35) \citet{KrausHillenbrand2009}, 
      36) \citet{Herczeg2014}, 
      37) \citet{Carpenter2008}
\tablefoot{
\tablefoottext{a}{We have reversed the signs of the width of the H$\alpha$ line taken from \citet{Rebull2010} and \citet{Winston2009}, as they list a negative value for emission and positive for absorption.}
\tablefoottext{b}{The accretion properties have been derived using a full X-shooter spectrum rather than only fitting the H$\alpha$ line.}
\tablefoottext{c}{The accretion properties have been derived using other lines (e.g. Br$\gamma$).}
\tablefoottext{d}{'Y' means accreting, 'N' means non-accreting', 'U' means unknown.}
}

\newpage

{\small
\begin{longtable}{lllllllllll}
\caption{Results of disk fitting procedure and classification}
\label{tbl:diskparams}\\
\hline
\hline
ID&Name&$r_{\rm cav}$&$\Sigma_c$&$M_{\rm disk,fit}$\tablefootmark{a}&$\delta_{\rm dust}$&$h_c$&$\psi$&$M_{\rm disk,mm}$\tablefootmark{a}&$L_{\rm disk}/L_*$&Classification\tablefootmark{b}\\
&&(AU)&(g cm$^{-2}$)&($M_{\rm Jup}$)&&&&($M_{\rm Jup}$)&&\\
\hline
\endfirsthead

\caption{Fitting results continued}\\
\hline
\hline
ID&Name&$r_{\rm cav}$&$\Sigma_c$&$M_{\rm disk,fit}$\tablefootmark{a}&$\delta_{\rm dust}$&$h_c$&$\psi$&$M_{\rm disk,mm}$\tablefootmark{a}&$L_{\rm disk}/L_*$&Classification\tablefootmark{b}\\
&&(AU)&(g cm$^{-2}$)&($M_{\rm Jup}$)&&&&($M_{\rm Jup}$)&&\\
\hline
\endhead

\hline
\endfoot

1&TCha&     140$^{+    10}_{   -10}$ &7E-02&7.8&1E-02&0.10&  1/7.&3.7&0.467&   ML\\
2&RECX11&       6$^{+     4}_{    -4}$ &3E-03&1.1&1E-02&0.01&  1/7.&$<$  1.5&0.067&   LS\\
3&RECX5&      10$^{+     2}_{    -2}$ &3E-04&0.1&1E-04&0.01&  1/7.&-&0.038&   LL\\
4&CHXR22E&      45$^{+    15}_{    -5}$ &3E-04&0.1&1E-06&0.01&  2/7.&-&0.178&   LL\\
5&ISO52&      30$^{+    50}_{   -18}$ &3E-03&0.9&1E-01&0.05&  2/7.&-&0.195&   LL\\
6&CSCha&      60$^{+    10}_{   -10}$ &8E-02&20.9&1E-06&0.05&  1/7.&20.6&0.105&   ML\\
7&11094742-7726290&       8$^{+     2}_{    -2}$ &3E-03&1.1&1E-02&0.15&  1/7.&-&0.685&   LS\\
9&T54&     120$^{+    20}_{   -10}$ &3E-04&0.0&1E-03&0.01&  1/7.&$<$  1.6&0.008&   DD\\
10&T21&     190$_{   -10}$ &3E-04&0.0&1E-02&0.01&  1/7.&$<$  2.1&0.001&   DD\\
11&SZCha&      30$^{+    10}_{   -10}$ &5E-02&15.8&1E-04&0.05&  2/7.&32.8&0.329&   ML\\
12&T35&      15$^{+     5}_{    -5}$ &3E-03&1.0&1E-03&0.05&  1/7.&$<$ 22.2&0.183&   LL\\
13&ISO-ChaII29&     190$_{   -10}$ &1E-02&0.2&1E-04&0.01&  1/7.&-&0.007&   DD\\
14&T56&      10$^{+     4}_{    -4}$ &3E-02&10.6&1E-04&0.05&  1/7.&2.5&0.201&   ML\\
15&CRCha&       1$^{+     1}$ &2E-01&55.6&1E+00&0.03&  1/7.&43.9&0.083&   MS\\
16&WWCha&      50$^{+    30}_{   -49}$ &4E-01&106.1&1E-01&0.20&  1/7.&-&0.934&   ML\\
17&11062554-7633418&      15$^{+    15}_{    -5}$ &3E-03&1.0&1E-06&0.05&  2/7.&-&0.172&   LL\\
18&T25&      30$^{+     5}_{    -5}$ &3E-03&0.9&1E-06&0.05&  1/7.&2.1&0.139&   LL\\
20&MPMus&       1$^{+     3}$ &1E-01&37.1&1E+00&0.04&  1/7.&15.9&0.17&   MS\\
21&HD142527&     110$^{+    10}_{   -20}$ &6E-01&100.6&1E-01&0.10&  1/7.&216&0.366&   ML\\
22&J16232807-4015368&      50$^{+    40}_{   -10}$ &5E-03&1.4&1E-06&0.05&  1/7.&-&0.465&   LL\\
23&Sz111&      60$^{+    10}_{   -10}$ &1E-01&26.1&1E-06&0.05&  2/7.&25&0.152&   ML\\
24&Lup60&      16$^{+     2}_{    -4}$ &3E-03&1.0&1E-06&0.02&  1/7.&$<$  3.8&0.151&   LL\\
25&J160830.3-390611&       4$^{+     4}_{    -2}$ &3E-03&1.1&1E-04&0.01&  1/7.&-&0.1&   LS\\
26&Sz104&       2$^{+     2}$ &3E-04&0.1&1E-02&0.05&  1/7.&$<$  5.0&0.307&   LS\\
27&Sz91&     120$^{+    30}_{   -20}$ &2E-02&3.4&1E-06&0.05&  1/7.&5.7&0.17&   LL\\
28&J160855.5-390234&       2$^{+     2}$ &3E-04&0.1&1E-02&0.10&  1/7.&$<$  3.4&0.434&   LS\\
29&Sz84&      70$^{+    40}_{   -10}$ &6E-02&14.5&1E-08&0.02&  1/7.&2.8&0.136&   ML\\
31&16225309-3724373&     170$^{+    10}_{   -20}$ &3E-04&0.0&1E-02&0.01&  1/7.&-&0.193&   DD\\
33&HD135344&      80$^{+    10}_{   -10}$ &2E-01&33.5&1E-01&0.05&  1/7.&49.6&0.233&   ML\\
34&CrA-466&      10$^{+     4}_{    -4}$ &3E-04&0.1&1E-02&0.20&  1/7.&-&0.25&   LL\\
35&Sz76&       2$^{+     2}$ &9E-02&33.2&1E-02&0.02&  1/7.&$<$  2.8&0.104&   MS\\
36&J154508.9-341734&1&2E-02&5.6&1E+00&0.05&  1/7.&6.7&0.337&   NH\\
38&RXJ1615.3-3255&      10$^{+    10}_{    -2}$ &5E-01&159.2&1E-06&0.03&  1/7.&60.1&0.094&   ML\\
39&V4046Sgr&      16$^{+     4}_{    -6}$ &1E-01&34.3&1E-04&0.03&  1/7.&16.8&0.117&   ML\\
40&J163154.7-250324&1&3E-03&1.1&1E+00&0.10&  1/7.&$<$  1.3&0.434&   NH\\
41&J163205.5-250236&1&3E-04&0.1&1E+00&0.05&  1/7.&$<$  1.3&0.107&   NH\\
43&J163023.4-245416&      45$^{+     5}_{   -20}$ &1E-03&0.3&1E-01&0.15&  2/7.&$<$  1.1&0.257&   LL\\
44&WSB63&       4$^{+     2}_{    -2}$ &7E-03&2.6&1E-04&0.05&  1/7.&2.4&0.146&   LS\\
45&SR24S&      50$^{+    40}_{   -20}$ &6E-02&16.8&1E-01&0.20&  1/7.&35.1&0.764&   ML\\
46&RXJ1633.9-2442&      20$^{+    10}_{    -5}$ &3E-02&10.1&1E-06&0.05&  1/7.&10.5&0.114&   ML\\
47*&ISO-Oph43&       1$^{+     1}$ &3E-03&1.1&1E+00&0.05&  1/7.&-&0.27&   LS\\
48&WSB60&       8$^{+     6}_{    -7}$ &3E-02&10.7&1E-02&0.05&  1/7.&-&0.35&   MS\\
49&J163115.7-243402&      20$^{+    30}_{   -10}$ &3E-03&1.0&1E-01&0.01&  1/7.&$<$  0.8&0.048&   LL\\
50&J162245.4-243124&       2$^{+     2}$ &1E-03&0.4&1E-04&0.01&  1/7.&$<$  0.7&0.11&   LS\\
51&IRS48&     120$^{+    10}_{  -100}$ &2E-02&3.0&1E-04&0.10&  1/7.&11.5&0.31&   LL\\
52&DoAr44&      80$^{+    10}_{   -20}$ &3E-02&5.6&1E-02&0.10&  1/7.&13.4&0.345&   ML\\
53&J162435.2-242620&1&3E-02&11.1&1E+00&0.01&  1/7.&-&0.054&   NH\\
54&SR21&      60$^{+    20}_{   -15}$ &3E-02&7.8&1E-01&0.10&  2/7.&34.3&0.356&   ML\\
55*&J162309.2-241705&1&2E-02&7.8&1E+00&0.03&  1/7.&5.9&0.227&   NH\\
56*&J163136.8-240420&       4$^{+     6}_{    -2}$ &1E-02&3.6&1E-06&0.20&  1/7.&-&0.842&   LS\\
58&J162648.6-235634&1&8E-04&0.3&1E+00&0.15&  1/7.&$<$  1.1&0.418&   NH\\
59&J162802.6-235504&       2$^{+     2}$ &3E-04&0.1&1E-04&0.01&  1/7.&-&0.115&   LS\\
60&oph62&       2$^{+     2}$ &3E-03&1.1&1E-06&0.01&  2/7.&$<$  1.5&0.062&   LS\\
61&J162532.5-232626&       5$^{+    85}_{    -3}$ &3E-04&0.1&1E-02&0.05&  1/7.&-&0.192&   LS\\
62&J162218.5-232148&1&1E-02&4.4&1E+00&0.05&  1/7.&4&0.238&   NH\\
63&DoAr28&      20$^{+     5}_{    -5}$ &4E-02&11.7&1E-06&0.03&  1/7.&6.1&0.087&   ML\\
64&J160421.7-213028&      70$^{+    20}_{   -30}$ &5E-02&12.1&1E-04&0.08&  1/7.&20.4&0.359&   ML\\
65&18015423-0437531&     190$_{   -10}$ &3E-04&0.0&1E-06&0.01&  1/7.&-&0.001&   DD\\
66&18044921-0436413&      30$^{+    10}_{   -26}$ &3E-03&0.9&1E-03&0.05&  1/7.&-&0.276&   LL\\
67*&18270980-0414297&     110$^{+    10}_{   -10}$ &3E-04&0.1&1E-06&0.01&  1/7.&-&0.004&   DD\\
69&18273408-0403247&     190$_{   -10}$ &3E-04&0.0&1E-06&0.01&  1/7.&-&0.001&   DD\\
70&18273858-0402289&      80$^{+    30}_{   -10}$ &6E-02&13.4&1E-04&0.03&  1/7.&8.3&0.077&   ML\\
71*&18255765-0357040&     180$^{+    10}_{   -20}$ &3E-04&0.0&1E-04&0.01&  1/7.&-&0.003&   DD\\
73*&18291383-0342355&     190$_{   -10}$ &3E-02&0.6&1E-04&0.01&  1/7.&-&0.011&   DD\\
74*&18284156-0341507&     140$^{+    10}_{  -100}$ &3E-02&3.4&1E-04&0.01&  1/7.&-&0.025&   DD\\
75*&18283439-0339371&     190$_{   -20}$ &3E-02&0.6&1E-04&0.01&  1/7.&-&0.013&   DD\\
76&18272161-0314158&     190$_{   -10}$ &3E-04&0.0&1E-06&0.01&  1/7.&-&0.002&   DD\\
77&18222604-0304383&     190$_{   -10}$ &3E-04&0.0&1E-06&0.01&  1/7.&-&0.002&   DD\\
79&18324685-0243273&     190$_{   -10}$ &3E-04&0.0&1E-06&0.01&  1/7.&-&0&   DD\\
81*&18324783-0239401&      70$^{+    30}_{   -10}$ &3E-03&0.7&1E-02&0.05&  1/7.&-&0.129&   LL\\
83&18292883-0221157&       2$^{+    16}$ &3E-03&1.1&1E-04&0.01&  2/7.&-&0.053&   LS\\
89&18323005-0204130&     110$^{+    40}_{   -10}$ &3E-05&0.0&1E-06&0.10&  1/7.&$<$  9.0&0.004&   LL\\
91*&18293961-0202414&      30$^{+    50}_{   -25}$ &3E-02&9.5&1E-02&0.10&  1/7.&-&0.272&   ML\\
94*&18312875-0159125&     190$_{  -150}$ &3E-04&0.0&1E-04&0.10&  1/7.&-&0.01&   LL\\
96*&18313657-0157320&      60$^{+    50}_{   -30}$ &3E-04&0.1&1E-04&0.05&  1/7.&-&0.039&   LL\\
99&J18303321-0152563&     190$_{   -10}$ &6E+00&111.7&1E-10&0.01&  2/7.&91.8&0.025&   ML\\
100&18295741-0151541&     190$_{   -10}$ &6E-01&11.2&1E-06&0.01&  1/7.&$<$  9.4&0.016&   ML\\
101*&18294721-0148301&       5$^{+    85}_{    -2}$ &1E-01&36.3&1E-06&0.01&  1/7.&-&0.045&   MS\\
102*&18293368-0145103&      45$^{+   115}_{   -40}$ &3E-02&8.7&1E-01&0.20&  1/7.&-&0.702&   ML\\
103*&18290819-0139215&      40$^{+     5}_{    -5}$ &3E-02&8.9&1E-01&0.10&  1/7.&-&0.338&   ML\\
105*&18290391-0115357&     190$_{   -10}$ &3E-04&0.0&1E-04&0.01&  1/7.&-&0.002&   DD\\
106&18371575-0026561&     190$_{   -10}$ &3E-04&0.0&1E-04&0.01&  1/7.&-&0.001&   DD\\
107&18381010-0023452&      70$^{+    40}_{   -20}$ &3E-04&0.1&1E-06&0.10&  1/7.&$<$  4.4&0.047&   LL\\
108*&18371444-0023261&     180$^{+    10}_{   -40}$ &3E-04&0.0&1E-02&0.05&  1/7.&-&0.011&   LL\\
111&J182813.5+000-249&1&3E-01&111.2&1E+00&0.05&  1/7.&-&0.207&   NH\\
112*&J182821.6+000016&1&3E-03&1.1&1E+00&0.10&  1/7.&-&0.304&   NH\\
113*&18384257+0001324&      60$^{+    20}_{   -20}$ &3E-03&0.8&1E-06&0.05&  2/7.&-&0.167&   LL\\
114*&18392594+0006382&     180$^{+    10}_{  -130}$ &3E-03&0.1&1E-06&0.01&  1/7.&-&0.01&   DD\\
115&J182850.2+000950&       5$^{+    35}_{    -4}$ &5E-03&1.8&1E-01&0.20&  1/7.&$<$ 12.4&0.889&   LS\\
117*&18385571+0014431&      60$^{+    40}_{   -10}$ &3E-04&0.1&1E-06&0.01&  1/7.&-&0.019&   LL\\
118&18394048+0014497&       2$^{+     2}$ &3E-04&0.1&1E-02&0.15&  1/7.&-&0.397&   LS\\
120&J182911.5+002039&      10$^{+     5}_{    -8}$ &3E-03&1.1&1E-06&0.05&  1/7.&$<$  5.4&0.17&   LL\\
124*&serp22&      10$^{+    10}_{    -5}$ &3E-03&1.1&1E-04&0.05&  1/7.&-&0.118&   LL\\
125&18401205+0029276&      50$^{+    10}_{   -49}$ &3E-02&8.4&1E-01&0.20&  1/7.&-&0.927&   ML\\
127&Serp127&      80$^{+    20}_{   -10}$ &5E-02&10.1&1E-05&0.03&  2/7.&$<$  7.5&0.086&   ML\\
128&J182935.6+003504&      25$^{+    45}_{   -24}$ &3E-03&1.0&1E-02&0.05&  1/7.&-&0.086&   LL\\
130&18401486+0037042&      30$^{+    40}_{   -25}$ &3E-04&0.1&1E-06&0.10&  1/7.&-&0.077&   LL\\
131*&Serp111&      10$^{+    10}_{    -5}$ &1E-02&3.5&1E-03&0.05&  1/7.&1.2&0.257&   LL\\
132&J182955.3+004939&     180$^{+    10}_{  -176}$ &5E-01&18.6&1E-04&0.05&  1/7.&27.8&0.146&   ML\\
133&J183008.6+005847&1&3E-02&11.1&1E+00&0.15&  1/7.&-&0.582&   NH\\
134&RXJ0432.8+1735&     190$_{   -10}$ &3E-04&0.0&1E-04&0.01&  1/7.&-&0.006&   DD\\
135&DMTau&       4$^{+     2}_{    -2}$ &9E-02&32.8&1E-06&0.05&  1/7.&16.8&0.121&   MS\\
136&UXTauA&      50$^{+    40}_{   -10}$ &3E-02&7.0&1E-04&0.05&  1/7.&12&0.206&   ML\\
137&043339.0+222720&       4$^{+     2}_{    -2}$ &4E-02&14.6&1E-01&0.20&  2/7.&5.1&1.236&   MS\\
138&043326.2+224529&     180$^{+    10}_{   -20}$ &3E-03&0.1&1E-06&0.01&  1/7.&-&0.1&   DD\\
139&J04390525+2337450&       2$^{+     2}$ &3E-03&1.1&1E-03&0.10&  1/7.&-&0.603&   LS\\
140&043649.1+241258&     190$_{   -10}$ &3E-04&0.0&1E-05&0.01&  1/7.&$<$  2.2&0.001&   DD\\
142&MWC758&      25$^{+    15}_{    -5}$ &3E-02&9.8&1E-01&0.05&  1/7.&29.4&0.203&   ML\\
144&044555.7+261858&     110$^{+    40}_{   -10}$ &3E-03&0.5&1E-06&0.01&  1/7.&-&0.037&   DD\\
145&DHTauAB&       1$^{+     1}$ &2E-02&5.6&1E+00&0.10&  1/7.&3.9&0.328&   MS\\
146&043044.7+263308&     190$_{   -10}$ &3E-04&0.0&1E-04&0.01&  1/7.&-&0.001&   DD\\
147&J04214631+2659296&       6$^{+     9}_{    -2}$ &3E-02&10.8&1E-01&0.05&  2/7.&-&0.426&   MS\\
148&IPTau&     100$^{+    10}_{   -30}$ &1E-02&2.2&1E-01&0.10&  1/7.&2.7&0.235&   LL\\
149&V892Tau&      10$^{+     8}_{    -8}$ &3E-02&10.6&1E-06&0.03&  1/7.&51&0.153&   ML\\
150&V410X-ray6&      15$^{+     5}_{    -5}$ &3E-04&0.1&1E-06&0.10&  2/7.&$<$  0.6&0.291&   LL\\
151&042254.6+282354&     190$_{   -10}$ &1E-04&0.0&1E-06&0.05&  1/7.&-&0.001&   LL\\
152&V819Tau&     150$^{+    20}_{   -20}$ &3E-04&0.0&1E-04&0.01&  1/7.&$<$  0.7&0.003&   DD\\
153&RYTau&       2$^{+     2}$ &3E-02&11.1&1E-02&0.10&  1/7.&44.8&0.385&   MS\\
154&V410X-ray2&       6$^{+     2}_{    -2}$ &6E-02&21.7&1E-04&0.02&  1/7.&-&0.082&   MS\\
155&041542.7+290959&      50$^{+    10}_{    -5}$ &1E-02&3.4&1E-06&0.05&  1/7.&4.4&0.141&   LL\\
156&041332.3+291726&      10$^{+    60}_{    -5}$ &1E-02&3.5&1E-06&0.01&  2/7.&$<$  2.5&0.033&   LL\\
157&J032800.1+300847&       1$^{+     4}$ &3E-03&1.1&1E+00&0.05&  1/7.&-&0.325&   LS\\
158&LkCa19&     190$_{   -10}$ &3E-04&0.0&1E-06&0.01&  1/7.&-&0.001&   DD\\
159&ABAur&       1$^{+     5}$ &2E-02&8.2&1E+00&0.05&  1/7.&28.7&0.152&   MS\\
160&J033341.3+311341&1&1E-01&37.1&1E+00&0.20&  1/7.&-&1.191&   NH\\
161&ASR118&       2$^{+     4}$ &8E-03&2.9&1E-03&0.20&  1/7.&$<$ 29.4&0.785&   LS\\
162&MBO22&       2$^{+     2}$ &2E-02&5.5&1E-06&0.05&  2/7.&3.4&0.321&   MS\\
163&J032856.6+311836&       1$^{+     1}$ &3E-02&11.1&1E+00&0.20&  1/7.&$<$ 29.9&0.978&   MS\\
164*&J034219.3+314327&       1$^{+     3}$ &1E-03&0.4&1E+00&0.20&  1/7.&-&0.505&   LS\\
165&J034227.1+314433&       8$^{+     4}_{    -4}$ &1E-03&0.4&1E-04&0.02&  1/7.&$<$  0.8&0.052&   LS\\
166*&J034109.1+314438&     140$^{+    20}_{   -10}$ &2E-01&16.8&1E-01&0.15&  1/7.&42.5&0.424&   ML\\
167*&J034355.2+315532&       8$^{+    42}_{    -4}$ &3E-02&10.7&1E-04&0.05&  1/7.&-&0.238&   MS\\
168&J034434.8+315655&       5$^{+     20}_{    -3}$ &3E-03&1.1&1E-04&0.05&  1/7.&$<$  0.8&0.177&   LL\\
169&IC348LRL190&       2$^{+     2}$ &3E-04&0.1&1E-04&0.02&  1/7.&-&0.194&   LS\\
171&J034520.5+320634&1&1E-02&4.4&1E+00&0.10&  1/7.&4.4&0.343&   NH\\
172&IC348-67&       2$^{+     1}$ &3E-02&11.8&1E-06&0.10&  1/7.&7.9&0.384&   MS\\
173&LkH-alpha330&     120$^{+    10}_{   -20}$ &3E-01&44.7&1E-04&0.05&  1/7.&53.6&0.331&   ML\\
174&J04300424+3522238&      18$^{+    10}_{    -6}$ &9E-02&30.5&1E-06&0.05&  1/7.&$<$ 22.2&0.257&   ML\\
175&J04303235+3536133&       4$^{+     8}_{    -2}$ &2E-03&0.5&1E-04&0.15&  2/7.&0.8&0.542&   LS\\
176&04300980+3540355&       2$^{+     2}$ &3E-03&1.1&1E-06&0.05&  2/7.&-&0.195&   LS\\
177&J04304004+3542101&      25$^{+    10}_{   -10}$ &6E-03&2.0&1E-06&0.01&  1/7.&$<$  0.4&0.019&   LL\\
179&J160044.5-415531&       1$^{+    59}$ &2E-02&7.4&1E+00&0.05&  1/7.&9.2&0.141&   MS\\
180&J190058.1-364505&      14$^{+     4}_{    -4}$ &3E-04&0.1&1E-06&0.05&  1/7.&$<$  1.8&0.071&   LL\\
181&03445614+3209152&       6$^{+    12}_{    -5}$ &3E-03&1.1&1E+00&0.01&  1/7.&$<$  0.6&0.067&   LS\\
182&03442156+3215098&       2$^{+    23}$ &1E-02&3.7&1E-04&0.01&  1/7.&-&0.044&   LS\\
183&03442257+3201536&       4$^{+     1}_{    -2}$ &3E-03&1.1&1E-06&0.05&  1/7.&$<$  0.8&0.119&   LS\\
184&04330422+2921499&     160$^{+    20}_{   -10}$ &1E-04&0.0&1E-04&0.01&  1/7.&$<$  1.5&0.003&   DD\\
185&J160825.76-390601.1&1&1E-02&3.7&1E+00&0.12&  1/7.&$<$  9.4&0.582&   NH\\
186&RXJ1556.1-3655&1&3E-04&0.1&1E+00&0.10&  2/7.&-&0.274&   NH\\
187&043150.5+242418&       4$^{+     2}_{    -2}$ &3E-02&9.9&1E-01&0.20&  1/7.&10&1.076&   MS\\
188&041413.5+281249&1&5E-03&1.9&1E+00&0.15&  1/7.&2.6&0.452&   NH\\
189&041841.3+282725&      40$^{+    80}_{   -30}$ &3E-02&8.9&1E-01&0.07&  1/7.&-&0.302&   ML\\
190&042025.5+270035&       2$^{+     2}$ &2E-02&5.5&1E-04&0.05&  1/7.&1.4&0.147&   MS\\
191&042921.6+270125&       2$^{+     2}$ &1E-03&0.4&1E-02&0.05&  1/7.&-&0.449&   LS\\
192&043249.1+225302&      25$^{+    10}_{   -10}$ &3E-03&1.0&1E-02&0.20&  1/7.&$<$  2.3&0.617&   LL\\
193&044221.0+252034&      10$^{+    10}_{    -8}$ &8E-03&2.8&1E-04&0.05&  1/7.&2.6&0.125&   LL\\
194&041539.1+281858&      10$^{+     2}_{    -5}$ &7E-02&24.8&1E-05&0.02&  1/7.&2.3&0.157&   ML\\
195&042318.2+264115&       2$^{+     2}$ &3E-04&0.1&1E-02&0.05&  1/7.&-&0.264&   LS\\
196&041414.5+282758&       2$^{+     2}$ &1E-02&4.1&1E-04&0.05&  1/7.&5.5&0.219&   LS\\
197&041915.8+290626&1&3E-02&11.1&1E+00&0.03&  1/7.&10.4&0.114&   NH\\
198&042155.6+275506&      90$^{+    50}_{   -10}$ &2E-02&4.1&1E-02&0.15&  1/7.&7.2&0.314&   LL\\
200&J160710.08-391103.5&1&2E-02&7.4&1E+00&0.10&  1/7.&$<$  6.5&0.397&   NH\\
201&J032741.47+302016.8&       1$^{+     3}$ &7E-02&25.9&1E+00&0.05&  1/7.&-&0.34&   MS\\
203*&J182815.26-000243.3&       2$^{+    98}$ &4E-02&14.7&1E-01&0.10&  2/7.&-&0.9&   MS\\
301&J130455.7-773949&       2$^{+     2}$ &3E-02&11.1&1E-02&0.05&  1/7.&-&0.242&   MS\\
303*&J160115.5-415235&       4$^{+     6}_{    -2}$ &7E-02&25.5&1E-04&0.20&  2/7.&-&2.82&   MS\\
307*&16083070-3828268&     160$^{+    10}_{   -20}$ &5E-02&3.4&1E-06&0.05&  2/7.&-&0.107&   LL\\
309&TWHya&      10$^{+     2}_{    -2}$ &3E-02&10.6&1E-04&0.02&  1/7.&-&0.066&   ML\\
310*&15395742-3414567&      20$^{+    60}_{   -14}$ &6E-01&201.1&1E-02&0.20&  2/7.&-&1.477&   ML\\
314*&16281385-2456113&      20$^{+    10}_{   -10}$ &1E-02&3.4&1E-06&0.05&  1/7.&-&0.146&   LL\\
318&DoAr21&      70$^{+    30}_{   -10}$ &3E-04&0.1&1E-02&0.05&  2/7.&-&0.048&   LL\\
319&J162740.3-242204&       1$^{+    17}$ &3E-03&1.1&1E+00&0.05&  1/7.&-&0.116&   LS\\
321&Serp48&     190$_{   -10}$ &3E-03&0.1&1E-05&0.01&  1/7.&$<$  5.1&0.018&   DD\\
322*&18302986+0035004&      70$^{+    10}_{   -10}$ &3E-01&72.6&1E-08&0.05&  1/7.&-&0.225&   ML\\
325&LkCa15&      80$^{+    40}_{   -35}$ &1E-01&29.0&1E-01&0.05&  1/7.&32.8&0.159&   ML\\
326&CoKu-Tau-4&       6$^{+     2}_{    -2}$ &3E-02&10.8&1E-06&0.10&  1/7.&-&0.46&   MS\\
329&GMAur&      30$^{+    10}_{    -5}$ &2E-01&50.6&1E-04&0.05&  2/7.&51.2&0.211&   ML\\
333*&03370363+3039291&      50$^{+    30}_{   -10}$ &1E-02&2.8&1E-05&0.10&  1/7.&-&0.533&   LL\\
334*&03401579+3055047&       8$^{+   112}_{    -7}$ &3E-01&107.3&1E-01&0.20&  2/7.&-&1.528&   MS\\
335&J033234.0+310056&       2$^{+     8}_{    -1}$ &3E-01&110.6&1E-01&0.05&  2/7.&-&0.28&   MS\\
348&UScoJ155837.1-225724&      20$^{+    10}_{    -8}$ &3E-03&1.0&1E-01&0.10&  1/7.&-&0.328&   LL\\
349&RXJ1842.9-3532&     160$^{+    10}_{   -10}$ &3E-01&22.3&1E-04&0.05&  1/7.&8.9&0.201&   ML\\
350&RXJ1852.3-3700&      10$^{+     2}_{    -2}$ &6E-02&21.2&1E-06&0.03&  1/7.&-&0.228&   ML\\
\end{longtable}
}
\tablefoot{
\tablefoottext{*}{Fit results are uncertain due to unknown spectral type.}
\tablefoottext{a}{Disk masses refer to the full disk mass, computed assuming a gas-to-dust ratio of 100.}
\tablefoottext{b}{NH = disks without holes, ML = massive disks with large holes, MS = massive disks with small holes, LL = low-mass disks with large holes, LS = low-mass disks with small holes, DD = low-mass disks with very low scale heights. See also definition in the text.}
}

\section{SEDs}

\begin{figure*}[!ht]\begin{center}
\includegraphics[scale=0.72,trim=0 -50 0 -50]{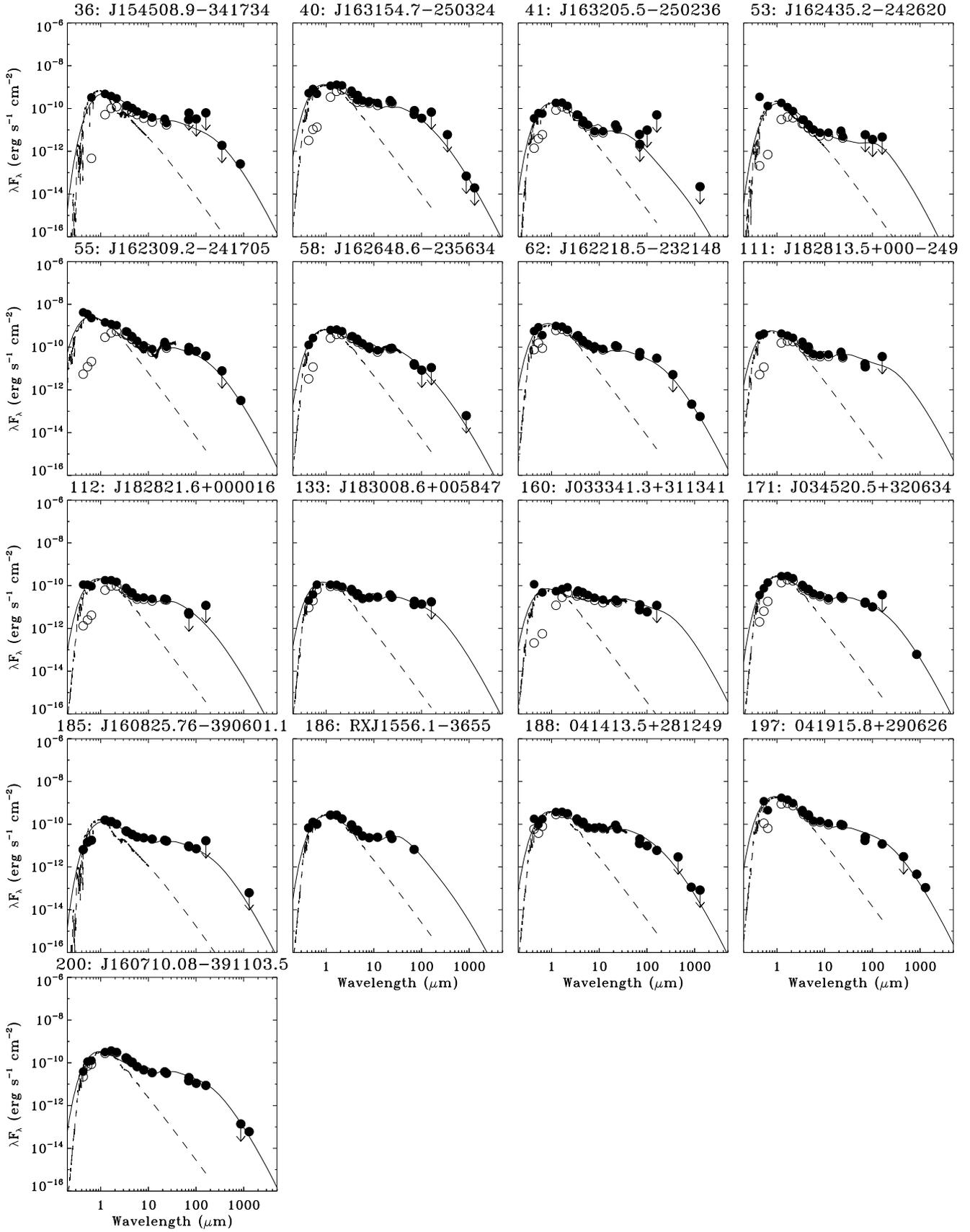}
\caption{SEDs of disks without holes. In this and subsequent Figures the dashed line indicates the stellar spectrum. Open circles denote observed fluxes before extinction correction, filled circles after extinction correction.}
\label{fig:SEDs1}
\end{center}\end{figure*}

\begin{figure*}[!ht]\begin{center}
\includegraphics[scale=0.72,trim=0 -50 0 -50]{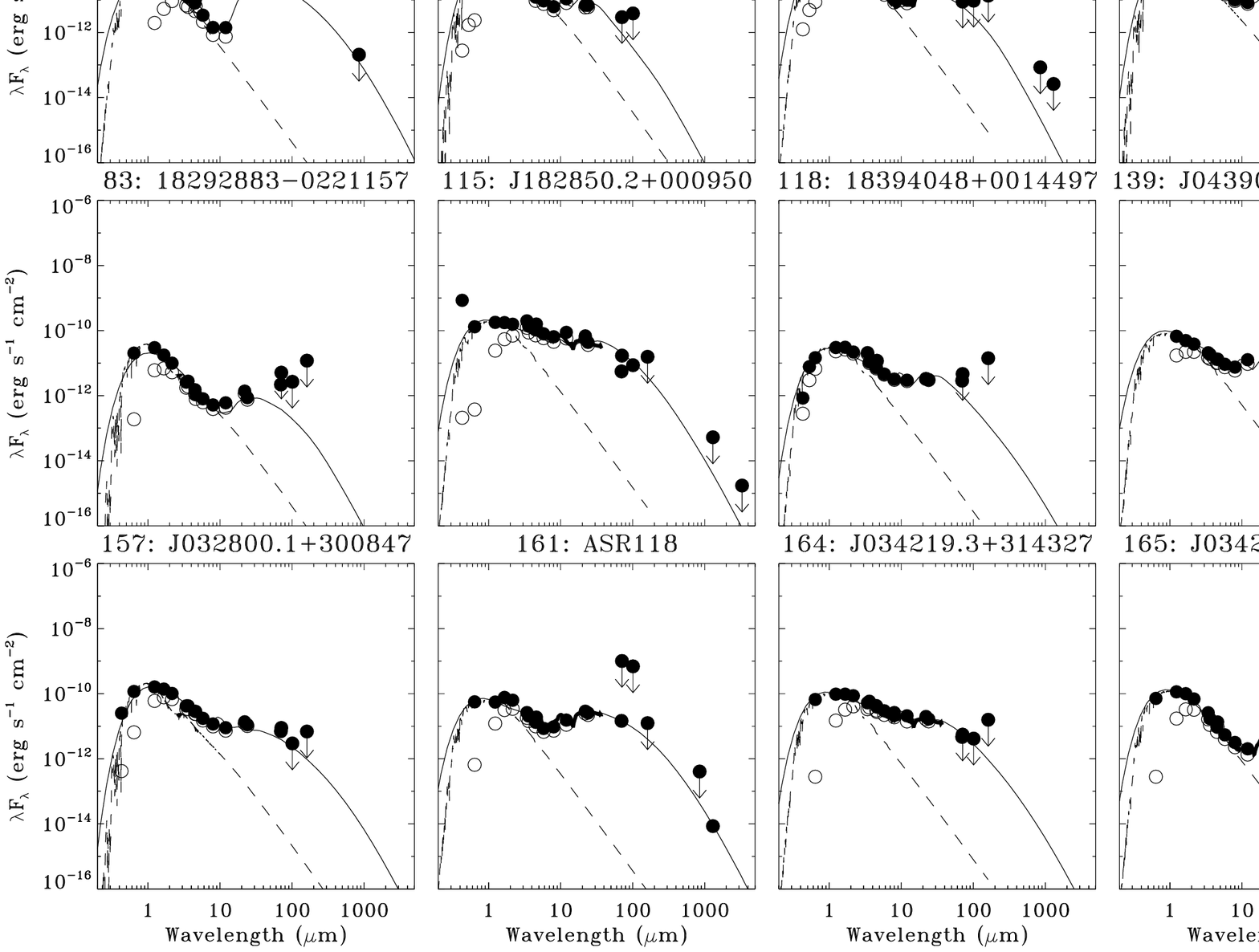}
\captcont{SEDs of low-mass disks with small holes.}
\label{fig:SEDs2}
\end{center}\end{figure*}

\begin{figure*}[!ht]\begin{center}
\includegraphics[scale=0.72,trim=0 -50 0 -50]{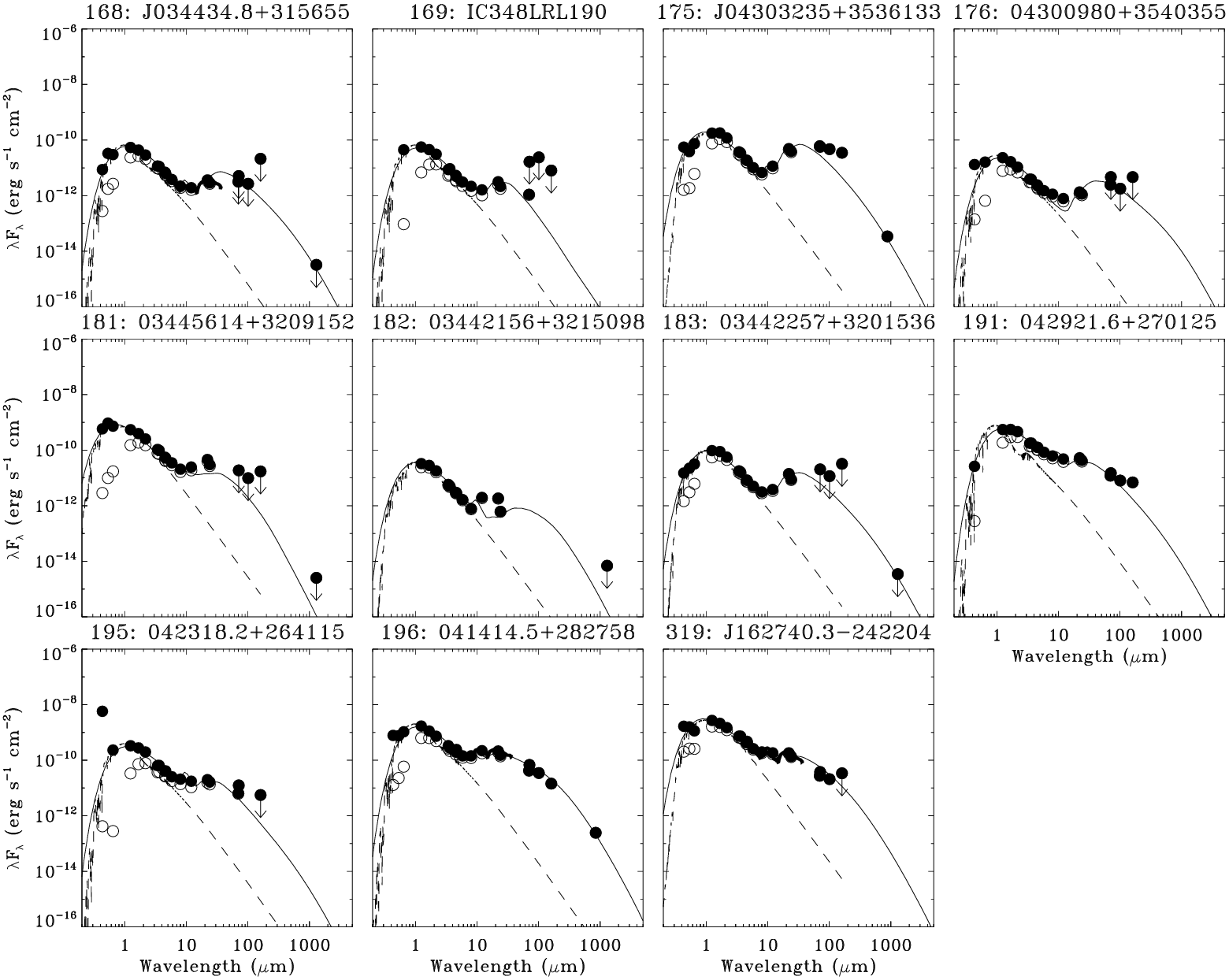}
\caption{SEDs of low-mass disks with small holes.}
\end{center}\end{figure*}

\begin{figure*}[!ht]\begin{center}
\includegraphics[scale=0.72,trim=0 -50 0 -50]{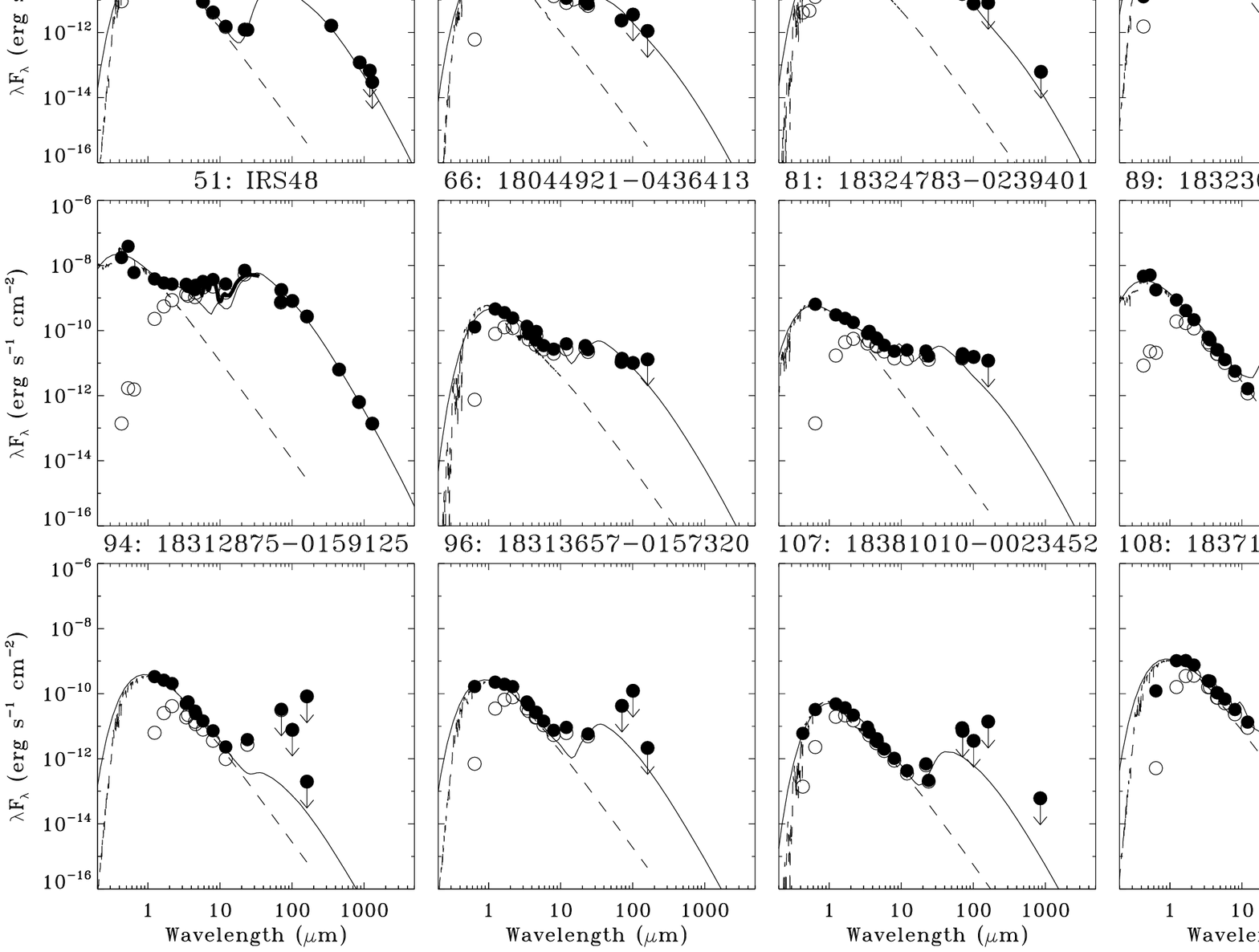}
\captcont{SEDs of low-mass disks with large holes.}
\label{fig:SED3}
\end{center}\end{figure*}

\begin{figure*}[!ht]\begin{center}
\includegraphics[scale=0.72,trim=0 -50 0 -50]{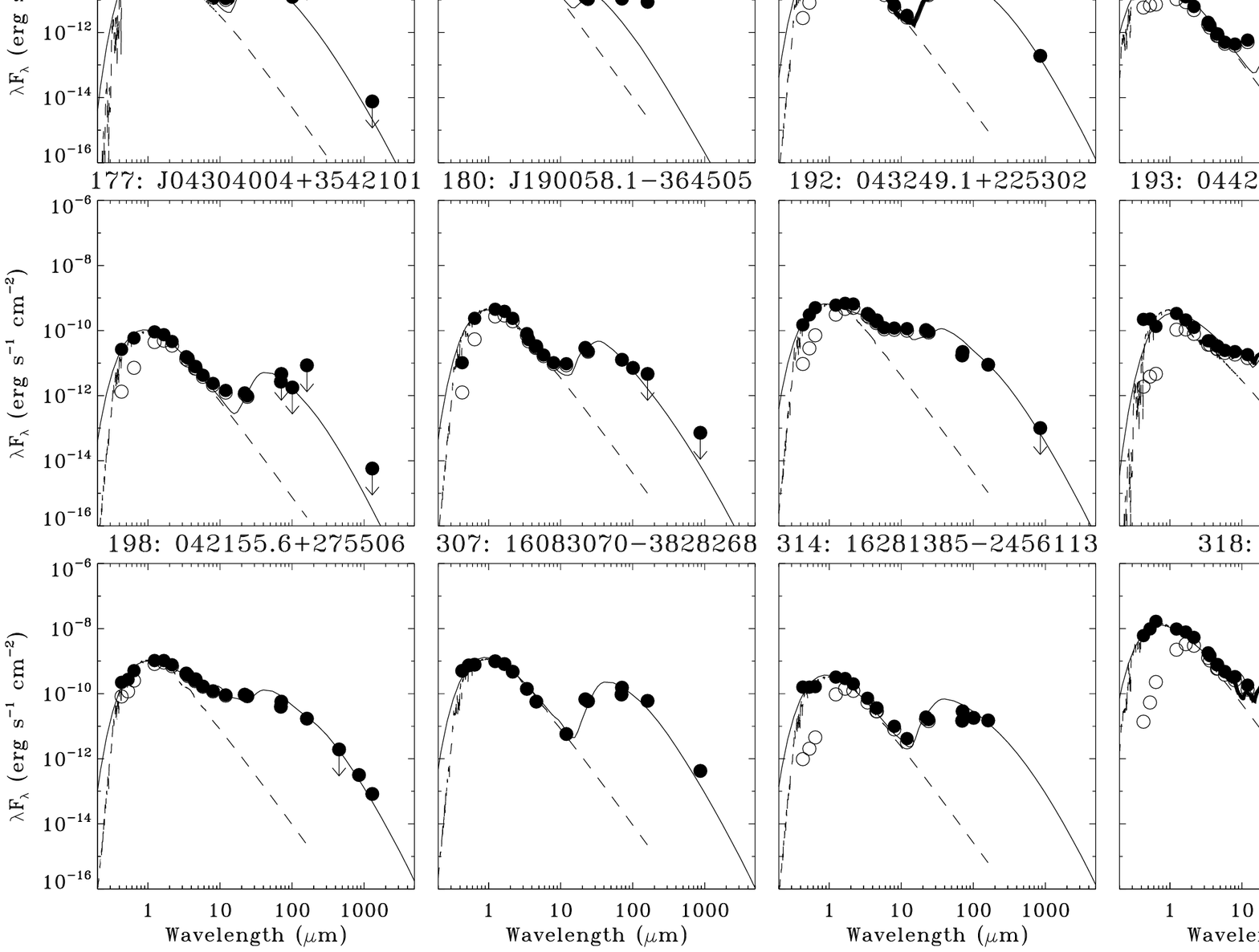}
\captcont{SEDs of low-mass disks with large holes.}
\end{center}\end{figure*}

\begin{figure*}[!ht]\begin{center}
\includegraphics[scale=0.72,trim=0 -50 0 -50]{Figures/Plots_gr2_p2.eps}
\caption{SEDs of low-mass disks with large holes.}
\end{center}\end{figure*}

\begin{figure*}[!ht]\begin{center}
\includegraphics[scale=0.72,trim=0 -50 0 -50]{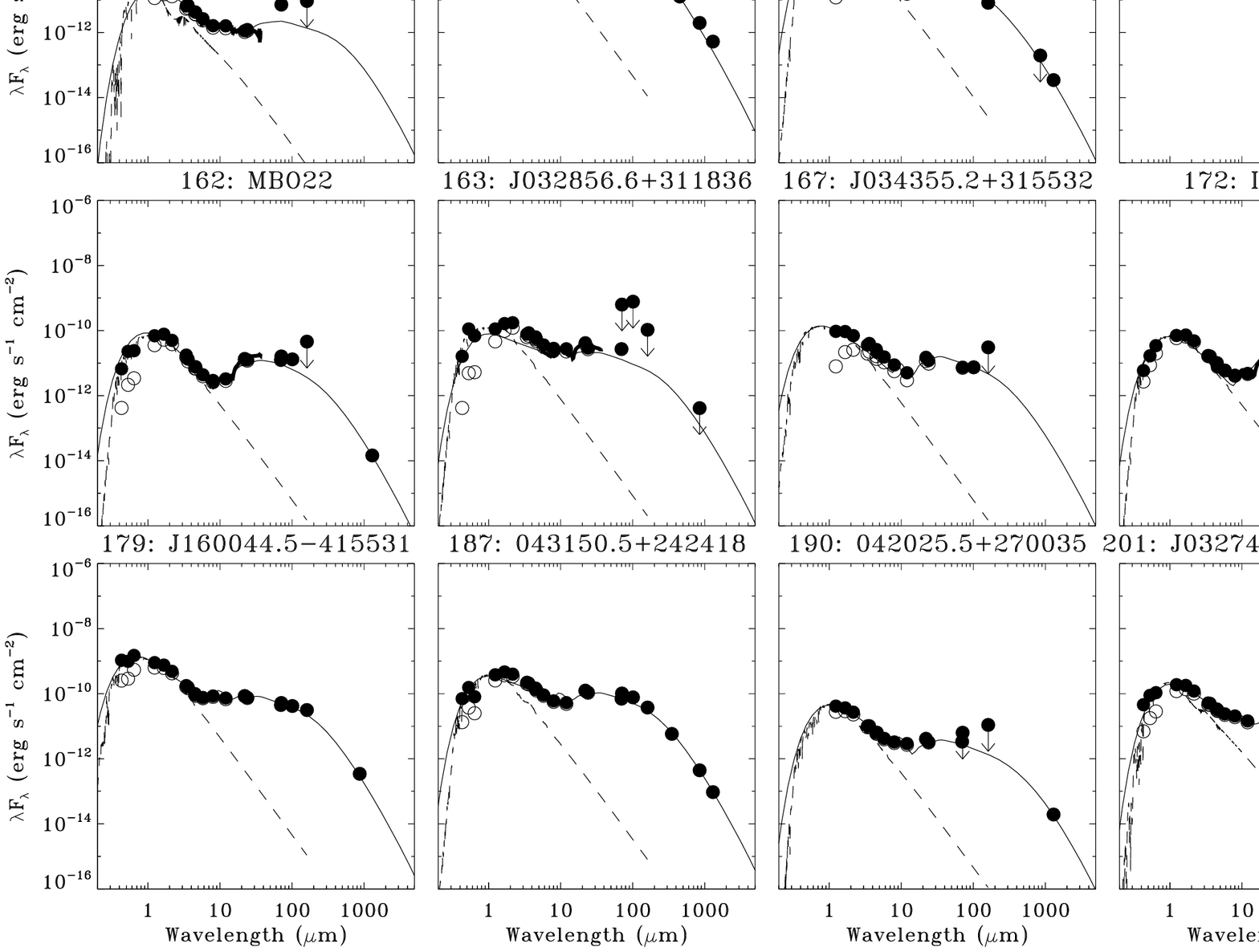}
\captcont{SEDs of massive disks with small holes.}
\label{fig:SEDs4}
\end{center}\end{figure*}

\begin{figure*}[!ht]\begin{center}
\includegraphics[scale=0.72,trim=0 -50 0 -50]{Figures/Plots_gr3_p1.eps}
\caption{SEDs of massive disks with small holes.}
\end{center}\end{figure*}

\begin{figure*}[!ht]\begin{center}
\includegraphics[scale=0.72,trim=0 -50 0 -50]{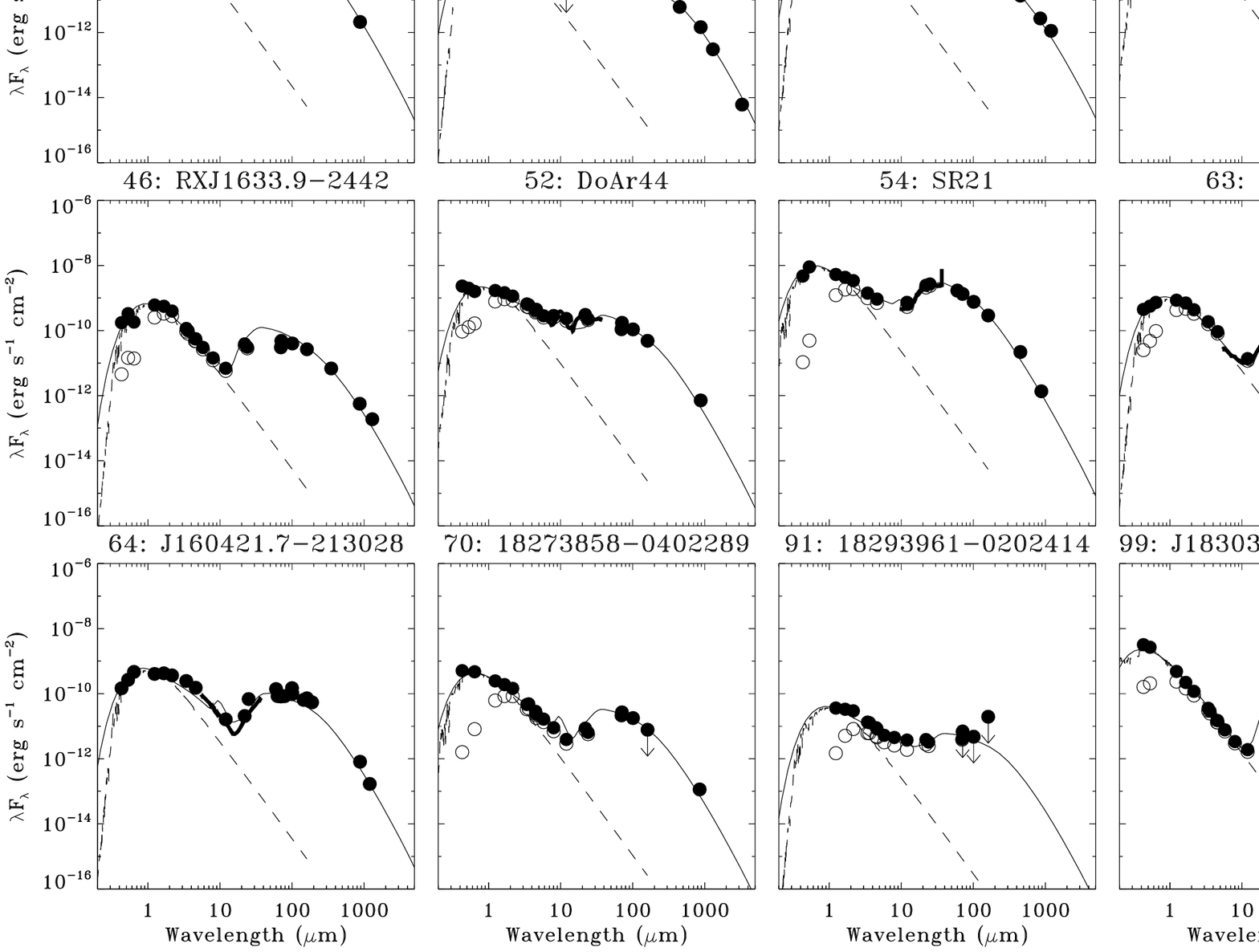}
\captcont{SEDs of massive disks with large holes.}
\label{fig:SEDs5}
\end{center}\end{figure*}

\begin{figure*}[!ht]\begin{center}
\includegraphics[scale=0.72,trim=0 -50 0 -50]{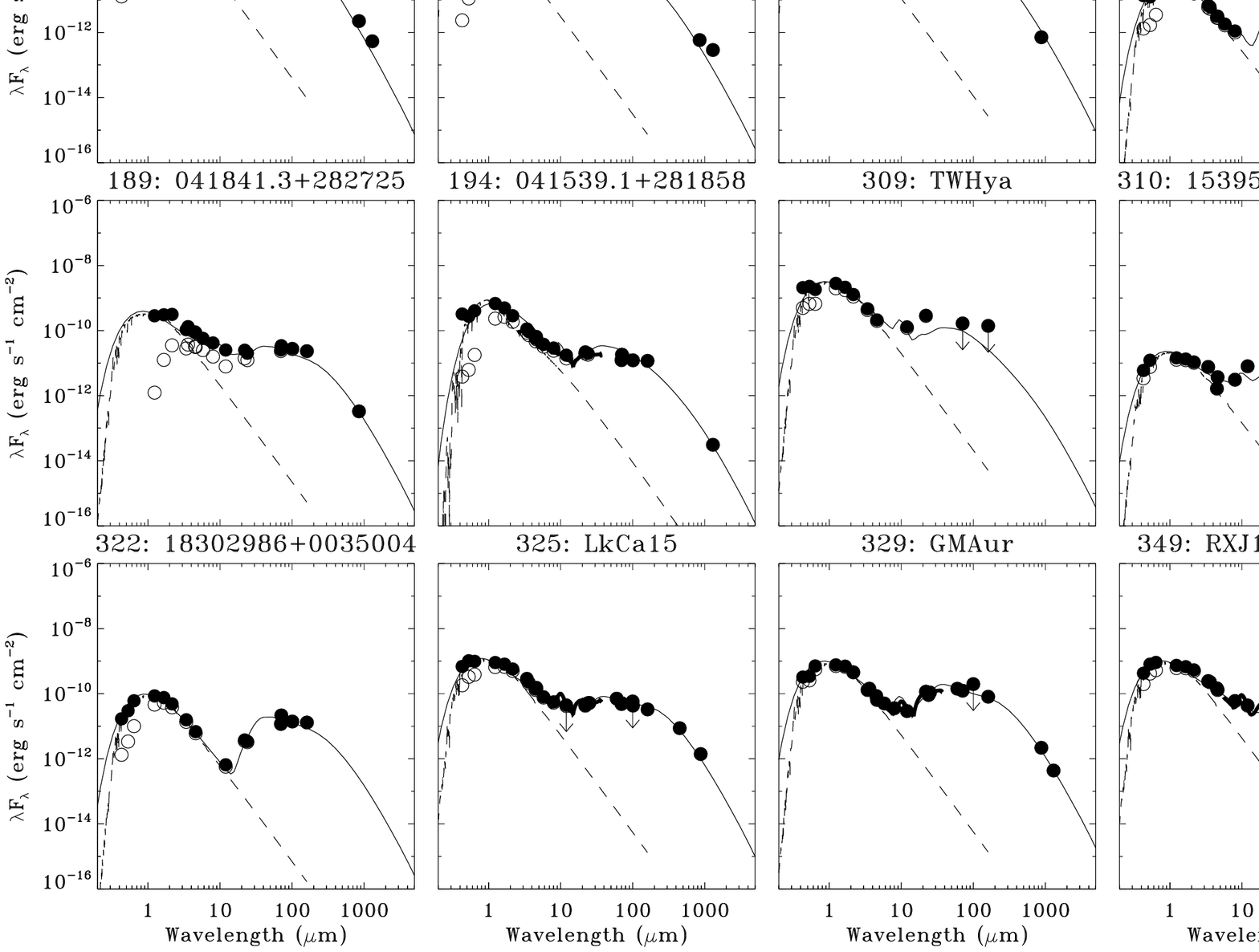}
\captcont{SEDs of massive disks with large holes.}
\end{center}\end{figure*}

\begin{figure*}[!ht]\begin{center}
\includegraphics[scale=0.72,trim=0 -50 0 -50]{Figures/Plots_gr4_p2.eps}
\caption{SEDs of massive disks with large holes.}
\end{center}\end{figure*}

\begin{figure*}[!ht]\begin{center}
\includegraphics[scale=0.72,trim=0 -50 0 -50]{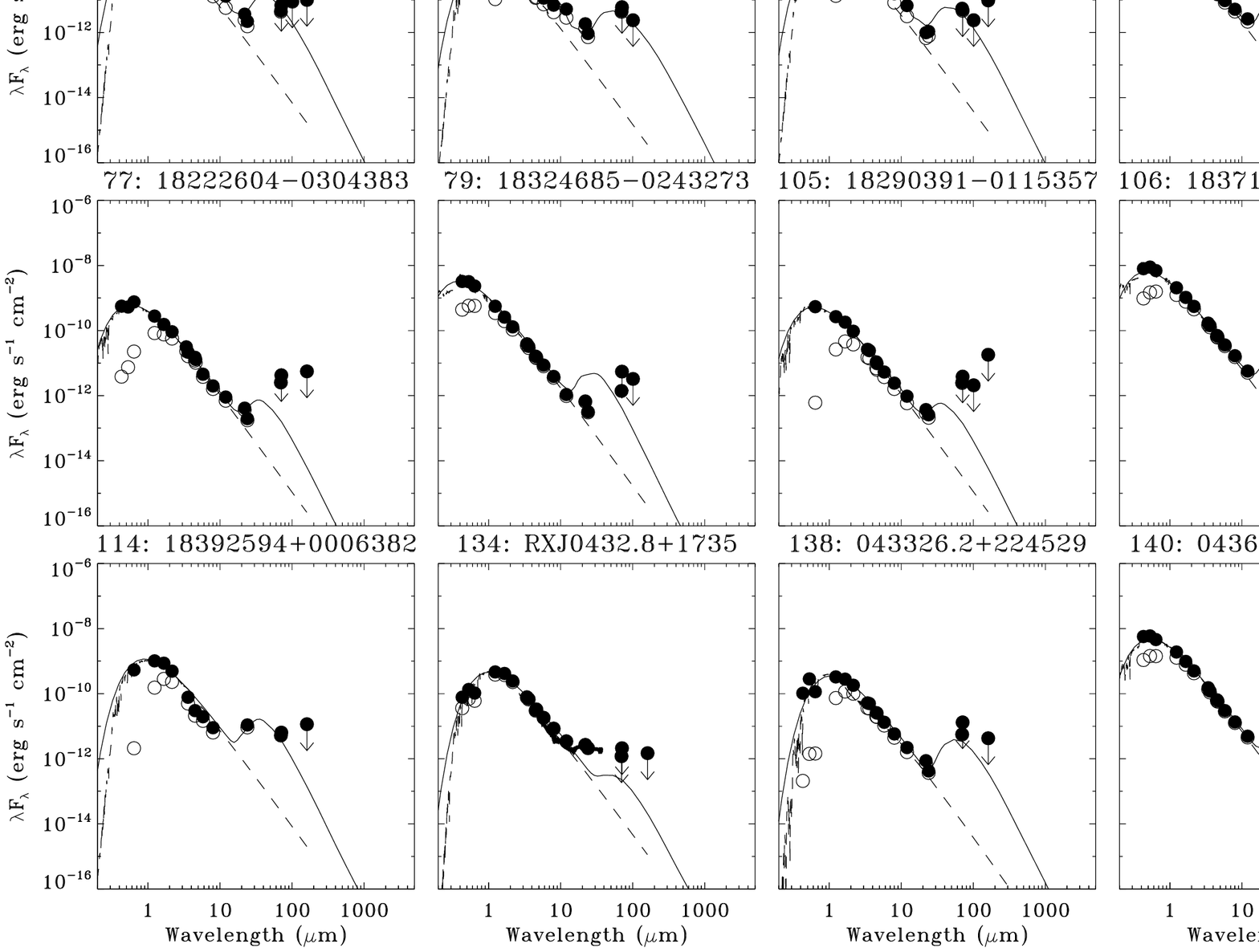}
\captcont{SEDs of low scale height disks.}
\label{fig:SEDs6}
\end{center}\end{figure*}

\begin{figure*}[!ht]\begin{center}
\includegraphics[scale=0.72,trim=0 -50 0 -50]{Figures/Plots_gr5_p1.eps}
\caption{SEDs of low scale height disks.}
\end{center}\end{figure*}

\clearpage
\newpage

\section{Submillimeter photometry}
\begin{longtable}{llllllllll}
\caption{Submillimeter photometry}
\label{tbl:submmphotometry}\\
\hline
\hline
ID&SABOCA&LABOCA&SCUBA&SCUBA&SMA&230 GHz&110 GHz&Ref\\
&[350 $\mu$m]&[870 $\mu$m]&[450 $\mu$m]&[850 $\mu$m]&[880 $\mu$m]&[1.3mm]&[3.3mm]&\\
&(mJy)&(mJy)&(mJy)&(mJy)&(mJy)&(mJy)&(mJy)&\\
\hline
\endfirsthead

\caption{Submillimeter photometry continued}\\
\hline
\hline
ID&SABOCA&LABOCA&SCUBA&SCUBA&SMA&230 GHz&110 GHz&Ref\\
&[350 $\mu$m]&[870 $\mu$m]&[450 $\mu$m]&[850 $\mu$m]&[880 $\mu$m]&[1.3mm]&[3.3mm]&\\
&(mJy)&(mJy)&(mJy)&(mJy)&(mJy)&(mJy)&(mJy)&\\
\hline
\endhead

\hline
\endfoot

1&2400$\pm$200&210$\pm$20&&&&105$\pm$150&7$\pm$1&1,2\\
2&&$<$40&&&&&&1\\
6&690$\pm$180&197$\pm$12.2&&&&&8.8$\pm$1.5&1,3\\
9&&$<$15&&&&$<$118&&1,2\\
10&&$<$20&&&&$<$143&&1,2\\
11&2400$\pm$500&314$\pm$12&&&&77.5$\pm$20.3&2.3$\pm$0.4&1,4\\
12&&&&&&$<$100&&2\\
14&&24$\pm$6&&&&47.8$\pm$15.5&&1,2\\
15&4000$\pm$200&420$\pm$50&&&&124.9$\pm$24&6.2$\pm$1.5&1,2\\
16&9800$\pm$2400&1500$\pm$10&&&&&33.1$\pm$1.2&1,3\\
18&&20$\pm$6&&&&$<$105&&1,2\\
20&1000$\pm$200&390$\pm$10&&&&224$\pm$8&22$\pm$3.3&1,5\\
21&&&&&2700$\pm$270&1190$\pm$30&47$\pm$6&6,7\\
22&$<$300&&&$<$31.4&&&&1\\
23&&&&153.2$\pm$11.5&&49$\pm$4.8&5.7$\pm$0.7&1,4\\
24&&$<$23.2&&&&$<$13&&8,4\\
25&&$<$30&&&&$<$57&&1,9\\
26&&&&$<$30.5&&$<$42&&1,9\\
27&190$\pm$40&35$\pm$3&&&&$<$27&&1,4\\
28&&$<$21&&&&&&8\\
29&&&&30.8$\pm$6.6&&$<$11.4&&1,10\\
32&$<$82&&&$<$36.8&&&&1\\
33&&&4200$\pm$840&&620$\pm$62&&&11,12\\
35&&$<$30&&&&$<$45&&1,4\\
36&$<$220&&&72.6$\pm$18.5&&&&1\\
38&&&919$\pm$184&&430$\pm$43&132$\pm$3.9&6.7$\pm$0.6&13,12\\
39&3154$\pm$419&&2042$\pm$111&770$\pm$39&&451$\pm$20&&14\\
40&$<$700&$<$20&&&&$<$8.4&&1,10\\
41&&&&&&$<$9.6&&10\\
43&&$<$18&&&&&&1\\
44&&&&37.8$\pm$11&&9.3$\pm$3&&1,10\\
45&&&1900$\pm$380&&550$\pm$55&&&13,12\\
46&800$\pm$200&164$\pm$14&&&&81.8$\pm$2.7&&1,10\\
47&&&&$<$50.1&&&&1\\
49&&&&$<$13&&&&10\\
50&&&&&&$<$5.4&&15\\
51&&&950$\pm$200&180$\pm$18&&60$\pm$6&&16,17\\
52&&&&&210$\pm$21&&&12\\
54&&&3300$\pm$660&&400$\pm$40&&&13,12\\
55&$<$900&92$\pm$6&&&&&&1\\
56&&&&$<$59.1&&&&1\\
58&&$<$18&&&&&&1\\
60&&&&$<$24.2&&$<$11.4&&1,10\\
62&$<$600&62$\pm$9&&&&24.5$\pm$3.1&&1,10\\
63&&&&95.2$\pm$16.4&&&&1\\
64&&&&&238$\pm$24&67.5$\pm$1.4&&18\\
70&&&&32.4$\pm$10.7&&&&1\\
82&&&&$<$57.1&&&&1\\
86&&&&$<$107.1&&&&1\\
88&&&&$<$137.3&&&&1\\
89&&&&$<$35.4&&&&1\\
92&&&&$<$79.3&&&&1\\
95&&49$\pm$11&&&&&&1\\
98&&63$\pm$8&&&&&&1\\
99&&360$\pm$30&&&&&&1\\
100&&&&$<$36.8&&&&1\\
107&&&&$<$17.3&&&&1\\
113&&&&$<$18.2&&&&1\\
115&&&&&&$<$22.8&$<$1.9&4,4\\
120&&&&$<$21.1&&&&1\\
124&&&&$<$19.2&&24$\pm$0.6&&1,15\\
127&&&&$<$29.2&&6.3$\pm$0.6&&1,15\\
131&&&&&&2.3$\pm$0.6&&15\\
132&&109$\pm$11&&&&&&1\\
135&&&&&210$\pm$21&&&12\\
136&&&&&150$\pm$15&&&12\\
137&&&&63.2$\pm$18&&31$\pm$2&&1,19\\
140&&&&$<$27&&&&20\\
142&&&&&180$\pm$18&&&12\\
145&&&&&49$\pm$4.9&&&21\\
148&&&$<$516&34$\pm$5&&16$\pm$5&&22\\
149&&&2570$\pm$350&638$\pm$54&&234$\pm$19&&22\\
150&&&&&&$<$3.3&&20\\
152&&&$<$317&$<$9&&$<$5.4&&22\\
153&&&1920$\pm$160&560$\pm$30&&229$\pm$17&&22\\
154&&&&$<$55.8&&15$\pm$1&&1,19\\
155&&&&54.5$\pm$17.7&&&&1\\
156&&&&$<$31.4&&&&1\\
159&&&3820$\pm$570&359$\pm$67&&103$\pm$18&&22\\
161&&&&$<$115.1&&3.7$\pm$0.9&&1,15\\
162&&&&&&6.3$\pm$1.1&&15\\
163&&&&$<$117.4&&&&1\\
165&&&&&&$<$1.4&&15\\
166&&&&166.7$\pm$14&&126$\pm$12&&1,23\\
168&&&&&&$<$1.4&&15\\
171&&&&17.3$\pm$5.7&&&&1\\
172&&&&31$\pm$6&&&&20\\
173&&&&&210$\pm$21&&&12\\
174&&&&$<$26.9&&9.7$\pm$1.5&&1,20\\
175&&&&&10$\pm$2&&&20\\
177&&&&&&$<$2.5&&20\\
178&&&&222.4$\pm$16.3&&&&1\\
179&&100$\pm$5&&&&&&8\\
180&&$<$21&&&&&&8\\
181&&&&&&$<$1.1&&20\\
182&&&&&&$<$3&&20\\
183&&&&&&$<$1.5&&20\\
184&&&&&&$<$2.8&&20\\
185&&&&&&$<$27&&9\\
187&680$\pm$114&&&125.5$\pm$18.3&&41$\pm$5&&22,1\\
188&&&$<$442&32$\pm$8&&$<$36&&22\\
189&&&&93.4$\pm$15.7&&&&1\\
190&&&&&&8.4$\pm$1.4&&19\\
192&&&&$<$28.6&&&&1\\
193&&&&31.9$\pm$9.4&&&&1\\
194&&&&&&13.4$\pm$1.4&&19\\
196&&&&69.3$\pm$18.8&&&&1\\
197&&&$<$456&130$\pm$7&&47$\pm$0.7&&22\\
198&&&$<$291&90$\pm$7&&36$\pm$5&&22\\
200&&$<$40&&&&26$\pm$9&&1,24\\
202&&&&$<$55.5&&&&1\\
203&&55$\pm$10&&&&&&1\\
303&&136$\pm$7&&&&&&1\\
307&&123$\pm$14&&&&&&1\\
316&&98$\pm$13&&&&&&1\\
321&&$<$20&&&&&&1\\
325&&&1310$\pm$260&&410$\pm$41&&&13,12\\
329&&&&&640$\pm$64&189$\pm$15&&12,25\\
333&&&&$<$56.1&&&&1\\
349&&&&&&49$\pm$9&&26\\
350&&&&&&60$\pm$8&&26\\
\end{longtable}
{\bf Refs.}
       1) This work.
       2) \citet{Henning1993}, 
       3) \citet{Lommen2007}, 
       4) \citet{Lommen2010}, 
       5) \citet{Graefe2013}, 
       6) \citet{Fukagawa2013}, 
       7) \citet{Verhoeff2011}, 
       8) \citet{Romero2012}, 
       9) \citet{Merin2008}, 
      10) \citet{Cieza2010}, 
      11) \citet{Perez2014}, 
      12) \citet{Andrews2011}, 
      13) \citet{vanderMarel2015-12co}, 
      14) \citet{Jensen1996}, 
      15) \citet{Merin2010}, 
      16) \citet{vanderMarel2013}, 
      17) \citet{Brown2012a}, 
      18) \citet{Mathews2012}, 
      19) \citet{Andrews2013}, 
      20) \citet{Cieza2012}, 
      21) \citet{AndrewsWilliams2007tau}, 
      22) \citet{AndrewsWilliams2005}, 
      23) \citet{Enoch2006}, 
      24) \citet{Nuernberger1997}, 
      25) \citet{Isella2009}, 
      26) \citet{Hughes2010}

\section{Herschel photometry}
This section presents the fluxes and cut out maps of the Herschel PACS photometry.

\begin{longtable}{llll}
\caption{Herschel photometry}
\label{tbl:herschelphotometry}\\
\hline
\hline
ID&PACS 70&PACS 100&PACS 160\\
&(mJy)&(mJy)&(mJy)\\
\hline
\endfirsthead

\caption{Herschel photometry continued}\\
\hline
\hline
ID&PACS 70&PACS 100&PACS 160\\
&(mJy)&(mJy)&(mJy)\\
\hline
\endhead

\hline
\endfoot
1	&	4.98$\pm$0.5	&	3.74$\pm$0.37	&	3.17$\pm$0.32	\\
2	&	0.22$\pm$0.03	&	0.20$\pm$0.04	&	$<$0.27	\\
3	&	0.11$\pm$0.03	&	0.14$\pm$0.03	&	$<$0.26	\\
4	&	$<$0.10	&	$<$0.07\tablefootmark{a}	&	$<$0.41\tablefootmark{a}	\\
5	&	0.18$\pm$0.05	&	0.17$\pm$0.03	&	$<$0.33\tablefootmark{a}	\\
6	&	3.11$\pm$0.31	&	2.90$\pm$0.29	&	2.15$\pm$0.25	\\
7	&	0.21$\pm$0.04	&	0.21$\pm$0.03	&	$<$0.31	\\
9	&	$<$0.65\tablefootmark{a}	&	$<$0.71\tablefootmark{a}	&	$<$1.06\tablefootmark{a}	\\
10	&	$<$1.20\tablefootmark{a}	&	$<$1.34\tablefootmark{a}	&	$<$4.21\tablefootmark{a}	\\
11	&	3.86$\pm$0.39	&	3.80$\pm$0.38	&	3.65$\pm$0.37	\\
12	&	0.44$\pm$0.05	&	0.40$\pm$0.05	&	$<$0.39	\\
13	&	$<$0.11	&	0.14$\pm$0.03	&	$<$0.55\tablefootmark{a}	\\
14	&	0.69$\pm$0.08	&	0.55$\pm$0.06	&	0.41$\pm$0.07	\\
15	&	1.58$\pm$0.16	&	2.31$\pm$0.23	&	2.80$\pm$0.28	\\
16	&	26.06$\pm$2.92	&	36.06$\pm$3.9	&	38.45$\pm$6.0	\\
17	&	0.21$\pm$0.05	&	0.25$\pm$0.03	&	0.30$\pm$0.09	\\
18	&	0.56$\pm$0.07	&	-	&	0.45$\pm$0.1	\\
21	&	117.78$\pm$11.78	&	100.65$\pm$10.06	&	67.82$\pm$6.78	\\
23	&	1.38$\pm$0.14	&	-	&	1.82$\pm$0.2	\\
24	&	0.17$\pm$0.04	&	0.11$\pm$0.03	&	$<$0.09\tablefootmark{a}	\\
25	&	$<$0.34	&	$<$0.32	&	$<$0.38\tablefootmark{a}	\\
26	&	$<$0.26	&	$<$0.12\tablefootmark{a}	&	$<$0.02\tablefootmark{a}	\\
27	&	0.61$\pm$0.07	&	0.80$\pm$0.08	&	0.96$\pm$0.17	\\
28	&	$<$0.11	&	$<$0.15	&	$<$-0.11\tablefootmark{a}	\\
32	&	0.73$\pm$0.08	&	$<$1.61\tablefootmark{a}	&	$<$2.98\tablefootmark{a}	\\
33	&	29.8$\pm$2.98	&	28.3$\pm$2.83	&	21.13$\pm$2.11	\\
34	&	$<$2.09	&	$<$0.12	&	$<$0.06\tablefootmark{a}	\\
36	&	$<$1.44	&	$<$1.09\tablefootmark{a}	&	$<$3.36\tablefootmark{a}	\\
40	&	1.89$\pm$0.2	&	1.18$\pm$0.15	&	$<$3.65	\\
41	&	$<$0.05\tablefootmark{a}	&	$<$0.33\tablefootmark{a}	&	$<$2.70\tablefootmark{a}	\\
43	&	0.38$\pm$0.05	&	0.26$\pm$0.05	&	$<$0.44\tablefootmark{a}	\\
44	&	0.54$\pm$0.06	&	0.74$\pm$0.09	&	$<$2.02\tablefootmark{a}	\\
45	&	10.58$\pm$1.06	&	9.74$\pm$0.98	&	8.96$\pm$0.96	\\
46	&	1.14$\pm$0.12	&	1.35$\pm$0.14	&	1.43$\pm$0.27	\\
47	&	1.00$\pm$0.22	&	1.15$\pm$0.37	&	$<$0.49\tablefootmark{a}	\\
48	&	0.86$\pm$0.09	&	1.16$\pm$0.12	&	$<$1.41	\\
49	&	0.26$\pm$0.04	&	$<$0.14	&	$<$0.57\tablefootmark{a}	\\
50	&	0.23$\pm$0.05	&	-	&	$<$-0.06\tablefootmark{a}	\\
51	&	41.14$\pm$4.12	&	27.3$\pm$2.77	&	14.53$\pm$4.18	\\
52	&	4.10$\pm$0.41	&	3.66$\pm$0.37	&	2.60$\pm$0.5	\\
53	&	$<$0.14	&	$<$0.12\tablefootmark{a}	&	$<$0.25\tablefootmark{a}	\\
54	&	31.58$\pm$3.16	&	25.75$\pm$2.58	&	15.61$\pm$1.88	\\
55	&	2.26$\pm$0.23	&	2.15$\pm$0.22	&	2.06$\pm$0.35	\\
56	&	1.26$\pm$0.39	&	$<$2.78	&	$<$6.05\tablefootmark{a}	\\
58	&	0.34$\pm$0.05	&	$<$0.28	&	$<$0.59\tablefootmark{a}	\\
59	&	$<$0.07	&	$<$0.13	&	$<$-0.06\tablefootmark{a}	\\
60	&	$<$0.21\tablefootmark{a}	&	$<$0.32	&	$<$0.73\tablefootmark{a}	\\
61	&	0.12$\pm$0.04	&	$<$0.08	&	$<$0.56	\\
62	&	1.32$\pm$0.14	&	-	&	1.62$\pm$0.31	\\
63	&	0.97$\pm$0.11	&	0.90$\pm$0.1	&	1.06$\pm$0.26	\\
64	&	-	&	3.48$\pm$0.35	&	3.88$\pm$0.84	\\
65	&	$<$0.10	&	-	&	$<$0.27	\\
66	&	0.32$\pm$0.05	&	0.34$\pm$0.04	&	$<$0.70\tablefootmark{a}	\\
67	&	$<$0.12	&	$<$0.07	&	$<$0.30	\\
68	&	$<$0.13	&	$<$0.08	&	$<$0.34	\\
69	&	$<$0.19	&	$<$0.22	&	$<$0.49	\\
70	&	0.63$\pm$0.07	&	0.60$\pm$0.06	&	$<$0.42	\\
71	&	$<$0.14	&	$<$0.11	&	$<$0.44	\\
73	&	$<$0.16	&	$<$0.30	&	$<$0.54\tablefootmark{a}	\\
74	&	$<$0.14	&	$<$0.08	&	$<$-0.02\tablefootmark{a}	\\
75	&	$<$0.11	&	$<$0.08	&	$<$0.52	\\
76	&	0.23$\pm$0.04	&	0.18$\pm$0.04	&	$<$0.42	\\
77	&	$<$0.10	&	-	&	$<$0.30\tablefootmark{a}	\\
78	&	$<$0.14	&	$<$0.10	&	$<$0.88	\\
79	&	$<$0.13	&	$<$0.11	&	$<$-0.08\tablefootmark{a}	\\
80	&	$<$0.18	&	$<$0.24	&	$<$0.59	\\
81	&	0.44$\pm$0.06	&	0.52$\pm$0.06	&	$<$0.64	\\
82	&	$<$0.33\tablefootmark{a}	&	$<$0.10\tablefootmark{a}	&	$<$2.61\tablefootmark{a}	\\
83	&	$<$0.12	&	$<$0.09	&	$<$0.64	\\
84	&	$<$0.12	&	$<$0.07	&	$<$0.42	\\
85	&	$<$0.27	&	$<$0.53	&	$<$0.05\tablefootmark{a}	\\
86	&	$<$1.05\tablefootmark{a}	&	$<$1.40\tablefootmark{a}	&	$<$1.55\tablefootmark{a}	\\
88	&	$<$32.9\tablefootmark{a}	&	$<$26.25\tablefootmark{a}	&	$<$59.51\tablefootmark{a}	\\
89	&	0.66$\pm$0.09	&	$<$0.74\tablefootmark{a}	&	$<$0.59\tablefootmark{a}	\\
90	&	$<$0.22	&	$<$0.22	&	$<$1.98	\\
91	&	$<$0.16	&	$<$0.16	&	$<$1.05	\\
92	&	$<$0.31\tablefootmark{a}	&	$<$0.47\tablefootmark{a}	&	$<$2.55\tablefootmark{a}	\\
93	&	$<$8.45\tablefootmark{a}	&	$<$5.27\tablefootmark{a}	&	$<$10.49\tablefootmark{a}	\\
94	&	$<$0.74\tablefootmark{a}	&	$<$0.26\tablefootmark{a}	&	$<$4.44\tablefootmark{a}	\\
95	&	$<$2.27\tablefootmark{a}	&	$<$2.78\tablefootmark{a}	&	$<$9.57\tablefootmark{a}	\\
96	&	$<$1.00\tablefootmark{a}	&	$<$4.13\tablefootmark{a}	&	$<$-4.57\tablefootmark{a}	\\
97	&	$<$2.49\tablefootmark{a}	&	$<$6.83\tablefootmark{a}	&	$<$8.58\tablefootmark{a}	\\
98	&	$<$1.46\tablefootmark{a}	&	$<$3.07\tablefootmark{a}	&	$<$10.72\tablefootmark{a}	\\
99	&	3.13$\pm$0.33	&	$<$3.84\tablefootmark{a}	&	$<$4.62\tablefootmark{a}	\\
100	&	0.56$\pm$0.07	&	0.55$\pm$0.06	&	0.56$\pm$0.16	\\
101	&	$<$0.13	&	0.20$\pm$0.05	&	$<$27.08	\\
102	&	0.13$\pm$0.04	&	0.31$\pm$0.04	&	$<$1.05	\\
103	&	0.27$\pm$0.06	&	0.28$\pm$0.08	&	$<$6.29	\\
104	&	$<$0.14	&	$<$0.07	&	$<$0.22	\\
105	&	$<$0.09	&	$<$0.07	&	$<$0.97	\\
106	&	$<$0.12	&	-	&	$<$0.66	\\
107	&	$<$0.17	&	$<$0.12	&	$<$0.75	\\
108	&	$<$0.14	&	-	&	$<$0.41\tablefootmark{a}	\\
110	&	$<$0.14	&	-	&	$<$0.54	\\
111	&	0.28$\pm$0.06	&	-	&	$<$1.93\tablefootmark{a}	\\
112	&	$<$0.11	&	-	&	$<$0.64	\\
113	&	$<$0.13	&	-	&	$<$0.38	\\
114	&	0.15$\pm$0.04	&	-	&	$<$0.62	\\
115	&	0.40$\pm$0.06	&	0.29$\pm$0.08	&	$<$0.84	\\
116	&	$<$0.22	&	-	&	$<$1.02	\\
117	&	$<$0.13	&	-	&	$<$0.64	\\
118	&	$<$0.11	&	-	&	$<$0.76	\\
119	&	$<$0.13	&	-	&	$<$0.02\tablefootmark{a}	\\
120	&	$<$0.13	&	$<$0.08	&	$<$2.04	\\
121	&	$<$0.59\tablefootmark{a}	&	-	&	$<$0.45\tablefootmark{a}	\\
122	&	$<$0.10	&	-	&	$<$0.07\tablefootmark{a}	\\
123	&	0.13$\pm$0.04	&	$<$0.07	&	$<$0.38	\\
124	&	$<$0.11	&	$<$0.07	&	$<$0.55	\\
125	&	1.42$\pm$0.15	&	-	&	$<$1.14	\\
126	&	$<$16.95	&	$<$33.54	&	$<$3.04\tablefootmark{a}	\\
127	&	0.34$\pm$0.05	&	0.18$\pm$0.03	&	$<$8.11	\\
128	&	$<$0.12	&	$<$0.10	&	$<$1.26	\\
129	&	$<$0.12	&	-	&	$<$0.10\tablefootmark{a}	\\
130	&	$<$0.15	&	-	&	$<$0.55	\\
131	&	$<$0.12	&	0.16$\pm$0.03	&	$<$0.46	\\
132	&	3.03$\pm$0.31	&	2.64$\pm$0.27	&	2.41$\pm$0.28	\\
133	&	0.44$\pm$0.06	&	0.46$\pm$0.07	&	$<$0.94	\\
134	&	$<$0.05	&	-	&	$<$0.08	\\
135	&	0.77$\pm$0.08	&	0.84$\pm$0.09	&	0.85$\pm$0.17	\\
136	&	3.21$\pm$0.32	&	-	&	2.68$\pm$0.28	\\
137	&	0.29$\pm$0.04	&	-	&	0.42$\pm$0.09	\\
138	&	$<$0.31	&	-	&	$<$0.23\tablefootmark{a}	\\
139	&	0.46$\pm$0.07	&	-	&	$<$0.37	\\
140	&	0.18$\pm$0.05	&	$<$0.18\tablefootmark{a}	&	$<$0.41	\\
142	&	20.39$\pm$2.04	&	17.59$\pm$1.76	&	12.65$\pm$2.57	\\
145	&	0.44$\pm$0.06	&	0.45$\pm$0.05	&	0.35$\pm$0.09	\\
146	&	$<$0.13	&	-	&	$<$0.25	\\
147	&	0.17$\pm$0.06	&	-	&	$<$0.50	\\
148	&	0.48$\pm$0.08	&	-	&	$<$0.43	\\
149	&	53.78$\pm$5.39	&	39.92$\pm$4.05	&	22.87$\pm$3.33	\\
150	&	0.49$\pm$0.07	&	0.42$\pm$0.08	&	$<$3.51	\\
151	&	0.26$\pm$0.05	&	-	&	0.47$\pm$0.12	\\
152	&	$<$0.16	&	$<$0.08	&	$<$0.25	\\
153	&	13.82$\pm$1.38	&	-	&	8.36$\pm$0.86	\\
154	&	0.66$\pm$0.09	&	0.64$\pm$0.07	&	0.44$\pm$0.11	\\
155	&	1.26$\pm$0.14	&	1.34$\pm$0.14	&	1.11$\pm$0.13	\\
156	&	$<$0.15	&	-	&	$<$0.36	\\
157	&	0.21$\pm$0.05	&	$<$0.10	&	$<$0.37	\\
158	&	$<$0.12	&	$<$0.13	&	$<$0.23	\\
159	&	131.87$\pm$13.19	&	-	&	67.75$\pm$6.79	\\
160	&	0.30$\pm$0.05	&	0.20$\pm$0.03	&	$<$0.64	\\
161	&	$<$23.85	&	$<$23.35	&	$<$0.67\tablefootmark{a}	\\
162	&	0.38$\pm$0.08	&	0.44$\pm$0.09	&	$<$2.46	\\
163	&	$<$14.92	&	$<$26.05	&	$<$5.67\tablefootmark{a}	\\
164	&	$<$0.13	&	$<$0.14	&	$<$0.85	\\
165	&	$<$0.14	&	$<$0.18	&	$<$1.01	\\
166	&	3.20$\pm$0.32	&	3.15$\pm$0.32	&	2.28$\pm$0.43	\\
167	&	0.17$\pm$0.05	&	0.25$\pm$0.05	&	$<$1.64	\\
168	&	$<$0.12	&	$<$0.09	&	$<$1.14	\\
169	&	$<$0.39	&	$<$0.80	&	$<$0.43\tablefootmark{a}	\\
171	&	0.40$\pm$0.06	&	0.34$\pm$0.07	&	$<$2.01	\\
172	&	0.37$\pm$0.06	&	0.32$\pm$0.05	&	$<$0.77	\\
173	&	11.41$\pm$1.14	&	10.25$\pm$1.03	&	7.38$\pm$0.74	\\
174	&	$<$0.58\tablefootmark{a}	&	$<$1.03\tablefootmark{a}	&	$<$3.80\tablefootmark{a}	\\
175	&	1.35$\pm$0.14	&	1.58$\pm$0.16	&	1.87$\pm$0.27	\\
176	&	$<$0.11	&	$<$0.06	&	$<$0.25	\\
177	&	$<$0.11	&	$<$0.06	&	$<$0.46	\\
178	&	2.47$\pm$0.36	&	4.41$\pm$0.76	&	10.19$\pm$1.65	\\
179	&	1.23$\pm$0.13	&	1.41$\pm$0.14	&	1.69$\pm$0.19	\\
180	&	0.30$\pm$0.04	&	0.24$\pm$0.03	&	$<$0.25	\\
181	&	$<$0.44\tablefootmark{a}	&	$<$0.33\tablefootmark{a}	&	$<$0.92\tablefootmark{a}	\\
182	&	$<$-0.07\tablefootmark{a}	&	$<$-0.03\tablefootmark{a}	&	$<$-0.72\tablefootmark{a}	\\
183	&	$<$0.48	&	$<$0.39\tablefootmark{a}	&	$<$1.73\tablefootmark{a}	\\
185	&	0.21$\pm$0.06	&	0.24$\pm$0.05	&	$<$0.91	\\
187	&	2.41$\pm$0.24	&	2.60$\pm$0.26	&	2.02$\pm$0.22	\\
188	&	0.48$\pm$0.1	&	0.33$\pm$0.11	&	0.32$\pm$0.08	\\
189	&	0.75$\pm$0.09	&	0.91$\pm$0.09	&	1.25$\pm$0.16	\\
190	&	$<$0.15	&	-	&	$<$0.59	\\
191	&	0.35$\pm$0.07	&	0.27$\pm$0.04	&	0.37$\pm$0.11	\\
192	&	0.52$\pm$0.09	&	-	&	0.48$\pm$0.16	\\
193	&	0.33$\pm$0.05	&	0.35$\pm$0.04	&	$<$0.57	\\
194	&	0.43$\pm$0.06	&	0.41$\pm$0.05	&	0.63$\pm$0.18	\\
195	&	0.29$\pm$0.06	&	-	&	$<$0.30	\\
196	&	1.60$\pm$0.17	&	1.16$\pm$0.12	&	0.77$\pm$0.1	\\
197	&	0.60$\pm$0.08	&	-	&	0.64$\pm$0.15	\\
198	&	1.35$\pm$0.15	&	-	&	0.92$\pm$0.14	\\
200	&	0.48$\pm$0.06	&	0.37$\pm$0.05	&	0.47$\pm$0.15	\\
201	&	0.22$\pm$0.05	&	0.21$\pm$0.03	&	$<$0.26	\\
202	&	1.73$\pm$0.3	&	$<$1.95	&	$<$1.53\tablefootmark{a}	\\
203	&	2.39$\pm$0.25	&	-	&	3.30$\pm$0.38	\\
204	&	$<$73.51	&	$<$60.36	&	$<$9.49\tablefootmark{a}	\\
301	&	0.26$\pm$0.05	&	0.24$\pm$0.05	&	$<$0.73	\\
303	&	1.63$\pm$0.17	&	2.88$\pm$0.29	&	3.90$\pm$0.4	\\
307	&	3.63$\pm$0.37	&	-	&	3.24$\pm$0.33	\\
309	&	$<$3.93\tablefootmark{a}	&	-	&	$<$7.50\tablefootmark{a}	\\
310	&	0.62$\pm$0.07	&	-	&	1.55$\pm$0.18	\\
314	&	0.68$\pm$0.07	&	0.61$\pm$0.08	&	0.81$\pm$0.26	\\
317	&	1.08$\pm$0.11	&	0.69$\pm$0.07	&	$<$0.38	\\
318	&	$<$21.71\tablefootmark{a}	&	$<$17.06\tablefootmark{a}	&	$<$16.7\tablefootmark{a}	\\
319	&	0.89$\pm$0.1	&	0.70$\pm$0.12	&	$<$1.82\tablefootmark{a}	\\
321	&	1.02$\pm$0.11	&	$<$1.72\tablefootmark{a}	&	$<$2.26\tablefootmark{a}	\\
322	&	0.51$\pm$0.07	&	0.47$\pm$0.05	&	0.70$\pm$0.14	\\
325	&	1.20$\pm$0.13	&	1.45$\pm$0.15	&	1.77$\pm$0.38	\\
326	&	1.03$\pm$0.11	&	0.92$\pm$0.09	&	1.36$\pm$0.35	\\
329	&	2.93$\pm$0.3	&	-	&	4.34$\pm$0.44	\\
334	&	$<$0.10	&	-	&	0.28$\pm$0.09	\\
335	&	0.49$\pm$0.15	&	$<$0.99	&	$<$2.39	\\
336	&	$<$0.12	&	0.13$\pm$0.02	&	$<$0.18	\\
				
\end{longtable}				
\tablefoot{					
\tablefoottext{a}{Upper limit due to cloud confusion.}					
}

\end{document}